\documentclass[12pt]{spieman}  
\usepackage{amsmath}
\usepackage{amsfonts}
\usepackage{amssymb}
\usepackage{graphicx}
\usepackage{setspace}
\usepackage{tocloft}

\usepackage{amsmath, amssymb, color}

\usepackage{multirow,booktabs,colortbl,hhline}
\usepackage{subfig}
\usepackage{siunitx}
\usepackage{hyperref}

\newcommand{\HII}{H{\sc ~ii}}
\newcommand{\lya}{Ly$\alpha$}

\newcommand{\hst}{{\em HST}\/}

\newcommand{\jwst}{{\em JWST}\/}

\newcommand{\fuse}{{\em FUSE}\/}
\newcommand{\GALEX}{{\em GALEX}\/}

\definecolor{mblue}{rgb}{0.3,0.4,1}

\def\rvir{\ifmmode {R_{\rm vir}} \else $R_{\rm vir}$\fi}
\def\phflux{\ifmmode {\rm cm^{-2}~s^{-1}~sr^{-1}} \else ${\rm cm^{-2}~s^{-1}~sr^{-1}}$\fi}

\title{Ultraviolet Technology To Prepare For The Habitable Worlds Observatory}

\author[a,*]{Sarah Tuttle (co-chair)}
\author[b]{Mark Matsumura (co-chair)}
\author[c]{David R. Ardila}
\author[c]{Pin Chen}
\author[d]{Michael Davis}
\author[d]{Camden Ertley}
\author[e]{Emily Farr}
\author[e]{Brian Fleming}
\author[e]{Kevin France}
\author[d]{Cynthia Froning}
\author[f]{Fabien Gris\'e}
\author[g]{Erika Hamden}
\author[c]{John Hennessy}
\author[h]{Keri Hoadley}
\author[i]{Stephan R. McCandliss}
\author[c]{Drew M. Miles}
\author[c, j]{Shouleh Nikzad}
\author[b]{Manuel Quijada}
\author[i]{Isu Ravi}
\author[k]{Luis Rodriguez de Marcos}
\author[b]{Paul Scowen}
\author[l]{Oswald Siegmund}
\author[g]{Carlos J. Vargas}
\author[e]{Dmitry Vorobiev}
\author[e]{Emily M. Witt}
\author[j]{Ultraviolet Technology Working Group}

\affil[a]{University of Washington, Department of Astronomy, Seattle, WA, USA}
\affil[b]{Goddard Space Flight Center, Greenbelt, MD, USA}
\affil[c]{Jet Propulsion Laboratory, California Institute of Technology}
\affil[d]{Southwest Research Institute, San Antonio, TX, USA}
\affil[e]{University of Colorado, Boulder, CO, USA}
\affil[f]{The Pennsylvania State University, Department of Astronomy \& Astrophysics, University Park, PA, USA}
\affil[g]{Steward Observatory, University of Arizona, Tucson, AZ, USA}
\affil[h]{The University of Iowa, Iowa City, IA, USA}
\affil[i]{Johns Hopkins University, Department of Physics \& Astronomy, Center for Astrophysical Sciences}
\affil[j]{NASA Cosmic Origins Program Analysis Group}
\affil[k]{The Catholic University of America, Washington DC, USA}
\affil[l]{University of California - Berkeley, Berkeley, CA, USA}

\cftpagenumbersoff{figure}
\cftpagenumbersoff{table}

\begin{document} 
\maketitle

\begin{abstract}
We present here the current state of a collection of promising ultraviolet technologies in preparation for the Habitable Worlds Observatory. Working with experts representing a significant number of groups working in the ultraviolet, we summarize some of the leading science drivers, present an argument for a 100 nm blue wavelength cutoff, and gather current state of the art of UV technologies. We present the state of the art of contamination control, a crucial piece of the UV instrument plan. We explore next steps with individual technologies, as well as present paths forward with systems level testing and development. 
\end{abstract}


{\noindent \footnotesize\textbf{*}Sarah Tuttle,  \linkable{tuttlese@uw.edu} }

\newpage
\tableofcontents
\newpage

\section{Executive Summary}\label{ExSum}
The astronomical community has long been preparing for the next-generation UV flagship, recently revealed to be the Habitable Worlds Observatory (HWO). The main HWO science drivers, ``cosmic ecosystems" and ``worlds and suns in context" as defined in the Astro2020 decadal review, require robust technological performance in the ultraviolet bandpasses accessible via normal incidence optics (100 -- 320 nm)\footnote{The ultraviolet includes the EUV, those wavelengths below 100 nm that are accessed with grazing incidence optics. We do not address those technologies or wavelength regimes here}. In this report, we summarize the state of the art of UV component technologies and systems, identify future needs, and suggest pathways to meet the HWO science goals and launch timing. UV performance for HWO is not just possible but plausible and essential for delivering the science return envisioned by the Astro2020 Decadal Review. We identify key areas of development below. 

\begin{itemize}
  \item As we extend the bandpass deeper into the UV, we capture extended scientific returns. Coverage down to 250 nm is crucial for detecting biosignatures of Earth-like planets through their histories. Spectral lines in the 100 -- 1000 nm range help us to understand protoplanetary disks. UV spectral lines in the 100 -- 1100 nm range enable studies of the circumgalatic medium - crucial for understanding Cosmic Ecosystems. Continuum emission down to 100 nm elucidates the mechanism of reionization of the universe. This science is critical to understanding our cosmic origins. 

  \item Current coatings in the ultraviolet reach baseline reflectivity requirements and TRL levels, shown in Section \ref{sec-coatings}. Next, it is important to demonstrate the robustness of these coatings. The largest optics coated are not yet at the sizes required for an observatory the size of HWO. Coordinating with the coronagraph team to develop clear guidelines and test coatings for uniformity and polarization performance is underway. 

  \item UV grating technology has long been a challenge. Recent developments in the optical, such as volume phase holographic (VPH) approaches do not port into the UV well. There has been recent success using approaches such as electron-beam lithography (EBL) and other techniques developed initially for X-ray approaches or science applications outside of astronomy. We present the current state of this work in Section \ref{sec:gratings}. Having recently flown promising technology sub-orbitally, development efforts focus now on scaling up to larger and more complex optics.

  \item Several mature UV detector technologies are available and flight qualified, with several more likely available within 3-5 years (Section \ref{sec-detectors}). There is significant room for performance improvement, especially increasing sensitivity which will benefit transformational astrophysics focused on for HWO. 

  \item Contamination control remains a challenge for each of the three components described above. Contamination must be controlled at the component as well as at the systems level. We describe here in Section \ref{sec-contamination} and Appendix \ref{app:contamination} the wealth of knowledge and past experience available to successfully manage the required contamination control at component and systems level. 

  \item In the era of survey astrophysics, it is hard to argue for a single object spectrograph. Several multiplexing technologies have been used in space-based instruments, and along with the developments in coatings, provide a strong possibility if a multiplexed instrument is specified by the eventual HWO requirements. We explore the state of the art in Section \ref{sec:mplex}. 

\end{itemize}

Current UV technology provides a strong starting point for HWO design and requirement work. A key theme throughout these items is the need to scale up. Many of these approaches and technologies, as described in this paper, have been tested on smaller scales - many in space, on cubesats, balloons, or rockets. A flagship mission introduces larger physical scales and longer timescales (on the ground and in space). Bringing these technologies together to demonstrate the feasibility of the systems approach in a pathfinder mission will be a powerful intermediate step on the way to HWO. We believe that there are plausible development paths for all necessary components to be at TRL 5 within 5 years and that systems level TRL can be demonstrated with a precursor/pathfinder mission.

\section{Introduction}\label{sec:intro}
What is the path to a flagship mission? Many astronomers and engineers spent the years leading up to the 2020 decadal survey providing worked examples of what could come next\footnote{https://www.jpl.nasa.gov/habex/documents/}\textsuperscript{,}\footnote{https://asd.gsfc.nasa.gov/luvoir/}. The Astro2020 Decadal Survey prioritized a Large IR/optical/UV space telescope (``Habitable Worlds Observatory'', HWO) to pursue an ambitious program of exoplanetary discovery and characterization, as well as cosmic origins astrophysics \cite{Astro2020}. The decadal survey report recognized that many of the galactic ecosystem, exoplanet, and stellar science goals of the HWO require high-throughput imaging and spectroscopy at ultraviolet (UV) to optical wavelengths. The Astro2020 Panel on Electromagnetic Observations from Space 1 recommended that ``. . . [the] mission will also need focal plane instrumentation to acquire images and spectra over the range of 100 nm to 2 µms with parameters similar to cameras and spectrometers proposed for the Large Optical UV Infrared Telescope (LUVOIR)...''

The decadal survey endorsed an early program of technology maturation ahead of the pre-Phase-A start of HWO to reduce cost and schedule risks to the mission. This recommendation, the Great Observatories Mission and Technology Maturation Program, was adopted by NASA in the form of the Great Observatories Maturation Program (GOMAP)\footnote{https://science.nasa.gov/astrophysics/programs/gomap/} to finalize the science requirements from Astro2020 and prepare the requisite technologies for HWO (from stable telescopes to starlight suppression to ultraviolet-enabling mirror coatings).

In this white paper, we have collected key science cases that drive the UV baseline for our next flagship mission and captured the current state of UV technology development - combining the current state of the art with the steps required to reach the projected instrument capabilities of HWO. Our expectation is that the details will evolve as the work of the GOMAP progresses. However, this document provides a foundational understanding of UV development work, accelerating the ability of our community to move quickly --- a necessity with the tight HWO timelines. 

The reader might approach this paper in a variety of ways. First, we present a summary in Section \ref{ExSum}. This provides a map to the rest of the text for those who have a particular area of question or interest. The main body of the work gives a detailed overview of both the science (Section \ref{Science}) and the technological underpinnings (Sections \ref{sec-coatings} - \ref{sec-contamination}) supporting the Habitable Worlds Observatory. For those looking for deeper technical details, especially on understanding the development of contamination control, we provide Appendix \ref{app:contamination}: Contamination Control. It begins with an FAQ for those questions that often come up in discussions around UV technology development, especially addressing areas where there has been a perception of risk when in fact little exists. We follow with several technological deep dives into topics that are crucial, where the full details have often not been captured in previously published work. We intend this to be a valuable resource for upcoming development work, and provide a framework to grow the UV technology and astronomy community.

\section{Motivating Science}\label{Science}
At the heart of every flagship mission NASA builds and flies sits the scientific boundaries we plan to explore and surpass. HWO focuses on two key areas: discovery and characterization of habitable worlds, and the broad remit of transformational astrophysics. The primary purpose of HWO is to directly determine whether planets in nearby stellar systems exhibit potential for habitability and, if so, to search for signatures of extra-terrestrial life. Such an investigation begs answers to the question of what fundamental processes lead to the formation and evolution of life. Below we outline specific science questions in the two relevant topic areas from the Astro2020 Decadal: ``Worlds and Suns in Context" and ``Cosmic Ecosystems." Both areas will stretch our understanding of the formation and evolution of stars, planets, and galaxies.

Being able to identify multiple Earth-analog systems would change our understanding of our own Solar system and the many systems we have found over the last decade. Beyond discovery, which will be accomplished with coronagraphy, characterization is crucial for us to differentiate planets, especially those that might have atmospheres that could support life as we know it. Biosignature identification requires spectrographic capability and a spectrograph operating down into the deep UV provides access to crucial diagnostic spectral lines in exoplanet atmospheres. Besides biocompatability, UV spectroscopy will reveal the impact of UV radiation on the chemistry and heating of planetary atmospheres, an understanding that is critical as we move forward to truly understand the excitingly broad range of exoplanets in our nearby universe.

Our understanding of galaxy evolution has shifted drastically in the last decade. Work with line-of-sight absorption systems and a small number of integral field emission maps has exposed the circumgalactic medium (CGM) to be a highly active and interactive ecosystem, one which is seemingly crucial to our understanding of both the evolution of galaxies themselves as well as the larger cosmological history of the Universe. The CGM feeds gas into galaxies to fuel star formation while simultaneously receiving energy and metal enriched gas as explosions and winds drive escape from the galaxy. We are just beginning to understand the mechanisms that govern the CGM and the overall role that this dynamic regime plays in how the history of matter in our universe unfolds. The UV is crucial to understanding the CGM in local systems where we are able to resolve them on physical scales that match the dominant processes. 

In the sections below, we focus on particular science cases that drive ultraviolet observational requirements. We call out lower boundary wavelength ranges alongside each topic heading. We use the bandpass convention of extreme ultraviolet (EUV) defined as  10.0 -- 91.1 nm, far ultraviolet (FUV) as 91.2 -- 180.0 nm, and near ultraviolet (NUV) as 180.0 - 320.0.

\subsection{Worlds and Suns in Context}

Exoplanetary science has witnessed an explosion of discovery in the last three decades, with over 5000 exoplanets now known and planetary occurrence rates of 10 -- 60\% for F, G, and K stars and 80\% or more for M dwarf stars \cite{bryson20,dressing15}. Now that large numbers of planets with basic properties (e.g., mass and radius) are known, the next task is to characterize the composition and evolution of their atmospheres, and ultimately to determine which of these worlds may support habitable conditions. UV radiation from protoplanetary and exoplanetary systems provides unique tracers of atomic and molecular gas in these environments, making spectral coverage at UV wavelengths (100 – 320 nm) essential to completing the exoplanet science goals of the HWO. 

The Astro2020 panel on Exoplanets, Astrobiology, and the Solar System highlighted two key research questions where UV diagnostics and environmental context are critical: 
\begin{enumerate}
\item How does a planet’s interaction with its host star and planetary system influence its atmospheric properties over all time scales? (E-Q2d)
\item What is the range of potentially habitable environments around different types of stars? (E-Q3c)
\end{enumerate}

These questions are addressed by (1) comprehensive characterization of the radiation environment of exoplanet host stars as a function of stellar mass and age. UV photons and high-energy particle outputs regulate the chemical and physical state of exoplanetary atmospheres (Section \ref{section-hoststars}). (2) Measurements of the chemical inventory and radial distribution of molecular gas in protoplanetary disks set the initial conditions for the planets that we will characterize with HWO (Section \ref{section-protodisks}). In order to understand the full life cycle of habitable planets, we need to place observational constraints on the physical and chemical conditions where they are born, including mapping the molecular abundances and physical structure of the inner regions ($<$ 3 AU) around young stars. 

\begin{figure}
\centering 
\includegraphics[width=0.5\textwidth]{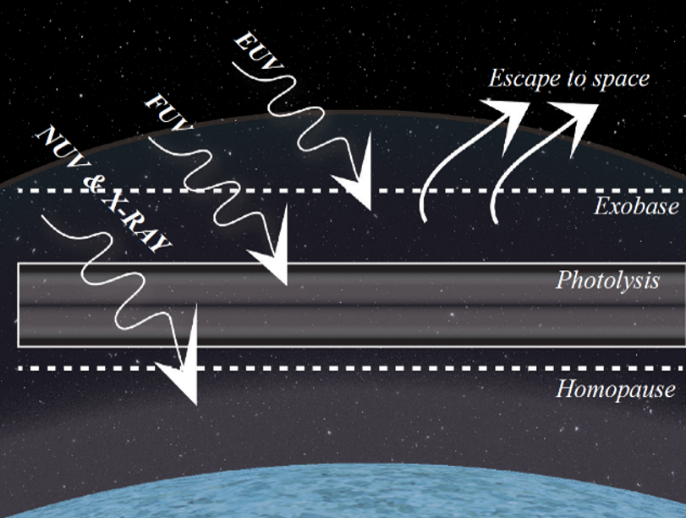}
  \caption{ A sketch demonstrating the interactions of UV light with different layers of a planetary atmosphere. The ultraviolet stellar spectrum drives photochemistry (FUV and NUV photons, 100 – 320 nm) and atmospheric escape (EUV photons, 10 – 90 nm) on orbiting planets. The EUV spectrum is calculated from chromospheric and coronal lines in the FUV spectrum, with the 100 – 115 nm band providing unique diagnostics for temperature and abundances. The host star spectra are used to predict the most promising habitable planet candidates and to interpret atmospheric spectra \cite{Harman2015, Zahnle2017}. }
  \label{fig:uvstellarspec}
\end{figure}

\subsubsection{UV emission from exoplanet host stars: 102 -- 160~nm}\label{section-hoststars}

In parallel with the discovery of thousands of planets, it has become clear that the planetary mass and effective surface temperature alone do not predict the characteristics of a planet’s atmosphere. The physics and chemistry of all types of planetary atmospheres are driven by the ultraviolet radiation (100 – 320~nm; see Figure \ref{fig:uvstellarspec}) from their parent stars. From the importance of the stellar FUV and NUV radiation on our ability to accurately predict and interpret biosignature gases on rocky exoplanets \cite{segura05,hu12,tian14,meadows18} to the young Neptune-mass planets being sculpted by their host star’s EUV radiation (a leading theory to explain the ``radius gap''; \cite{fulton18,owen18}), the stellar UV radiation inputs are critical. Indeed, with the very first James Webb Space Telescope (JWST) transiting planet observations, we are seeing direct evidence of UV-induced photochemistry on the Hot Jupiter WASP-39b \cite{tsai23}, moving this field from model predictions to empirical studies of the influence of stellar photons on exoplanetary atmospheres. 

\begin{figure}
\centering 
\includegraphics[width=0.5\textwidth]{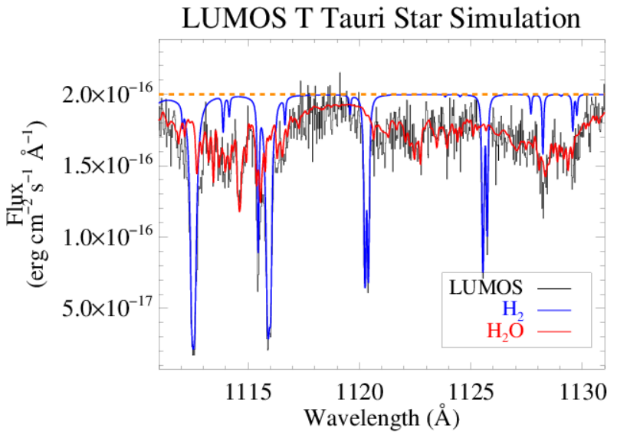}
  \caption{ Spectral simulation the 111.1 -- 113.2~nm\ spectrum of an edge-on protoplanetary disk, a spectral region containing strong lines of H$_2$ and H$_2$O \cite{france14,cauley21}. Multi-object spectroscopy from 100 -- 170~nm enables simultaneous detection of up to 30 young stars in a region like the Orion Nebula Cluster (ONC). A multiplexed instrument on a $\approx$6 m space observatory would eclipse the total UV disk archive of HST in 2 -- 4 pointings in the ONC. }
  \label{fig:uv_abs}
\end{figure}

Access to the FUV bandpass, specifically coverage from 102 -- 115~nm, is critical for calculating the EUV inputs into exoplanetary atmospheres. The EUV radiation is the key forcing function in determining the atmospheric escape timescale of all types of planets.  The 102 -- 115~nm range includes unique spectral tracers of portions of the host star chromosphere and coronae; this formation temperature (T$_{form}$) coverage greatly reduces the uncertainty of EUV reconstruction techniques \cite{duvvuri21}. As there is no EUV spectroscopy observatory on the horizon, these techniques will remain our best means of calculating atmospheric escape rates from temperate, terrestrial planets in the HWO era.  The 102 -- 115~nm region includes emission lines of S IV (106.3, 107.3 nm) [log(T$_{form}$) = 4.9], O VI (103.2, 103.8~nm) [log(T$_{form}$) = 5.5], Ne V (113.6, 114.5~nm) [log(T$_{form}$) = 5.4], and Fe XIX (111.8~nm) [log(T$_{form}$) = 7.0]. In concert with the intermediate- and high-temperature lines found in the FUV redward of 115nm: C II and Si II [log(Tform) = 4.5], Si III [log(T$_{form}$) = 4.7] Si IV [log(T$_{form}$) = 4.8], C IV [log(T$_{form}$) = 5.0], N V [log(T$_{form}$) = 5.2], O V [log(T$_{form}$) = 5.3], Fe XII [log(T$_{form}$) = 6.2], and Fe XXI [log(T$_{form}$) = 7.1], this suite of lines provides the comprehensive temperature cover to calculate the EUV irradiance from all nearby planet-hosting stars without the need for coordinated X-ray measurements \cite{Feinstein2022}.

\subsubsection{Characterizing Protoplanetary Disks: 100 -- 115 ~nm}\label{section-protodisks}

How do planets form? To answer this question, it is crucial to trace the composition, distribution, and evolution of planet-forming material in protoplanetary disks. UV spectroscopy is a unique tool for observing the molecular gas in the inner regions of protoplanetary disks: the strongest electronic band systems of H$_{2}$ and CO reside in the 100 – 170~nm wavelength range (e.g., \cite{herczeg02,france11}). UV-fluorescent H$_{2}$ spectra are sensitive to gas surface densities lower than 10$^{-6}$ g cm$^{-2}$, making them an extremely useful probe of remnant gas at r $<$ 10 AU during the disk dispersal stage after planet cores have formed. In cases where mid-IR CO spectra or traditional accretion diagnostics (e.g. H$\alpha$ equivalent widths) suggest that the inner gas disk has dissipated, far-UV H$_{2}$ observations can offer unambiguous evidence for the presence of a remnant molecular disk and ongoing protostellar mass accretion \cite{ingleby11,france12,arulanantham18,alcala19}. 

A high-sensitivity space observatory with a multi-object, high-resolution capability (R $\geq$ 30,000, $\geq$4 square arcminutes per field) enables transformative absorption-line studies of high-inclination ($i$ $>$ 60 degrees) disks. Absorption-line spectroscopy through high-inclination disks, currently limited on HST to a small number of bright stars (e.g., \cite{roberge00,roberge01,france14,cauley21}), is important because the strongest molecular absorption systems of key disk volatile species such as CO, OH, H$_{2}$O, CO$_{2}$, and CH$_{4}$ have their peak absorption cross-sections in the 100 – 170~nm range. Access to wavelengths from 100 – 115~nm is particularly critical as (1) absorption from cool H$_{2}$ (T $<$ 500 K) is restricted to $\lambda$ $<$ 111~nm, (2) the strongest absorption bands of H$_{2}$O reside between 111 – 113~nm (Figure \ref{fig:uv_abs}), and (3) the CO-dissociation bands reside at $\lambda$ $<$ 108~nm. UV absorption line spectroscopy is the only direct observational technique to characterize co-spatial populations of these molecules with H$_{2}$, offering direct measurement of the absolute abundance and temperature of inner disk gas without having to rely on molecular conversion factors or geometry-dependent model results (as is required with emission-line spectroscopy).

\subsubsection{Cometary Emissions: 100 -- 205 ~nm}

UV spectroscopy has been instrumental in analyzing cometary emissions, especially when detecting molecules with subtle emission features. A pivotal advancement using the Rosetta Alice FUV instrument (70 -- 205 nm) on comet 67P/C-G was the simultaneous measurement of both parent molecules like H$_2$O, CO, CO$_2$, and O$_2$ and their dissociative products (e.g., H, O, and C). Findings from Rosetta, such as the high O$_2$/H$_2$O ratios in a comet \cite{Keeney2017}, were groundbreaking. By assessing both the parent molecules and their subsequent products, we gain a deeper insight into comet composition, the conditions of the solar nebula where comets originated, and the dynamics within the comet's environment \cite{Hendrix2020}. This dual observation approach has significantly enriched our perspective on comets and their intricate molecular interactions. Once the Hubble Space Telescope is decommissioned, the astronomical community will face a significant void in its capability to observe cometary species in the FUV, which would be a profound loss. This highlights the pressing need for dedicated resources in this spectral domain to continue advancing our understanding of comets, a crucial pathway to studying the formation of our solar system.

\subsection {Cosmic Ecosystems}\label{section-cosmic_eco}

Astro2020 identified ``Cosmic Ecosystems'' as a key scientific priority for the 2020s. How do galaxies acquire the gas they use to form stars? How do they sustain star formation over billions of years when they appear to contain much less gas than this requires? How does feedback from star formation and active galactic nuclei (AGN) expel gas and metals, and to what extent is this feedback recycled into later star formation? What happens to a galaxy’s gas when it quenches? Is it used up, ejected, or hidden? Inflows and outflows of gas likely shape the evolution of star formation within a galaxy. All of these flows meet in the CGM, a diffuse gaseous medium spanning roughly 30 times the radius and 10,000 times the volume of the visible stellar disk \cite{tumlinson17}. This complex medium is diffuse, multi-temperture and multiphase medium, and encodes the histories of star formation, galaxy interactions, and other processes that shape galactic evolution. How can we measure this medium? It requires observations of the 100 -- 200 nm bandpass.

\begin{figure}
\centering 
\includegraphics[width=0.99\textwidth]{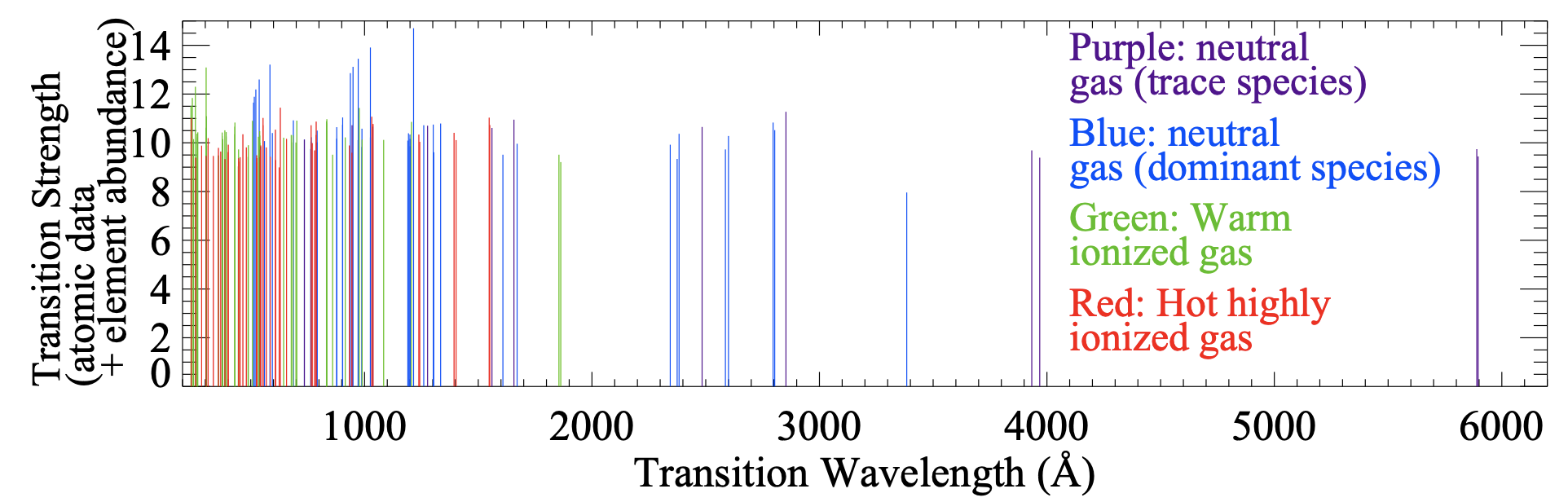}
  \caption{The distribution of resonance lines as a function of the rest-frame wavelength in the ultraviolet regime (figure adapted \cite{Tripp13}). The strength of the line represents the elemental abundance and strength of the transitions. The colors indicate the ionization state of the gas the lines are expected to trace. The line density increases as we go deeper into the UV, demonstrating a wealth of scientific targets there.}
  \label{fig:UVlinedensity}
\end{figure}

FUV wavelengths allow access to a host of species that trace gas with a wide range of ionization states and densities. These lines can be used to trace matter in emission or absorption and are ideal for probing the interstellar medium (ISM), CGM, and, intergalactic medium (IGM). The wide range of lines, especially at wavelengths $<$100~nm, provide strong diagnostics to trace the physical, chemical, kinematic, and ionization state of the gas and enable the study of the baryon cycle that supports cosmic ecosystems. Figure~\ref{fig:UVlinedensity} shows the abundance of lines and their expected strengths as a function of wavelength adapted from \cite{Tripp13}. The density of line tracing gas over four orders of magnitude in temperature ($\rm 10^{2-6}$K) makes this wavelength range ideal for probing a range of physical processes (e.g., wind-driven feedback, gas accretion into galaxies, turbulence in the CGM and IGM, etc.) that have remained elusive so far. Spectroscopic capabilities throughout the UV bandpass allow access to a significant range of local redshifts to better understand the evolution of the galaxy-CGM-IGM relationship over time.

\begin{figure}
\centering 
\includegraphics[width=0.88\textwidth]{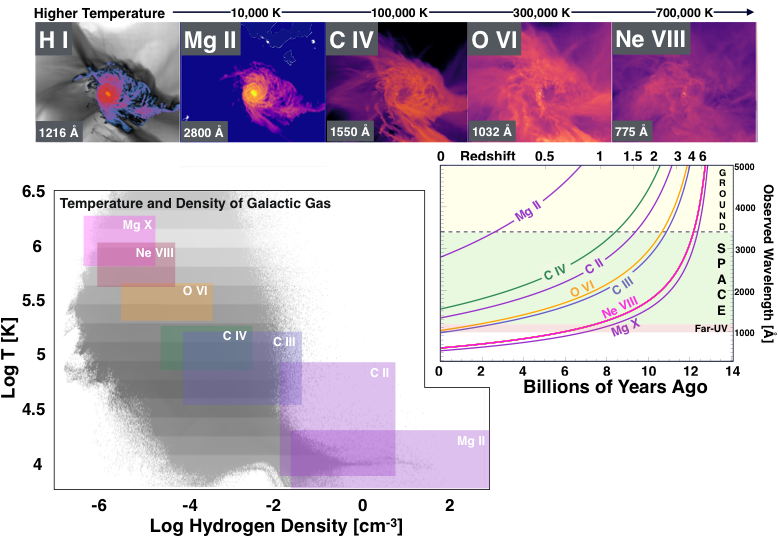}
  \caption{Diffuse gas in and around galaxies requires UV capability for most of cosmic time. The top row shows a simulated galaxy at z = 0.7 from the FOGGIE suite \cite{peeples19}, rendered in some key diagnostic ions. The temperature and density regimes probed by these ions are marked in the ``phase diagram'' of this galaxy’s gas. In upper right, we show how these lines, ranging from Mg X at 68nm to Mg II at 280~nm, vary in observed wavelength with redshift. Even with redshift, most of this diffuse gas is visible only in the UV for the last 10 Gyr of cosmic time. X-ray lines such as O VII and O VIII (both around 2~nm) probe gas at $\approx$1 million K but not the cooler phases where accretion and recycling occur. The 100 -- 120~nm range marked ``Far-UV'' is critically important to capture O VI (103.2~nm) at z $>$ 0.1 and the EUV ions Ne VIII and Mg X at z $>$ 0.5 rather than z $>$ 1. }
  \label{fig:IGM_Sims}
\end{figure}

\subsubsection{Mapping multiphase gas in the CGM: 100 -- 300~nm }\label{section-cgm}
Figure \ref{fig:IGM_Sims} shows why UV coverage is essential to understanding the intrinsically ``multiphase'' galactic gas flows. The shaded map at lower left shows the distribution of gas in a simulated L* galaxy from the EAGLE project \cite{oppenheimer16}, which spans multiple ``phases'' from 10$^{3}$-10$^{6}$ K and eight orders of magnitude in density. In such ionized gas, the quantum-mechanical rules of electron orbits dictate that the gas will emit and absorb energy predominantly at UV wavelengths, up to 80\% according to detailed simulations \cite{bertone13}. Many of the transitions appear as strong UV absorption and emission lines. This inescapable physics means that access to UV wavelengths in space is essential if we are to resolve questions about how galaxies acquire, process, eject, and recycle their gas over the last 10 Gyr of cosmic time. 

Solving these problems requires pushing the boundaries of CGM characterization far beyond the limits of today’s measurements, to z = 1 – 2, for two major reasons. First, this period 7 -- 10 Gyr ago encompasses the peak of cosmic star formation. Second, at z $>$ 0.5 we gain UV access to a wide range of extreme UV lines such as Ne VIII, O II to IV, and Mg X ($<$ 80~nm in the rest frame, see Figure \ref{fig:IGM_Sims}) that enable a much broader set of diagnostics of physical state and metal content that are only available with redshift. A large wavelength grasp (100 -- 300~nm) provides coverage of critically important rest-frame extreme-UV ions that redshift into the FUV for z $>$ 0.5. This includes nearly every ionization state of the most abundant heavy element, oxygen, from O I (cold gas), through O VI (warm ionized gas), which could be covered $simultaneously$ for sightlines at z $\sim$ 1. The OVI line (103 nm) is a particularly important diagnostic. As can be seen from Figure \ref{fig:IGM_Sims}, this line is a probe of environments between hot gas at T $>$ 10$^{6}$K (probed by Mg X and NE VIII) and cooler gas at T $<$ 10$^5$K probed by other lines. As such this line can elucidate the mechanisms that maintain these temperature differentials, and help us understand the detailed histories of galaxies. Fortunately, very high ionization lines like Ne VIII (77.5~nm), Mg X (61~nm), and Si XII (50~nm) become available in (optically thin) IGM and CGM gas at z $>$ 0.5, reaching a temperature regime (T $>$ 10$^{6}$ K) that is usually thought to be the exclusive domain of X-ray telescopes. 

The same broad wavelength coverage that enables a robust baryon census will also complete a multiphase metals census over a wide range of galaxy masses, star-formation rates, and environments. This is only possible with UV coverage and redshift that places EUV lines of O II, O III, and O IV into FUV \cite{Tripp13}. Using this capability, HWO’s users can complete the low-z metals census, constrain the physics of feedback from galaxies, and assess the importance of galactic recycling with robust statistics. 

The combination of large UV wavelength coverage and the potential to include a UV integral field unit (IFU) instrument allows for truly transformational studies of the CGM and IGM. Observational evidence for enormous circumgalactic halos has historically come in the form of absorption lines in the spectra of bright background objects (e.g., quasars) that intersect the halos of galaxies (e.g., \cite{werketal16,nicastroetal18}). These studies have defined the current paradigm and taught us a great deal about the physical properties of the CGM, including the average covering fraction of gas and the average ratios of gas phases. But because these studies require bright background objects from which absorption lines can be measured, we have only seen single, narrow sightlines through the halo gas for the vast majority of galaxies. Qualities derived from these studies are averages, not individual views of particular galaxies. A complete gas census can only be developed by eventually moving to volume-filled mapping of the ultra-diffuse gas residing in the CGM and IGM. Since the smallest spatial scales are most easily observed in nearby galaxies, it is advantageous to initiate CGM mapping campaigns in nearby galaxies that target the aforementioned suite of diagnostic lines at $\lambda < $ 200~nm.

\subsubsection{Lyman continuum observations across cosmic time: 100 -- 120 nm}\label{section-lycont}

\begin{figure}
\includegraphics[width=0.48\textwidth]{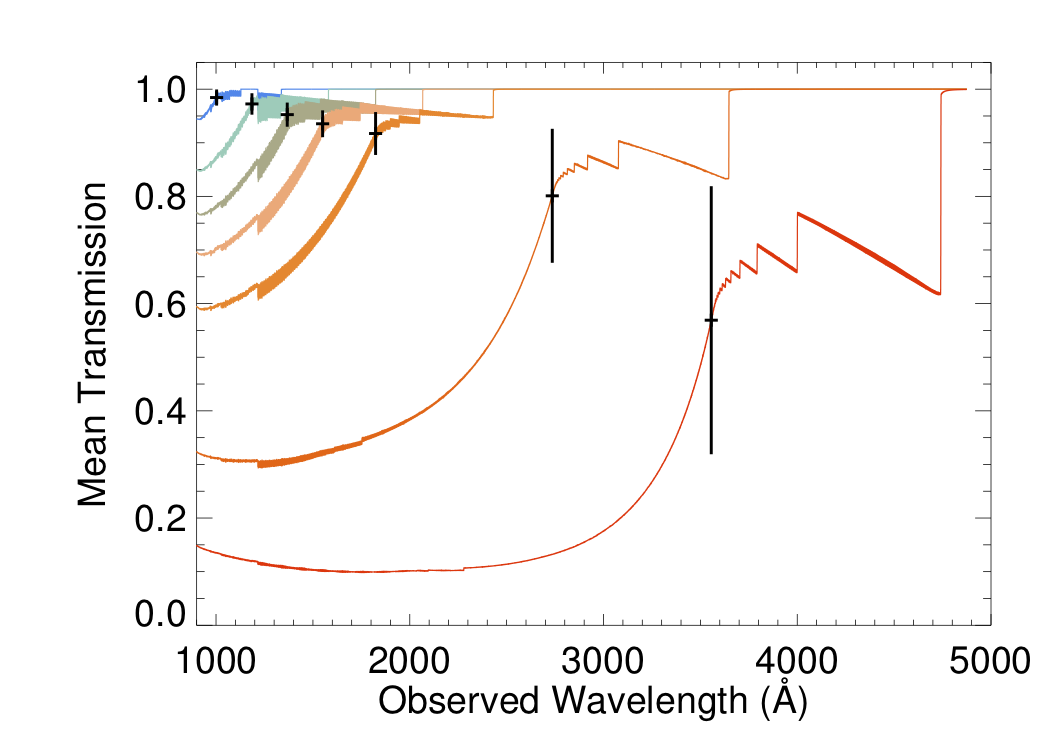} \hspace{\fill}
\includegraphics[width=0.42\textwidth]{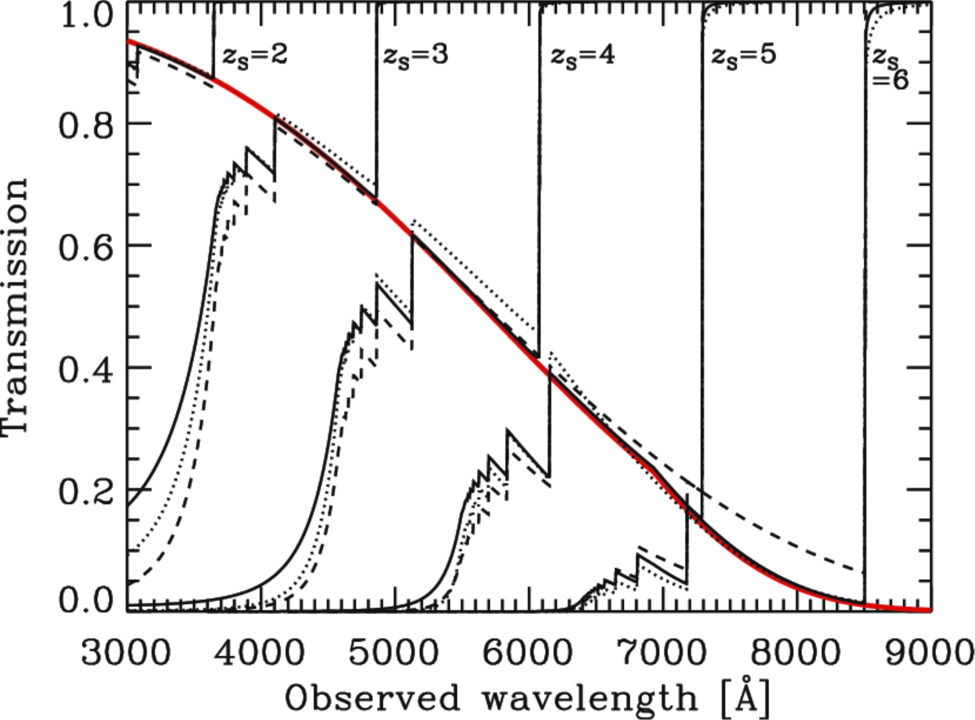}
\caption{\small Left - Mean IGM transmission for $z$ = (0.1, 0.3, 0.5, 0.7, 1.0, 2.0, 2.9) \cite{McCandliss2017}. Vertical bars mark the Lyman edge and indicate the level of expected variation found in a Monte Carlo study of IGM transmission by \cite{Inoue2008}. Right - same for $z$ = (2, 3, 4, 5, 6) \cite{Inoue2014}. The red line on the right marks the mean transmission at \lya. } \label{fi}
\end{figure}

Quantifying the physical conditions that allow radiation emitted shortward of the hydrogen ionization edge at 91.18~nm (the Lyman Continuum; LyC) to escape the first collapsed objects and ultimately reionize the universe is a compelling problem for astrophysics. The escape of LyC emission from star-forming galaxies and active galactic nuclei is intimately tied to the emergence and sustenance of the metagalactic ionizing background that pervades the universe to the present day and in turn is tied to the emergence of structure at all epochs. The James Webb Space Telescope (\jwst) was built in part to search for the source(s) responsible for reionization, but it cannot observe LyC escape directly because of the progressive increase in the mean transmission of the intergalactic medium towards the epoch of reionization (Figure~\ref{fi}).

\begin{figure}
\includegraphics[width=\textwidth]{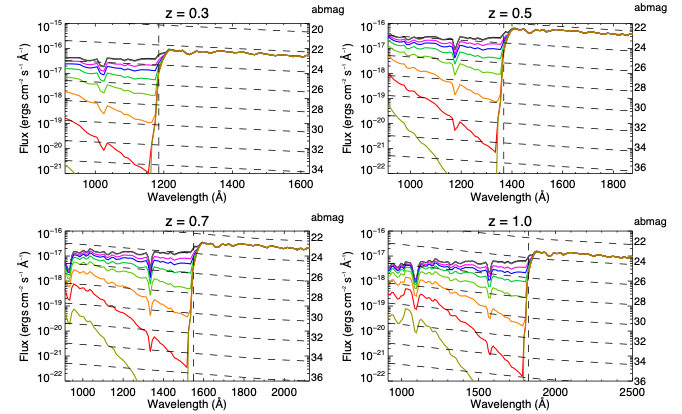} 
\caption{\small Left - flux estimates for $L^*_{150 nm}$ galaxies with escape fractions of $f^e_{LyC}$ =(1, 0.57, 0.36, 0.17, 0.04, 0.003, 0) at redshifts of $z$ = (0.3, 0.5, 0.7, 1.0) \cite{McCandliss2017}. The recovery of flux towards the shortest wavelengths suggests the possibility of LyC drop-ins. } \label{fii}
\end{figure}

The low-redshift universe (0.1 $< z <$ 1) is the ideal place to study the physics of LyC escape because corrections for the mean transmission of the IGM are modest. In particular, the very lowest redshifts offer the opportunity to examine direct escape from \HII\ regions within nearby galaxies. Direct quantification of the escape fraction of LyC photons $f^e_{LyC}$ at low- and intermediate-$z$ require space-based facilities with extreme sensitivity to the EUV restframe, as estimates for $L^*_{150~nm}$ in Figure~\ref{fii} show. Access to EUV wavelengths down to 100~nm is essential for this science.

\section{Reflective Coatings}\label{sec-coatings}

An environmentally robust and reflectively uniform broadband (UV/O/IR) reflective coating has been a goal of coating technology development for many years. The 2020 Astrophysics Decadal Survey states of HWO that ``The mission will also need focal plane instrumentation to acquire images and spectra over the range of 100 nm to 2 $\mu$m'' (Page I-2). The coating used to deliver this broadband performance must be resilient to the space environment, as well as compatible with the high-contrast imaging system and other potential science instruments for HWO. Recent advances in deposition technologies have provided new possibilities for coatings that not only provide high reflectively, but improve ease of handling and decrease environmental constraints. 

When we discuss broadband mirror coatings that include ultraviolet reflectance, two components must be considered - the reflective coating and the protective overcoat. There is also the possibility of applying a second overcoat or ‘capping’ layer, often a small number of individual layers to improve durability. The primary material provides the underlying reflectance, while the overcoat improves the durability and environmental stability of the coating (but also modifies it’s reflectance). Past and current broadband mirror coatings operating in the UV typically utilize reflective aluminum as the base material, as other common coatings like silver and gold become partially transparent in this range. This aluminum layer is capped with a thin dielectric fluoride that is transparent to the UV through NIR but prevents aluminum oxidization, which degrades far-UV reflectance. The choice of the dielectric capping layer sets the bandpass, reflectance, and environmental stability/resilience of the coating. Magnesium fluoride (MgF$_{2}$; \hst, \GALEX) and lithium fluoride (LiF; {\it Far Ultraviolet Spectroscopic Explorer} - \fuse) are both flight qualified options in this bandpass. Both of these approaches use physical vapor deposition (PVD) to coat optics. This requires a vacuum chamber to hold the optic while deposition material is heated and then evaporated onto the surface. The optic is usually unheated, but some techniques in development have found improved performance through heating the optic during deposition

\begin{figure}
  \begin{center}
  \begin{tabular}{c} 
  \includegraphics[height=5cm]{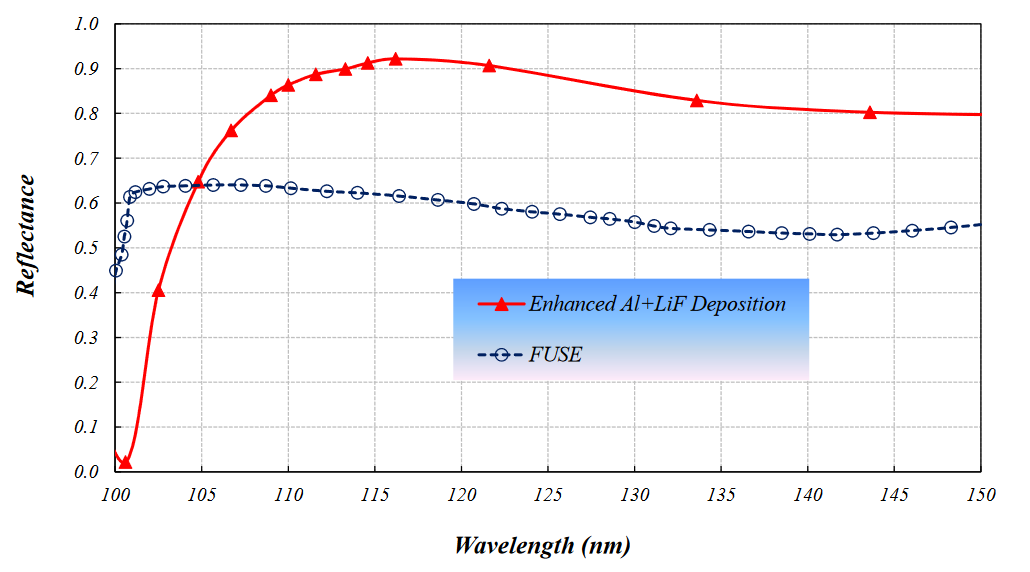}
  \end{tabular}
  \end{center}
  \caption{The measured reflectance of enhanced lithium fluoride (eLiF) deposited at elevated substrate temperature versus the performance of conventional LiF developed for the FUSE mission. \cite{Quijada2014}}
  \label{fig:eLiF} 
\end{figure}

A key technical challenge, historically, for wavelengths above 100 nm, the reflectance of protected Al is limited by the residual absorption and the eventual degradation of the fluoride overcoats. The Al/LiF technology was discovered about 60 years ago, and up until recently, there have been few but significant advances. \cite{Angel1961} reported the first aluminum mirrors protected with LiF in the 1960s, and a few years later \cite{Cox1968} were first in optimizing Al/LiF mirrors at the Hydrogen Lyman $\beta$ line (102.6 nm). Then, \cite{Hutcheson1972} studied the effect of deposition rate and substrate temperature for Al/LiF mirrors. The fact that high substrate temperatures enhanced the mirror reflectivity was further explored several decades later by \cite{Quijada2014}, who demonstrated the highest R at 122 nm wavelength with Al/LiF mirrors fabricated using the 3-step hot process \cite{Quijada2012}. The results of these hot LiF deposition are shown in  Figure~\ref{fig:eLiF}). Optimizations in pre- and post-annealing conditions have been explored, and these enhanced LiF (eLiF) coatings have been demonstrated on several sounding rocket missions.

There are multiple viable potential coating choices - as can be seen summarized in Table \ref{tbl-coatings}. There are a collection of coatings that have been flight qualified, and two newer approaches that have provided promising coatings with better FUV performance that still require investment to be physically scaled up. For example, the estimated TRL levels for Al+MgF$_2$ and Al+LiF are the highest given their HST and FUSE heritages respectively. Likewise, eLiF coatings have flown on the SISTINE sounding rocket and the SPRITE CubeSat and Aspera Astrophysics Pioneer mission will use an eLiF mirror coating that is capped with an ALD MgF$_2$ layer given the maximum sensitivity this coating provides to rest-frame O VI emission lines (103 nm; Section \ref{section-cosmic_eco}). The two emerging technologies (XeLiF and e-beam passivated Al+AlF$_3$) have been shown to be at the TRL 3 and 4 respectively.

\begin{figure}
  \begin{center}
  \begin{tabular}{c}  
  \includegraphics[height=6cm]{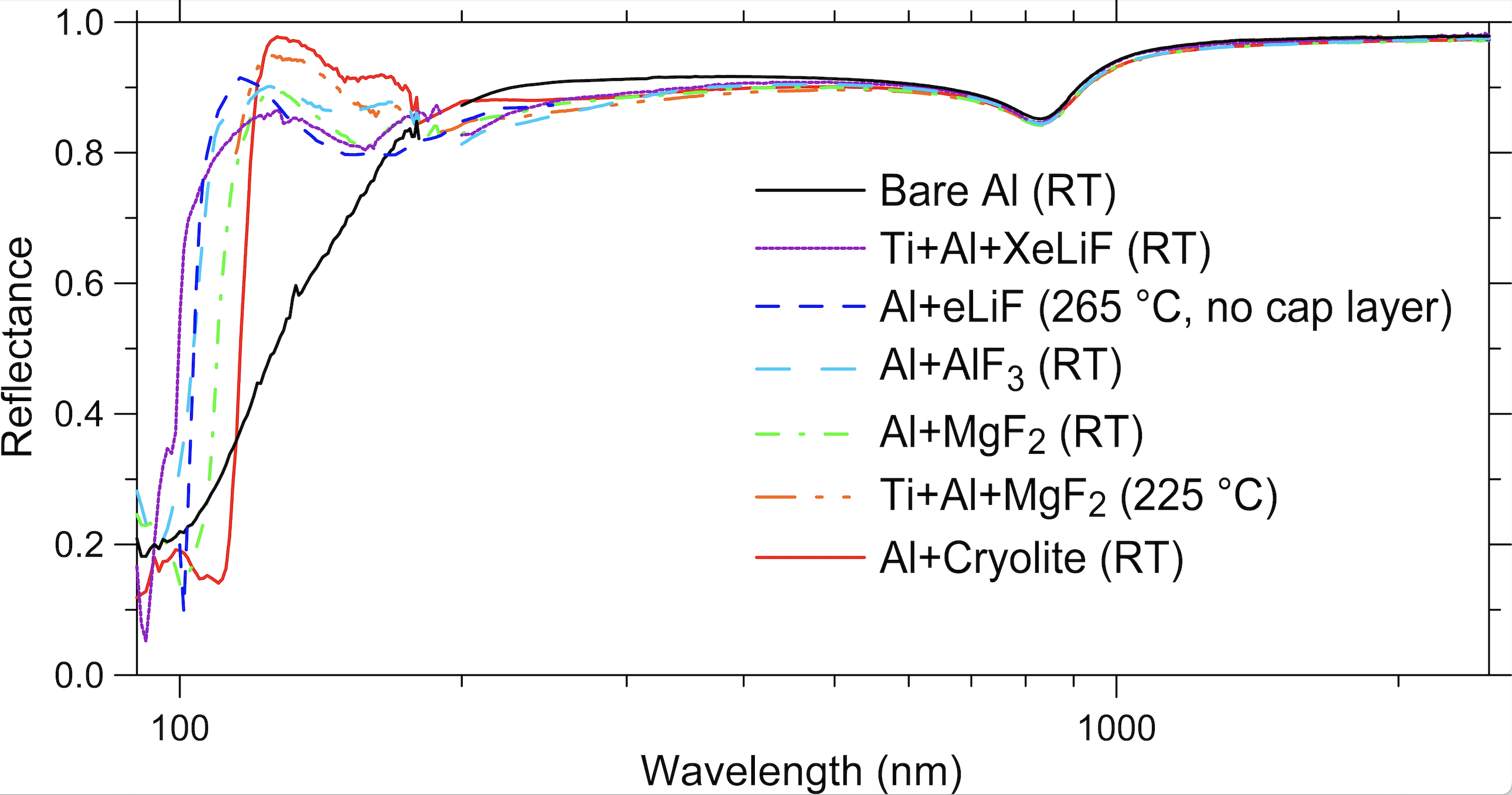}
  \end{tabular}
  \end{center}
  \caption{Comparison of experimental UV/Optical/IR reflectance of representative mirror coatings.  ``RT'' indicates a room temperature deposition process.}
  \label{fig:Coating-Technologies-UVOIR} 
\end{figure}

Figure~\ref{fig:Coating-Technologies-UVOIR} displays reflectance data for a variety of coatings across the NUV, optical and near-infrared (200~$\lesssim$~$\lambda$~$\lesssim$~2500 nm) bands. The main feature to note from this figure is that the reflectance performance of all these Al-based mirror coatings is very similar in the UV/O/IR spectral regions. The most notable exceptions are from the bare Al sample (which actually exhibits a slightly higher reflectance in the NUV) and the curve labeled as "Al+eLiF (265 C, no cap layer)" that shows a reduction in reflectance of 70\% at 200 nm (when compared to an average value of 80\% at this wavelength for all the other samples). The main reason for this reduced reflectance value for this sample is due to the fact that the thickness of the LiF layer was optimized to provide maximum performance at the shortest wavelength of interest for SPRITE of 103 nm \cite{Rodriguez2022}.

While the simplest description of a fluoride protected aluminum coating is that the fluoride provides a transmissive overcoat to the reflective aluminum, the nuances of that overcoat, especially the thickness, have a great deal of influence on the final coating performance. In general, a thinner overcoat of fluoride shifts the peak reflectance of the coating in the FUV to shorter wavelengths, while also suppressing the amplitude of that peak FUV reflectance. A thicker coating shifts the peak to longer wavelengths while also enhancing the amplitude of that peak. A mission that prioritizes very short wavelengths, such as Ly $\beta$ at 1026 \AA , for instance, would therefore have a reflectance at Ly $\alpha$ and longer FUV wavelengths lower than an identical coating formula with a thicker fluoride layer. This is most apparent when evaluating the $FUSE$ and SPRITE LiF and protected eLiF coatings (the SPRITE coating curves are included in Figure~\ref{fig:Coating-Technologies-UV1}, \cite{Bowen2023}). Beyond 2000 \AA\ all LiF+Al curves rapidly approach 90$\%$ reflectance. For this paper, it is essential that the reader consider that the comparison of coatings is not entirely straightforward, especially as enhanced performance over a narrow bandpass at short wavelengths is not always apparent in a plot.

\begin{table}[b]
  \begin{center}
  \begin{tabular}{c}
 \includegraphics[width=0.98\textwidth,angle=0,trim={.0in 0.0in 00.0in 0.0in},clip]{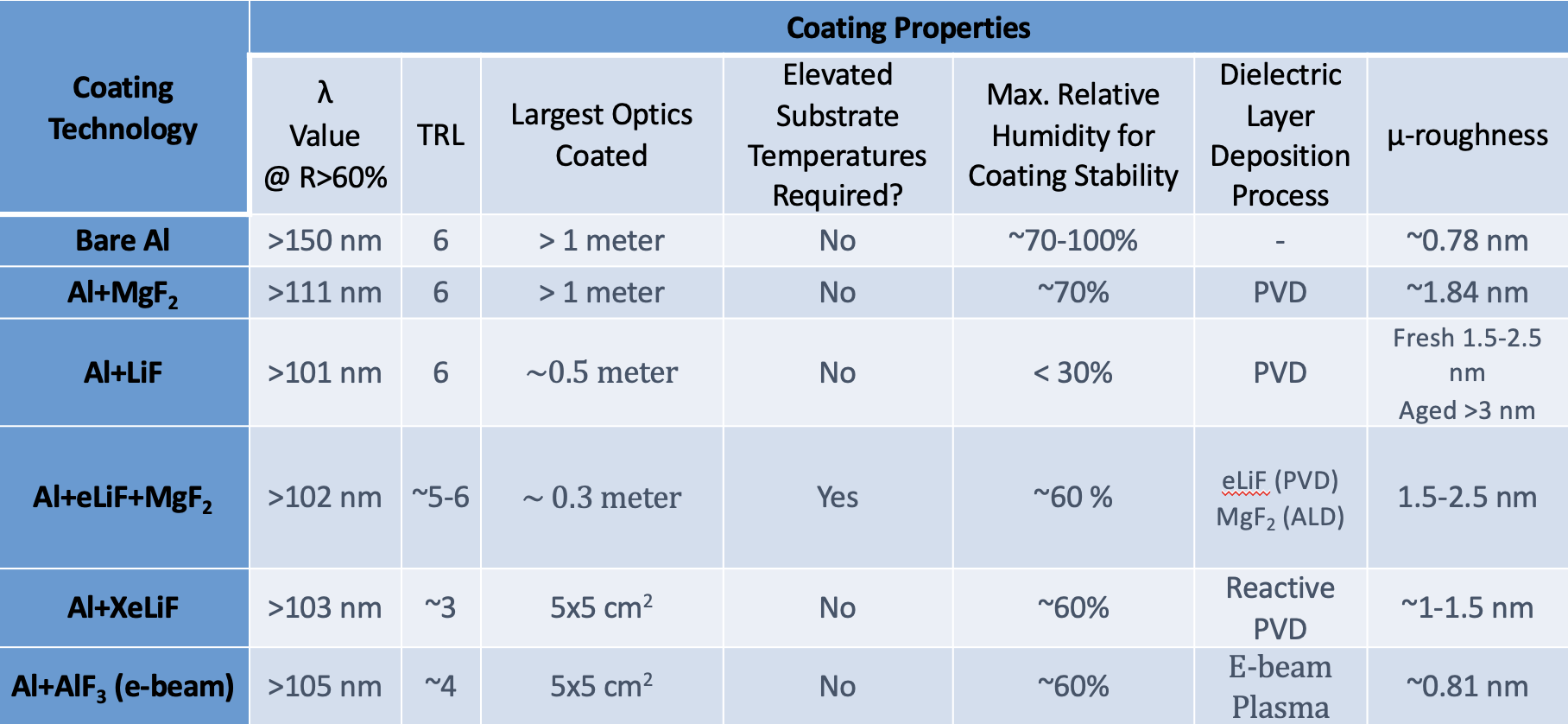} 
  \end{tabular}
  \end{center}
 \captionsetup{labelformat=empty}
\caption{Table \ref{tbl-coatings}: Parameter comparison for various protected-Al FUV coatings. \label{tbl-coatings}}
\vspace{-0.15in}
\end{table}

We present two NASA funded on-going efforts to advance ultraviolet reflective coating performance, sometimes used in combination. The first is improving physical layer deposition using either room temperature (eLiF) or elevated temperature (XeLiF) deposition, described in Section \ref{section-pvd}. The second is focused on atomic layer deposition (ALD), described in Section \ref{section-ald}. We then compare some example coatings.

\subsection{eLiF and XeLiF - Physical Vapor Deposition (PVD) Coatings}\label{section-pvd}
\paragraph{Enhanced Lithium Fluoride (eLiF)} 
Conventional deposition techniques for LiF have not achieved predicted theoretical performance, with reflectances that peak between 70-75$\%$ across the 100 -- 200~nm\ bandpass before rising to $>$85$\%$ from the NUV through NIR. Over the past decade, the community has developed several enhanced deposition techniques for LiF+Al that have drastically improved the reflectance in the UV and resilience to humidity with no known detrimental impact on the visible and NIR performance \cite{Quijada2014,Fleming2017,Hennessy2021,Quijada2022}. As mentioned previously, eLiF refers to deposition of LiF protecting layer onto a substrate at elevated temperature.

The most developed of these new deposition techniques is known as protected enhanced lithium fluoride, or ``protected eLiF'' \cite{Quijada2014,Fleming2017,Rodriguez2022}. This coating combines two recent advancements pioneered under NASA's APRA, SAT, and Roman Technology Fellowship (RTF) programs: post-coating annealing of the LiF+Al to enhance the reflectance, and then capping with an ultra-thin overcoat of MgF$_{2}$ to reduce hygroscopicity \cite{Angel1961,Hennessy2021}. The overcoat process was developed at JPL using atomic layer deposition, which enables ultra-thin and highly uniform capping layers that have a minimal impact on the reflectance of the underlying LiF-based coating. 

The eLiF process was pioneered at NASA-GSFC in the Thin Films Coating Lab (TFCL), first as a high-temperature deposition process where the mirror was heated to $\approx$250$^\circ$C prior to LiF deposition, but has since evolved to only require heating for post-coating annealing. This change significantly reduces the complexity of the process, while also increasing repeatability. The first usage of eLiF (as well as an early form of protected eLiF) was on the SISTINE sounding rocket \cite{Nell2020}, which employed the first-generation version of the coating (deposited onto a heated substrate) on optics as large as 0.5-m in diameter. While several SISTINE optics had excellent reflectance ($>$80$\%$ at 105~nm), the results were mixed, due most likely to non-uniform heating of the larger optics. The SISTINE grating was not coated with eLiF, as it was a replica grating with a photoresist that would not withstand such temperatures. The secondary mirror of SISTINE was capped with a thin layer of AlF$_{3}$ using the JPL ALD process for added resilience to humidity \cite{Fleming2017}. Over the course of three launches, five years of student-led development and testing in the laboratory, as well as over 6 months of over-ocean transit from Australia to the United States in 2022 in a partial pressure of nitrogen, the SISTINE secondary mirror had only minimal degradation from 100 -- 120~nm \cite{Nell23}. It has since been determined that an MgF$_{2}$ capping layer can provide even better protection, although there is a slight loss in reflectance of 2-7$\%$ between 100 -- 110 nm\ due to absorption in the MgF$_{2}$ that should be traded against added risk in AlF$_{3}$ capping or bare eLiF.

\begin{figure}[t]
  \begin{center}
  \begin{tabular}{c}
 \includegraphics[width=0.88\textwidth,angle=0,trim={.0in 0.0in 00.0in 0.0in},clip]{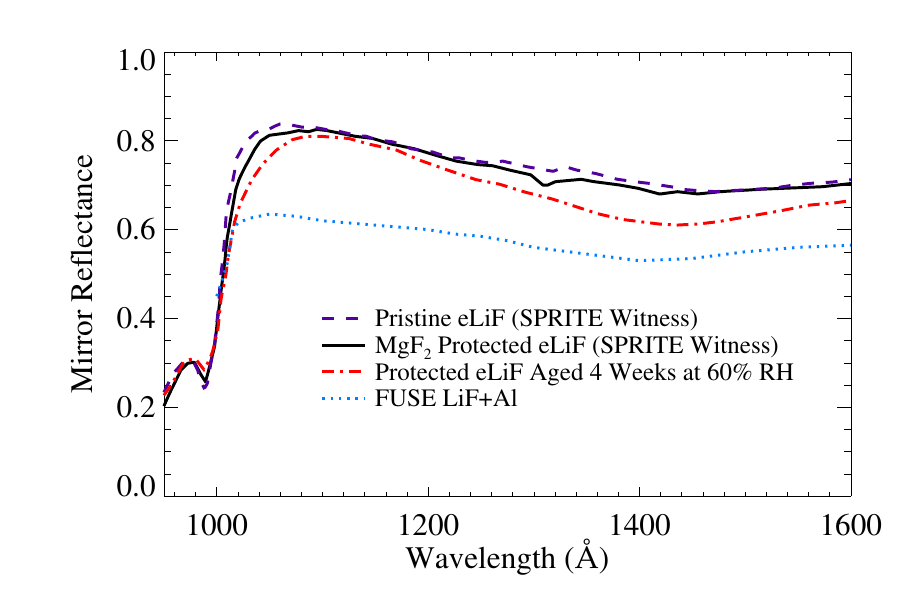} 
  \end{tabular}
  \end{center}
  \caption[example] 
 { \label{fig-SPRITE_Coatings} 
{\em The measured reflectance of a SPRITE test sample with just eLiF
 (dashed), after the addition of the capping layer (solid) and after
 four weeks of aging at 60$\%$ RH and elevated temperature \cite{Rodriguez2022}. It is important to note that
 SPRITE is specifically optimized for reflectance at 1050 \AA, which
 requires a thin LiF layer and therefore lower reflectance at
 $\lambda$ $>$ $1150 \AA$  than is possible with other
 formulations. FUSE (dotted) was also optimized for short wavelength
 reflectance, which is why its peak reflectance does not match other
 conventional LiF+Al optics.}}
\vspace{-0.15in}
\end{figure}

The second generation of protected eLiF is to be flight tested on the SPRITE
CubeSat, which is expected to launch in mid-2024 to low Earth orbit (LEO)
\cite{Indahl23}. SPRITE is designed as a HWO technology testbed and carries a
calibration channel that will aid in dissociating any on-orbit
degradation of the coatings from losses due to contamination or
detector degradation. Three of the SPRITE optics are coated in
MgF$_{2}$ protected eLiF, while the SPRITE primary mirror (PM) is coated in
MgF$_{2}$ protected conventional LiF+Al. The SPRITE PM was not coated
in eLiF as the annealing process has not yet been installed in the
large 2-m coating chamber at the GSFC TFCL. Testing carried out during
the SPRITE coating development effort shows that MgF$_{2}$ protected
eLiF is resilient to humidity exposure, with only minimal
degradation after four weeks of exposure to 60$\%$ RH
\cite{Rodriguez2022}. SPRITE has already demonstrated this resilience,
with the SPRITE telescope assembled over two weeks in a modestly
controlled environment in Maryland where the ambient humidity varied
from 25$\%$ to 50$\%$ and nitrogen purging was employed only overnight
and when the optics were not being handled. Witness sample testing
showed no significant degradation in the coating reflectance over this
time. The SPRITE grating has also been coated successfully with protected eLiF, as
it is a master grating and therefore has no photoresist.

Several other programs are now baselining eLiF and protected eLiF,
including the INFUSE (Integral Field Ultraviolet Spectroscope Experiment) sounding rocket (eLiF; launched October 2023) and the Aspera Pioneer
(protected eLiF; launch circa 2025).

\paragraph{XeLiF} \label{sec:XeLiF}

This section will describe a new reactive process that offers protection of Al coatings with a more stable and transparent LiF protection layer, along with unprecedented reflectivity. The process for preparing these Al+LiF mirror (or XeLiF) coatings utilizes exposure to XeF$_2$ gas before and after the Physical Vapor Deposition (PVD) of the LiF metal-fluoride dielectric coating is applied. The technical description of the process is as follows. First, bare optically smooth glass is coated with Al in a high vacuum chamber by PVD. This process, along with the chamber in which the process is carried out, has been extensively described elsewhere \cite{Rodriguez2021} where the only difference is that in the present process, the chamber was baked out for several days at 100$-$105 $^\circ$C before the deposition. Then, the bare Al mirror is immediately exposed to a reactive XeF$_2$ gas before and after the application of the final flash evaporated LiF layer by conventional PVD. Therefore, the Al is exposed to XeF$_2$ within a few seconds after the aluminum evaporation, and typically the exposure lasts for up to 3 mins. The quick exposure of bare Al to XeF$_2$ grants a thin ( $\approx~2.5-3.2$ nm, typical values measured through ellipsometry) protective layer of AlF$_3$. Once the mirror is complete, it is exposed again to a high partial pressure of XeF$_2$ for a few seconds. An important detail is that all are carried out at ambient temperature.

After the mirror coating fabrication, optical characterizations are performed with a spectrometer and measurement configurations that have been described elsewhere \cite{Quijada2018}. The estimated absolute reflectance error is $\pm~1$\%. FUV reflectance data were measured immediately after deposition, and then again after several weeks of storage in a desiccator with a relative humidity of 40\%. These samples were also characterized with ellipsometry to derive thickness and optical properties (in the 200$-$2500 nm range) of the ALF$_3$ layers as well as Atomic force microscopy (AFM) to study the surface roughness and topography of each sample in two fields of 500 nm $\times$ 500 nm and 5 $\mu$m $\times$ 5 $\mu$m size. 
\begin{figure}
  \begin{center}
  \begin{tabular}{c} 
  \includegraphics[height=6cm]{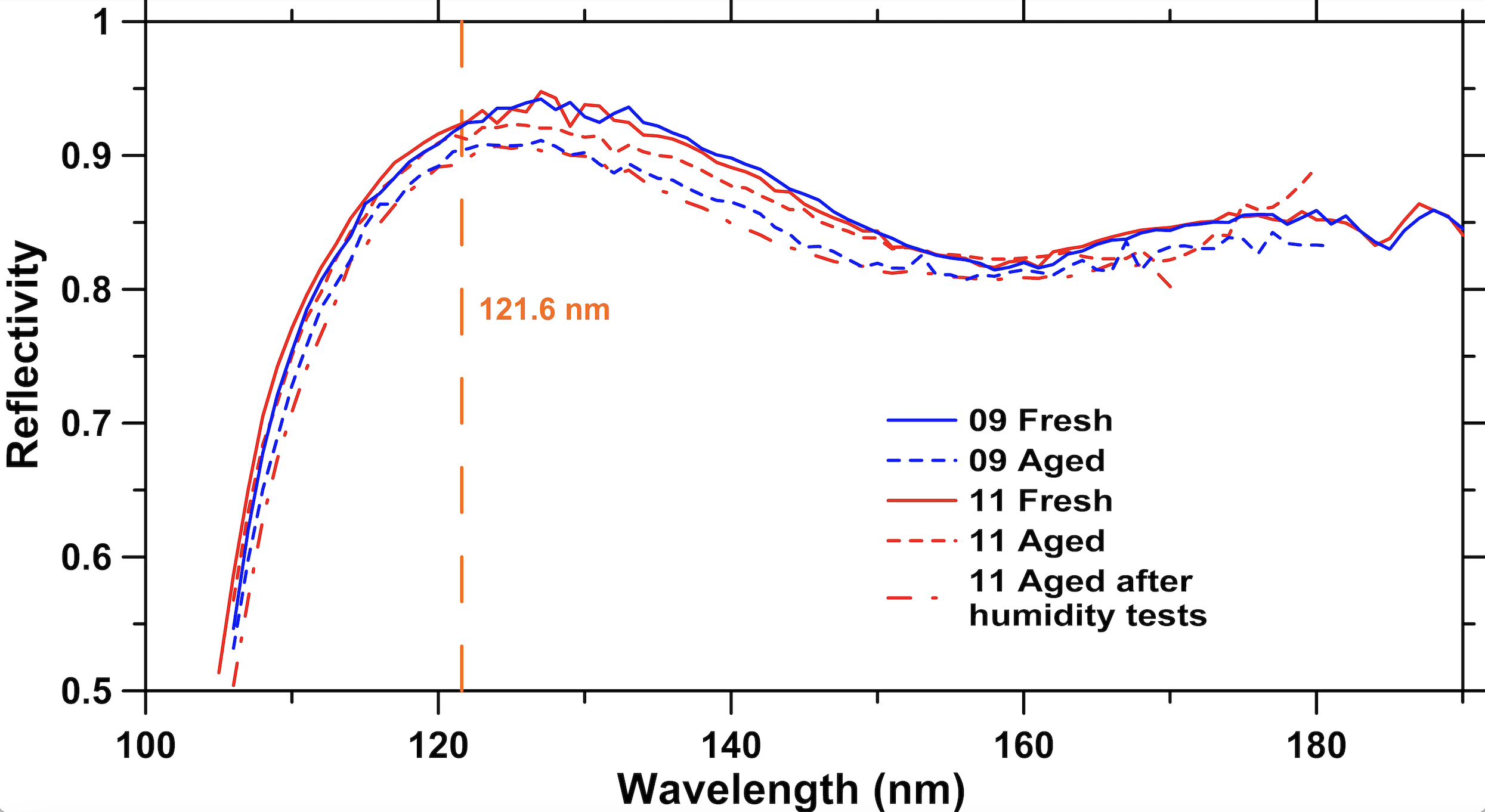}
  \end{tabular}
  \end{center}
  \caption{Reflectivity of two XeLiF samples (ID 09 and 11) both fresh and aged (18 and 12 weeks, respectively).}
  \label{fig:Samples_09-11} 
\end{figure} 
Figure~\ref{fig:Samples_09-11} displays FUV reflectance for two examples of Al/LiF mirrors fabricated with the rPVD described above. The LiF thicknesses of samples 09 and 11 are $\approx~22.9$ nm and are optimized to provide the higher reflectivity at the Hydrogen Lyman $\alpha$ line (121.6 nm), which is one important diagnostic for astronomy and often used as a reference. Among the highest reflectance values reported at 122 nm wavelength:  \cite{Quijada2012} and \cite{Rodriguez2018} demonstrated $\approx~0.90-0.91$ with Al/MgF$_2$,  \cite{Oliveira1999}, \cite{Stempfhuber2020}, and \cite{Larruquert2021}) demonstrated $\approx$ 0.90$-$0.91 with Al/AlF$_3$, and \cite{Quijada2014} showed $\approx~0.90$ with Al/LiF. Most of these works used high-substrate temperatures; \cite{Rodriguez2021} are the exceptions in which room-temperature e-beam generated plasma in SF$_6$/Ar gas mixtures was used to passivate bare Al and \cite{Quijada2021} in which the high reflectance was obtained through optimization of the PVD parameters such as a high deposition rate for both Al and MgF$_2$. The results shown in  Figure~\ref{fig:Samples_09-11} with the rPVD process are unprecedented reflectivity of 0.926 at 122 nm, which is the highest value ever reported at this wavelength for the Al/LiF mirror coatings.

\subsection{Atomic Layer Deposition (ALD) Coatings}\label{section-ald}

Additional improvements in LiF stability with respect to long-duration storage and possible exposure to elevated relative humidity levels have been explored by capping LiF mirror structures with thin layers of more stable fluoride materials like MgF\textsubscript{2} or AlF\textsubscript{3}. If the capping layer is thin enough the short wavelength performance of LiF can be maintained while sufficiently altering the surface chemistry to avoid deleterious interactions with water vapor. This general approach has been explored in a number of previous studies \cite{wilbrandt2014protected,balasubramanian2015}. \cite{Fleming2017} reported on samples of eLiF encapsulated with ALD AlF\textsubscript{3} that were subjected to long duration elevated humidity exposure where it was observed that the capping layer was able to reduce, but not completely eliminate, reflectance degradation over year-long exposure conditions. The secondary mirror of the SISTINE sounding rocket was coated with this eLiF/AlF\textsubscript{3} approach \cite{aguirre2021assembly} and showed $\lesssim$~5\% reflectivity degradation over 4.5 years including three rocket integration and launch campaigns and commercial freight shipping across the ocean.

Further stability enhancement demonstrations have been made by using ALD capping layers of MgF\textsubscript{2} on eLiF mirrors \cite{Rodriguez2022}. Such coatings have been deposited on the flight optics of the SPRITE CubeSat which anticipates a 2024 launch. SPRITE includes a calibration channel with MgF\textsubscript{2}/Al optics that will allow for a direct comparison of the on-orbit degradation of both systems. The same eLiF/MgF\textsubscript{2} coating is also planned to be implemented on the Aspera Pioneers Mission, with flight optic coatings expected to be completed by the end of 2023.

ALD methods can also be utilized to deposit the entire protective layer on aluminum mirror coatings. In addition to the MgF\textsubscript{2} and AlF\textsubscript{3} processes noted above, ALD processes have also been established for LiF coatings including protected Al mirrors \cite{Hennessy2018}. The possibility of using ALD methods for the deposition of the protective coating is largely related to the general ability of ALD for precise film thickness uniformity. For a protected-Al mirror coating, the uniformity of the protective layer largely determines the ultimate reflectance uniformity of the mirror coating. Achieving film thickness uniformity with physical vapor deposition (PVD) methods often requires motion control of the substrate and/or source, or requires physical masking of the vapor species. ALD is a non-line-of-sight technique and is therefore generally agnostic of both the spatial scale of the substrate but also the substrate's surface shape. This may be relevant for the coating of a many-segmented system with individual segments of varying figures.

Commercial ALD solutions exist for the coating of meter class substrates, although these have largely been focused on flat-panel substrate processing or batch processing \cite{putkonen2009ald}. Custom systems designed for the coating of telescope optics at the meter-scale and have been demonstrated for protected silver mirror coatings at longer wavelength applications \cite{fryauf2018scaling}. The key challenge associated with the adoption of ALD approaches for future UV mirrors centers around the combination of PVD Al with the ALD fluoride protective coating. There currently exists no promising method for the ALD of Al films with sufficient optical quality in the UV/vis. This restriction has been overcome by investigating dual-use coating chambers where evaporation and ALD can occur without breaking vacuum. It has also been explored via atomic layer etching methods that would allow coating to be performed in two separate chambers, for example by allowing the aluminum to oxidize and then remove the native oxide with a self-limiting chemical etching process prior to the deposition of the protective coating.

\subsection{Coating Performance}\label{section-coatingperformance}

The field of UV coatings has produced numerous solutions for materials and techniques capable of producing high-reflectivity and environmentally robust overcoats. The most common dielectric coating material for protected aluminum is MgF$_{2}$, as was used on {\it HST, GALEX}, and many other UV-sensitive missions. MgF$_{2}$ is a common overcoat available at many commercial coating vendors. It is resilient to most environmental impacts, including humidity, light touch/contact, and the space environment. It is birefringent, however, with angle-of-incidence dependent differences in the optical indices that could provide an extra factor for consideration in high-contrast imaging instrument design for HWO. Al+MgF$_{2}$ also has very limited reflectance in the bluest portion of the UV (91.2~nm $<$
$\lambda$ $<$ 120~nm ), a bandpass highlighted by the 2020 Decadal Survey as an objective for HWO sensitivity (\S\ref{section-hoststars} and \S\ref{section-lycont}). 

However, very thin layers of MgF$_{2}$ (and AlF$_{3}$; 1~--~2~nm) can now be applied as a second overcoat, or capping layer, to impart the protective and environmentally stable benefits of these fluorides on top of the primary reflectance and protective layers without imposing spectral restrictions on the final throughput. These thin depositions typically applied through an Atomic Layer Deposition (ALD) process \cite{Hennessy2018}, have been demonstrated to add strong environmental resiliency to the final optical surface at a modest cost in UV reflectance. These "capped overcoats" have been deployed in the lab and on suborbital missions and will be flown on small orbital astrophysics missions in 2024 and 2025, as described below. 

Lithium fluoride protected aluminum (LiF+Al) has the highest energy bandgap of any fluoride overcoat, with high reflectance to $\approx$102~nm and flight heritage on {\it OAO-3 Copernicus, FUSE}, and many sounding rockets \cite{Ohl2000,Moos2000,France2016}.

\begin{figure}
  \begin{center}
  \begin{tabular}{c}  
  \includegraphics[height=6cm]{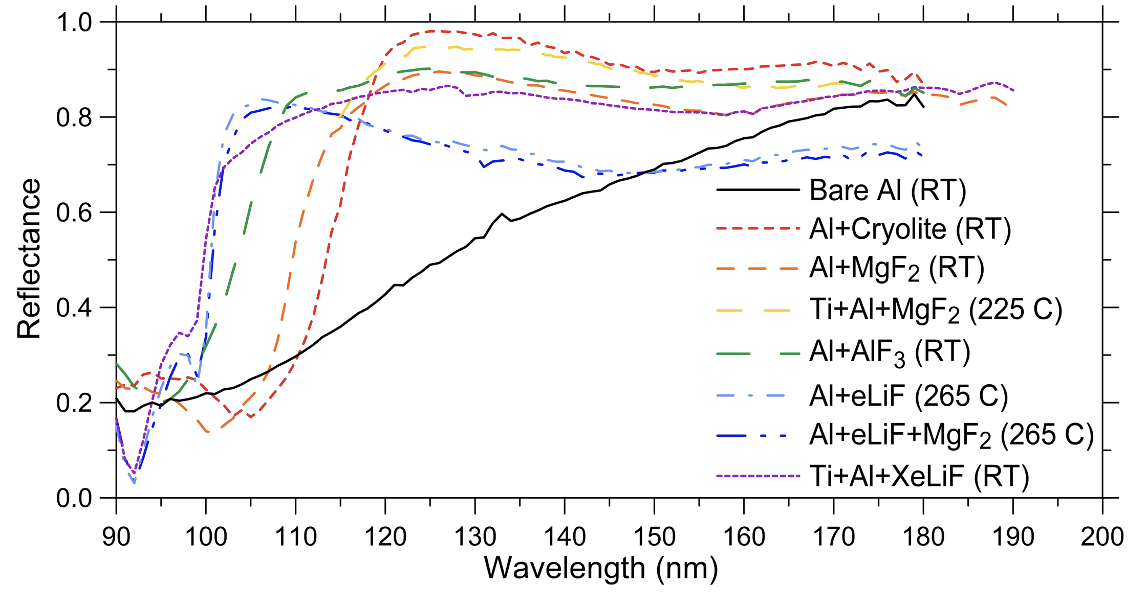}
  \end{tabular}
  \end{center}
  \caption{Comparison of experimental FUV reflectance of representative mirror coatings. This figure compares a broad range of coating materials and approaches. It is important to keep in mind that these coatings were not made to be compared - they are samples from different projects which each had their own sets of requirements and optimized bandpasses.}
  \label{fig:Coating-Technologies-UV1} 
\end{figure}

As can be seen in Figure \ref{fig:Coating-Technologies-UV1}, when one zooms in on the UV portion of the reflectance curves originally shown in Figure \ref{fig:Coating-Technologies-UVOIR}, there is a wide variation in reflectivity with material and application technique.

The curve that exhibits the highest reflectance ($>$ 96\% at the Lyman-alpha (Ly$\alpha$) wavelength of 121.6 nm is the one labeled as ``Al+Cryolite (RT)''. The cryolite overcoat is a salt that has a chemical composition of sodium, aluminum, and hexafluoride (Na$_3$AlF$_6$) that was deposited by a conventional PVD process done at ambient or room temperature (RT) \cite{DelHoyo2023}. This metal fluoride overcoat would seem as an attractive choice, given that it provides the highest reflectance at Ly$\alpha$ and longer wavelengths). However, it has the lowest band-gap energy (when compared to the other metal-fluoride materials) and this limit its use to wavelengths longer than about 118-120 nm. The cryolite salt is also a hygroscopic material that would have to be protected from degradation even in moderate levels of relative humidity ($>~\sim~30-40$\%). The curve with the second highest reflectance at Ly$\alpha$ ($\sim~91$\%) in  Figure~\ref{fig:Coating-Technologies-UV1} is the one labeled as "Ti+Al+MgF$_2$ (225 $^\circ$C)". This sample was made on a glass substrate that had a seed layer of titanium (Ti) under the Al coating and the MgF$_2$ deposition was done at an elevated temperature of 225 $^\circ$C \cite{Quijada2012}. It should be noted that this sample provides better performance when compared to the curve with legend "Al+MgF$_2$ (RT)" where the MgF$_2$ was deposited at ambient temperature. The third best sample (in terms of FUV reflectance performance) is labeled as "Al+AlF$_3$ (RT)". This sample consists of Al protected with AlF$_3$, and was done at room temperature by using the e-beam plasma process \cite{de2021room}. This sample exhibits a more balanced reflectance of around 90\% (at 121.6 nm), while it is over 80\% at wavelengths as short as 110 nm (on account of the larger band-gap energy of AlF$_3$ when compared to cryolite or MgF$_2$).

When performance is required at wavelengths as short as 100 nm, the metal-fluoride material of choice is LiF. The two sets of samples with the LiF protection shown in  Figure~\ref{fig:Coating-Technologies-UV1} are labeled as "Al+eLiF (265 C)" and "Al+eLiF+MgF$_2$ (265 C)". In reality, these two curves correspond to the same sample where the dashed blue line is the original sample (after the hot or enhanced LiF deposition done at 265 $^\circ$
C), whereas the solid blue curve is the reflectance after the application of a thin MgF$_2$ layer done via the ALD process \cite{Hennessy2018}. This thin ALD layer only reduces the reflectance by 1$-$2\% from the peak value of 80\% at 103 nm \cite{Rodriguez2022}. A most important consideration is the fact that the ALD overcoat of MgF$_2$ renders the sample much more environmentally stable when compared to the unprotected eLiF one \cite{Fleming2017}. 

An exciting recent coating possibility is shown in the "Ti+Al+XeLiF (RT)" curve. It was produced with the process described in Subsection~\ref{sec:XeLiF}. This XeLiF sample is made with the Al and LiF depositions followed by the fluorination of a XeF$_2$ precursor gas. This sample shows a modest reflectance of 70\% at 103 nm. However, the reflectance is over 80\% for wavelengths longer than 110 nm. Another remarkable feature of this sample is the fact that reflectance has been shown to be tolerant to relative humidity of around 60\% for over a 7-days period \cite{Quijada2022}.

\subsection{Polarization Sensitivity}\label{subsec:Pol-Tests}

\begin{figure}
 
  \begin{center}
  \begin{tabular}{c} 
  \includegraphics[height=5cm]{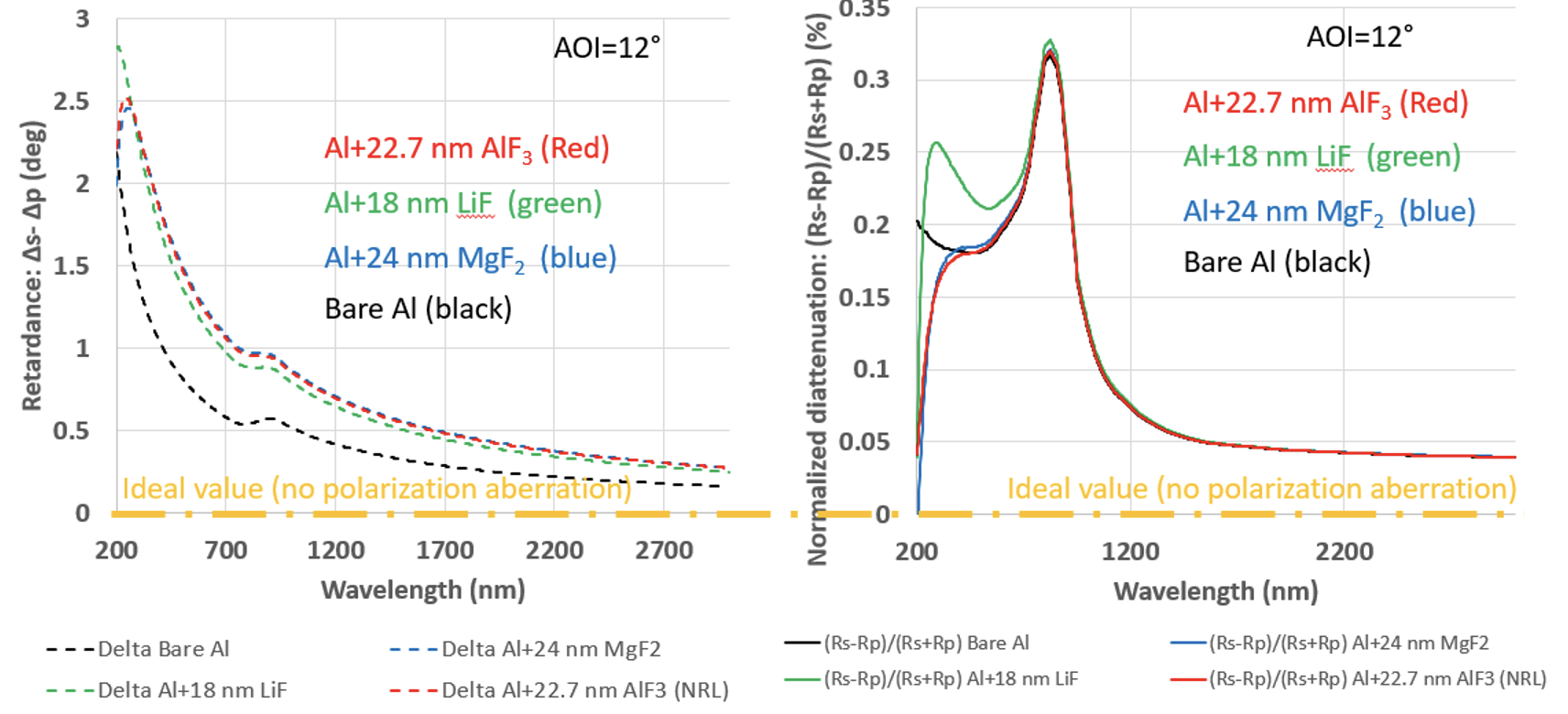}
  \end{tabular}
  \end{center}
  \caption{Polarization Sensitivity: Retardance (left) and normalized diattenuation (right) at AOI=12$^\circ$, which is the maximum acceptance angle stated in LUVOIR final report, of bare Al and Al protected with different fluorides (LiF, AlF$_3$, and MgF$_2$), calculated with experimentally derived optical constants.}
  \label{fig:Polarization} 
\end{figure}

The results of the calculations based on ellipsometric measurements of optical constants are shown in  Figure~\ref{fig:Polarization}. The left panel of this figure shows the calculated average phase retardance for bare Al coatings as well as protected with various metal fluoride overcoats (such as AlF$_3$, LiF, or MgF$_2$). These results show an average retardance at the angle of incidence, AOI=12$^\circ$ that increases with decreasing wavelength in the spectral range shown in the figure. It is worth pointing out that there were no requirements for this specified in the LUVOIR\footnote{https://asd.gsfc.nasa.gov/luvoir/reports/}/HabEx\footnote{https://www.jpl.nasa.gov/habex/pdf/HabEx-Final-Report-Public-Release-LINKED-0924.pdf} reports \cite{LUVOIR, HabEx}. Moreover, the average calculated coating-induced polarization is 0.11\% at AOI=12$^\circ$ in the 200$-$3000 nm spectral range. This number compares very favorably with the requirement specified in LUVOIR/HabEx reports of $<$~1\%. The coating-induced polarization measured at AOI=12$^\circ$ agrees with calculations. There are some concerns regarding differential polarization below 12$^\circ$, and this will represent a beam speed limit (and related packaging challenge) for the overall design. Overall, the choice of the protective metal-fluoride layer does not add much to the polarization aberration as long as this fluoride layer is not very thick.

\subsection{Coating Development towards HWO}

As has been demonstrated, several coatings have reasonable broadband performance in reflectance in the UV through NIR. To reach TRL 6 for HWO, these coatings will need to be successfully deposited onto a meter-class optic and then measured {\em on the optic itself} to demonstrate the required uniformity. This will require scaling the process through a focused development effort, as there are no other missions likely that will have meter-class UV optics on the horizon,
as well as the development of a test facility for mapping the reflectivity of a meter-class optic. These coatings should also then be deployed on feed optics for ground-based high-contrast imaging testbeds to demonstrate compatibility with the coronograph. We propose these paths while we await the work of GOMAP to define the exact requirements needed for the observatory. 

Any new deposition method should also require on-orbit testing for resilience in the relevant environment. CubeSats and SmallSats are ideal platforms for such testing; sounding rockets provide the most rapid path to space, but not long-term exposure. The following list presents current and recommended advancement steps for UV coatings in the next 3~--~5 years to advance the technology in support of a late-2020s Phase A start for HWO: 
\begin{itemize}
\item Scaling, SPRITE = 16 x 18 cm; the next steps for capped Al+eLiF coatings should be $\sim$~half-meter then meter-scale optics
\item Scaling of XeLiF process to 1 meter diameter (via a current SAT grant)
\item Extend current ALD work especially for larger optics (ongoing collaboration between JPL and UCSC for larger chamber (Private Communciation; J. Hennessy)
\item Demonstration of the most promising UV/O/IR coating candidate (to be evaluated in 2025-2026) in the space environment
\item Demonstration of repeatable uniformity segment-to-segment 
\item Demonstration of appropriate coating uniformity and polarization properties to meet the coronograph driven requirements
\item Demonstration of steps between state of the art and 1m, likely requiring new test facilities to demonstrate these from UV to NIR. 
\item Development of the characterization chain for HWO optical segments, including handling/shipping protocols and test facilities.

\end{itemize}

\section{Detectors}\label{sec-detectors}

Detectors are the heart of any telescope. In addition to being the end of the optical path, the sensor type, size of pixel, sensitivity, count rate limits, and other detector characteristics set many of the parameters for the instrument or telescope: e.g., a fast optical system typically requires small pixels, and vice versa. The limits and capabilities of detectors place strict limits on the science that can be achieved by any mission. UV detector development has lagged behind IR, due in part to a lack of commercial and military funding sources. In the last 15 years, NASA has invested in UV detector technology development via the APRA and SAT programs. As a result of these and other investments, there are several compelling UV detector technologies. Since at least the 2010 Decadal survey\footnote{https://nap.nationalacademies.org/catalog/12951/new-worlds-new-horizons-in-astronomy-and-astrophysics}, UV astrophysics has worked towards a detector that can achieve the ``triple crown'' of sensors: Large format, high quantum efficiency, and low noise/high SNR properties. We also consider power requirements, out of band sensitivity and rejection, For spectroscopy, this often translates to a photon counting device. For imaging, photon counting is typically not needed. Below, we detail the current state of the art for microchannel plates and solid state detectors. 

\subsection{Microchannel Plates}

Microchannel Plates (MCP) are widely used as an amplification stage in particle and photon detectors. They are comprised of parallel micron-diameter glass capillaries bundled into a two-dimensional array. Amplification occurs when a ``primary'' electron or other ionized particle, after being accelerated in an electric field, strikes the wall of a pore and excites multiple ``secondary'' electrons within the material. ``Secondary electrons'' escaping the pore surface are again accelerated in the electric field, eventually striking the wall, and generating a cascade effect as the process continues down the pore \cite{wizaMicrochannelPlateDetectors1979}. Typical MCP based photodetectors incorporate a photocathode to convert photons to photoelectrons, a stack of 2 or 3 MCPs to achieve gains $>10^6$, and a position sensitive anode. 

\begin{figure}[htb!]
\centering 
\includegraphics[width=0.45\textwidth]{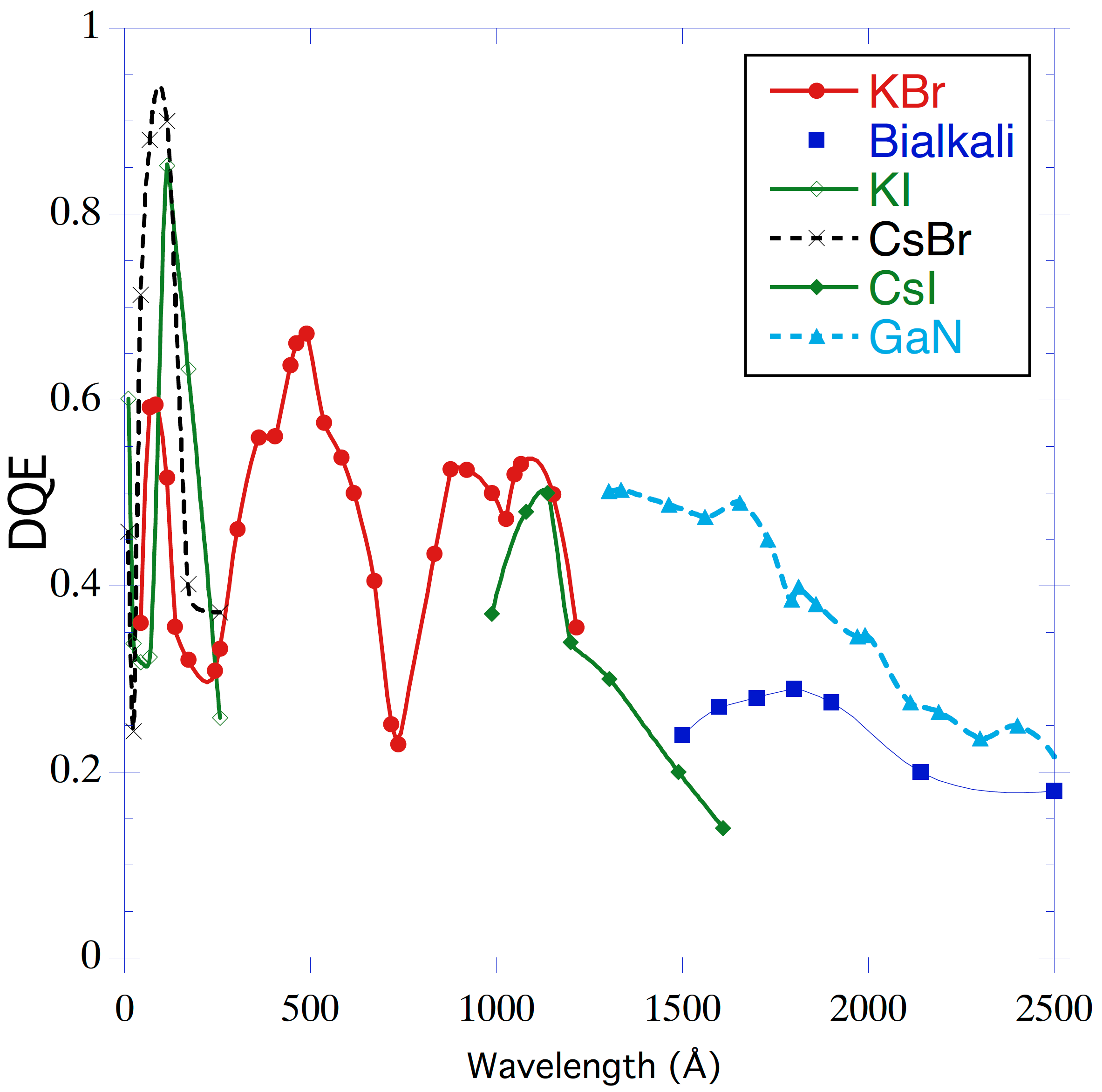}
\caption{Detective quantum efficiency (DQE) as a function of wavelength for different photocathode materials as shown in the key.}
\label{fig:photocathode_qe}
\end{figure}

Detectors based on MCPs have flown on dozens of missions including sub-orbital, shuttle, space station, low Earth orbit satellite, geosynchronous satellite, and planetary missions, with hundreds of successful operational years accumulated. MCP detectors have been implemented for many successful UV astronomy missions and instruments, such as {\it Extreme Ultraviolet Explorer (EUVE), Far Ultraviolet Spectroscopic Explorer (FUSE), Galaxy Evolution Explorer (GALEX), Hubble Space Telescope (HST) Space Telescope Imaging Spectrograph (STIS)}, and HST’s {\it Cosmic Origins Spectrograph (COS}) \cite{siegmundMicrochannelPlateEUV1984,siegmundPerformanceDoubleDelay1997,jelinskyPerformanceResultsGALEX2003,vallergaHSTCOSFarultravioletDetector2001}. They are also on the recently launched NASA {\it Ionospheric Connection Explorer (ICON)} SMEX mission \cite{korpelaInFlightPerformanceICON2023}, {\it Global-scale Observations of the Limb and Disk (GOLD)} Mission of Opportunity \cite{mcclintockGlobalScaleObservations2020b}, and are widely used on NASA UV suborbital sounding rocket and CubeSat investigations. SwRI's {\it Alice/UVS} line of FUV spectrographs have a cumulative 50+ years of failure-free operation, with five instruments in flight and a sixth due to launch in 2024. Much of this success is partially attributable to the robustness and durability of MCP detector construction, their adaptability to a wide range of formats (including curved focal planes), choice of the sensitivity bandpass (utilizing different photocathodes as seen in Figure \ref{fig:photocathode_qe}), lack of red end sensitivity, very low background noise and no requirements for cooling. MCP detector technology using imaging readouts, electronics, MCPs, and photocathodes, in open-faced and sealed-tube packages has continued to make significant advancements in the last few years. Considerable improvements in background rate, quantum efficiency, spatial resolution, event handling rates, maximum size formats, thermal and mechanical robustness, lifetime stability, and low power/mass electronic readouts are all in various stages of implementation and infusion.

A traditional MCP is implemented as a wafer of lead glass tubes forming a microcapillary array of semi-conductive tubes with a secondary emissive layer on the glass surface. Conventional manufacturing techniques are limited to formats less than $\sim$150 mm. The gain of traditional Pb-glass MCPs degrades with usage, requiring extended ``burn-in'' periods to stabilize the gain. \cite{darlingMicrochannelPlateLife2017a,mcphateLifeTestingConventional2019,martinFurtherScrubbingQuantum2003} Pb-glass MCPs also have background rates set by radioactive materials present in the glass (e.g., $^{40}$K) and have\char`\~ 2\% detection efficiency for gamma-rays because of the Pb in the glass \cite{schindhelmMicrochannelPlateDetector2017}. 

The advent of atomic layer deposition (ALD) has opened the door to new MCP manufacturing techniques which have shown improvements to MCP performance. ALD is ideal for coating high aspect ratio parts, such as MCPs, and is compatible with materials demonstrating high secondary electron yield \cite{jokelaCharacterizationSecondaryElectron2011}. These coatings have been applied to Pb-glass MCPs, for example, the MCPs used in the JUICE-UVS \cite{davisJUICEUltravioletSpectrograph2021}, to achieve better gain and detector lifetime, but the MCPs still retain the mechanical properties and background characteristics of conventional MCPs.

\begin{figure}[htb!]
\centering 
\includegraphics[width=0.45\textwidth]{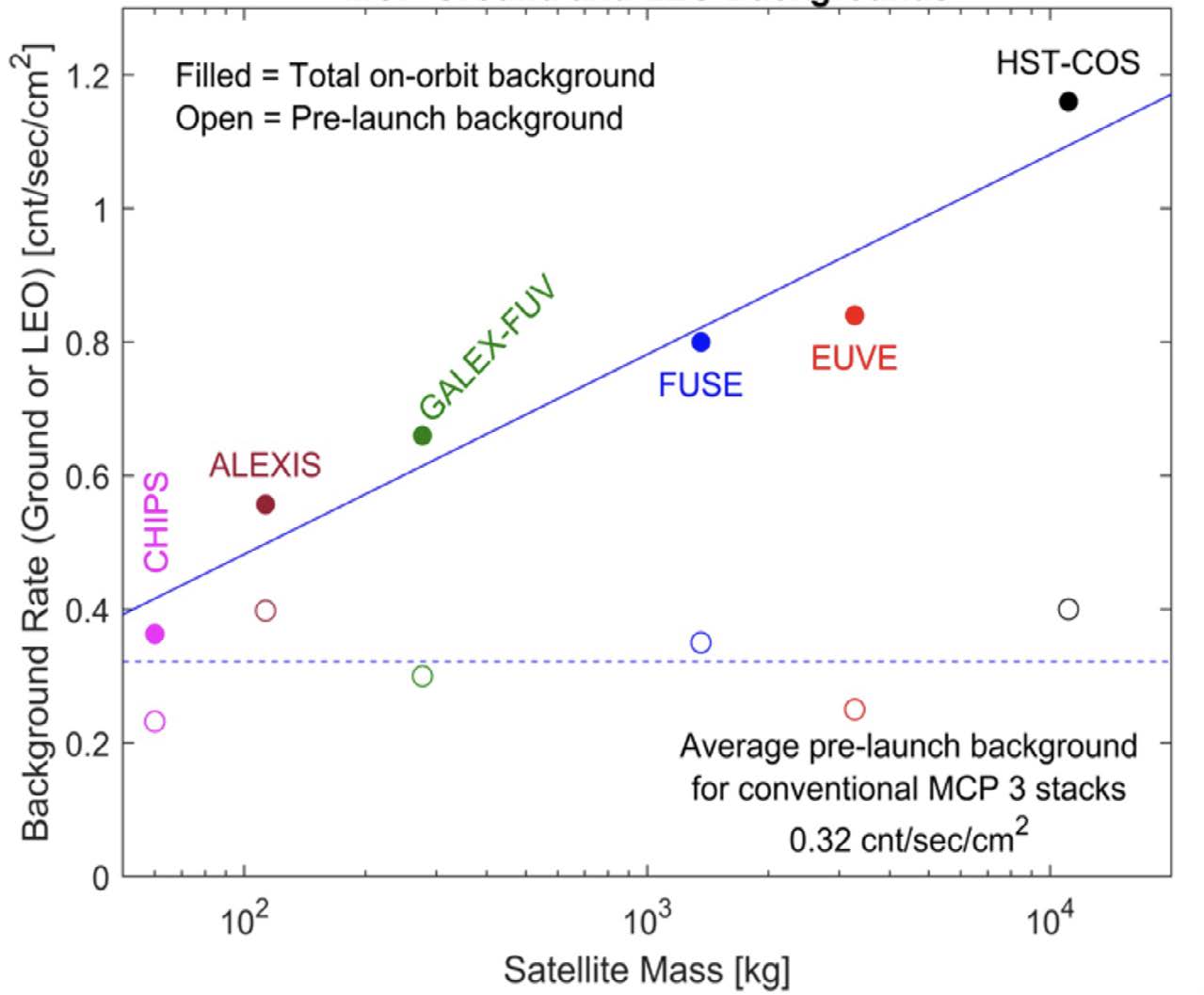}
\caption{Background for conventional MCP detectors in low Earth orbit. The trend suggests an on-orbit background rate increase consistent with the log of the satellite mass.\cite{siegmundDevelopmentUVImaging2020} ALD technology should significantly reduce this effect.}
\label{fig:mcp_flight_bkg}
\end{figure}

New technology Borosilicate glass MCPs utilize ALD to provide both resistive and emissive surfaces. They have many advantages over traditional Pb glass MCPs, including more mechanical robustness (larger formats of 200mm have been successfully flown), very low intrinsic background (minimal radioactive content), a factor of 3 reduction in gamma-ray detection efficiency (no lead), and significantly improved gain stability. ALD MCPs with large-area formats (\qtyproduct[product-units = power]{12 x 12}{\cm} with \qty{10}{\um} pores, up to \qtyproduct[product-units = power]{20 x 20}{\cm} with \qty{20}{\um} pores), and very low background rates (\qty{< 0.05}{events\per\square\cm\per\second}), extended lifetimes (\qty{> 4e13}{events\per\square\cm}) without degradation have been made \cite{siegmundPerformanceCharacteristicsAtomic2013,ertleyMicrochannelPlateDetectors2018}. Historical mission data shows that the in orbit background (for LEO) is strongly affected by the mass of the satellite converting galactic cosmic rays to local radiation background via Bremsstrahlung (See Figure \ref{fig:mcp_flight_bkg}) . ALD borosilicate MCP technology should also help reduce these contributions for both the pre-launch and in orbit backgrounds by a factor of 2.

\begin{figure}[htb!]
\centering 
\includegraphics[width=0.45\textwidth]{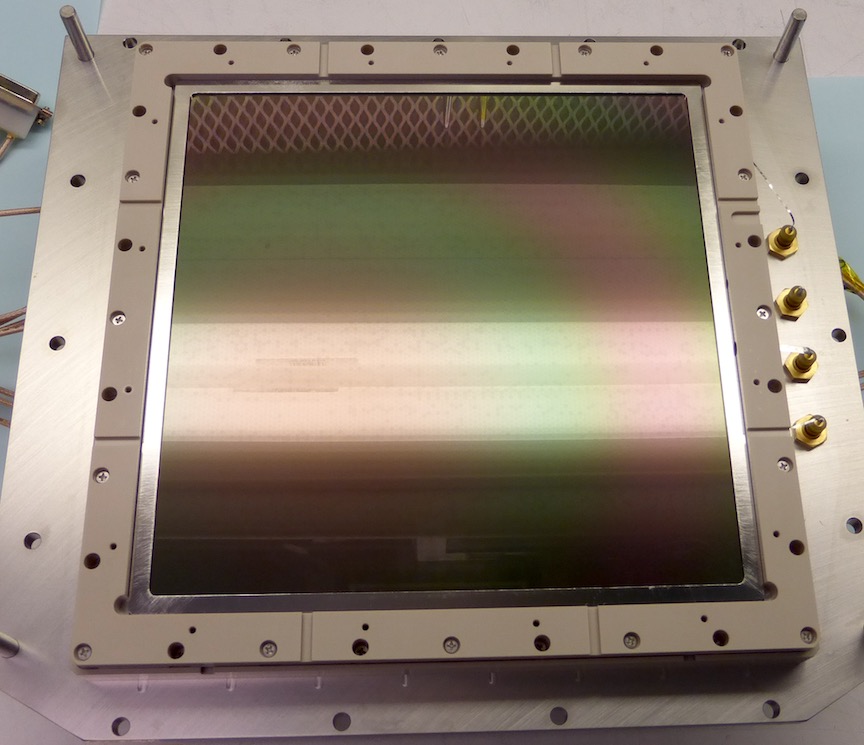}
\caption{20 x 20 cm Microchannel Plate with XDL readout for the DEUCE rocket program.}
\label{fig:deuce_detector}
\end{figure}

\begin{figure}[htb!]
\centering 
\includegraphics[width=0.45\textwidth]{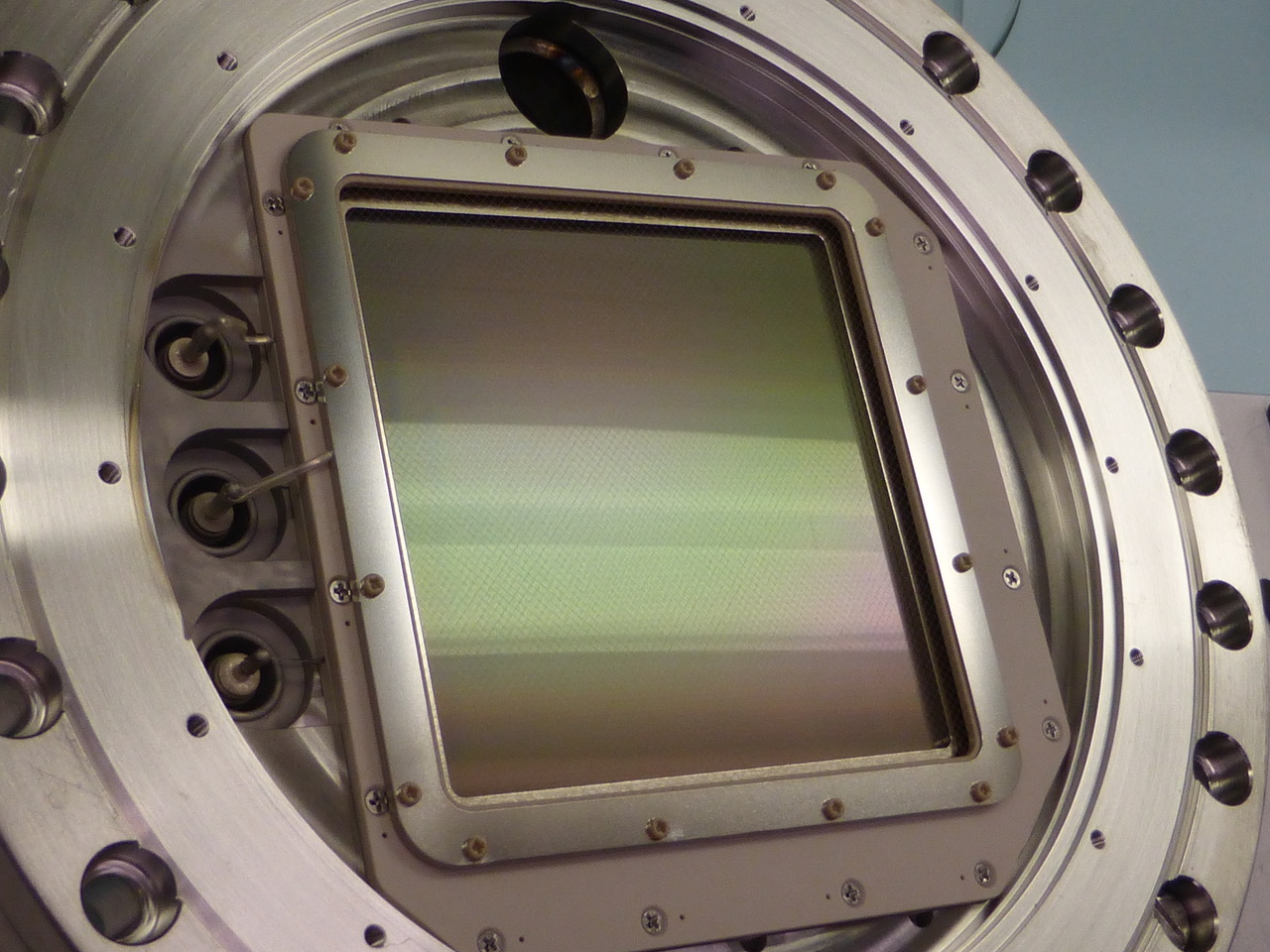}
\caption{INFUSE (2023) 100 mm XS Detector with \qty{25}{\micro\meter} resels.}
\label{fig:infuse_detector}
\end{figure}

\paragraph{Photocathodes}
The conversion of photons to electrons by the photocathode and photoelectron detection by the MCP determine the MCPs’ quantum efficiency (QE) \cite{siegmundNovelLargeFormat2011}. The photocathode can be deposited on a window directly in front of the MCP (a proximity-focused semitransparent cathode) or directly onto the MCP (opaque mode) \cite{jelinskyProgressSoftXray1996,tremsinUVRadiationResistance2001,larruquertOpticalPropertiesQuantum2002,tremsinQuantumEfficiencyStability2005,ertleyDevelopmentOpaquePhotocathodes2017}. Opaque alkali-halide photocathodes are widely used for EUV/UV sensors, as are semitransparent multi alkali photocathodes for NUV detectors. EUV/UV photocathodes have broadband response with efficiency peaks at wavelengths where photoelectron emission maximizes. These reach to more than 50 percent detective quantum efficiency (DQE) around 100~nm, more than 60 percent at about 500~nm, and over 70 percent around 12~nm depending on the material chosen. NUV DQE is somewhat lower at about 30 percent at 180~nm for bialkali photocathodes. Alkali-halide (CsI, KBr) opaque photocathodes on MCPs obtain high QE (50\% at \qty{\sim 110}{\nano\meter}) and have broadband sensitivity from \SIrange{10}{160}{\nano\meter}. We compare photocathode material performance in Figure \ref{fig:photocathode_qe}.

\paragraph{Readout Techniques}
Cross delay line, and more recently cross strip conductive patterned anodes are often the choice for UV MCP imaging detectors. These anode schemes derive photon event centroid positions from the charge distribution that an event produces across a set of strips in an MCP amplified detector. The incoming photon produces a primary electron(s) from the photocathode. Each photoelectron is then multiplied within the pores of a microchannel plate pair and the resulting electron cloud is collected on two orthogonal sets of metal strips that form the anode. To obtain an accurate event position the size of the electron cloud is optimized so that the charge impinges on several neighboring strips. The cross delay line derives positions from the difference in pulse arrival times at opposing ends of serpentine delay lines in each axis formed from connecting the strips. The cross strip anode provides signal amplitudes on each strip allowing sub-strip accuracy centroiding to be accomplished. The cross strip scheme achieves high spatial resolutions (sub 20 µm) at relatively low overall MCP gain (less than 1,000,000). Recently, cross strip anode formats of 100~mm have been implemented in both open face (Figure \ref{fig:deuce_detector}) and sealed vacuum tube compatible (UHV/500 deg C) configurations (Figure \ref{fig:infuse_detector} as a significant step towards large area devices suitable for HWO spectroscopy detectors.

\paragraph{Imaging electronics}
Candidate MCP detector photon event readout systems include both cross delay line (XDL) and cross strip (XS) anodes. Low power, high performance electronics for cross delay lines have been implemented in a number of forms for existing missions and can be finessed for future application using current state of the art components. However, XS readouts are more appropriate for large area, high performance detectors for HWO. XS electronics implemented with discrete components are working well for recent instruments, but recognizing the need for low mass/power electronics for future applications an effort is underway to produce compact ASIC electronics. Prototypes (GRAPH - Gigasample Recorder of Analog waveforms from a PHotodetector) have already demonstrated low power consumption and promise the ability to process events at high rates (more than \qty{10}{\mega\hertz}) with power consumption of less than \qty{7}{\watt} for a 50mm detector \cite{seljakASICsReadoutSystem2018}. GRAPH ASICs are also configured to be used in parallel so that larger format XS detectors are addressable with adapted firmware.

\subsection{Solid State Detectors}

Solid-state detectors offer significant advantages in size, mass, and manufacturability compared to the vacuum tube based technology of MCPs \cite{Nikzad2012}. Using solid-state detector arrays improves instrument compactness and reduces instrument complexity.

There are intrinsic advantages to the fabrication of solid-state UV detectors in wide bandgap material such as gallium nitride and its alloys. Their bandgaps, ranging from 4.6 eV (GaN) to 6.2 eV (AlN), offer a broad range of out of band rejection thresholds, while their direct bandgap for the entire range of alloys enables strong absorption of UV photons, and their wide bandgap could enable higher operating temperatures. There are, however, practical material challenges to fabricating high quality detectors in III-N materials. The lack of native substrate in the III-N family creates challenges for producing high quality crystalline material and doping of these semiconductors---the first step of creating a semiconductor junction and detectors---remains challenging.

Silicon carbide is another wide bandgap (3.26 eV) material with direct bandgap and intrinsic properties for visible rejection and higher operating temperatures. While SiC is not as tailorable as III-N material for the cutoff wavelength, it can be grown on silicon wafers, a clear advantage for manufacturability. 

There is enormous past and continuing investment in silicon imaging devices which can be leveraged for scientific UV imaging. Silicon imaging and detector arrays, especially charge coupled devices (CCDs) and complementary metal oxide semiconductor (CMOS) arrays, are ubiquitous in different fields of imaging applications. This is due in part to the steady advancement of silicon VLSI (very large-scale integration) technology and the consumer market for imaging. Shortly after the invention of CCDs in 1969 at Bell Labs, extensive programs were established at JPL to advance early CCDs for imaging systems aboard NASA’s Flagship mission {\it Galileo8} and its first Great Observatory, {\it HST}. For a more in-depth discussion of the history of CCDs and their use in space-based applications, we refer the reader to \cite{Janesick2001} and \cite{NikzadChapter2015}. CMOS-based imaging started around the same time in the 1960s, but it was not until the 1990s that their development began in earnest, due in part to fabrication advancement and to focused effort of CMOS-Active Pixel Sensor development at JPL and other CMOS images elsewhere. CMOS imaging has had an enormous impact on the consumer field and in recent years has become viable for scientific applications.

\begin{figure}
  \begin{center}
  \begin{tabular}{cc} 
  \includegraphics[width=0.49\textwidth]{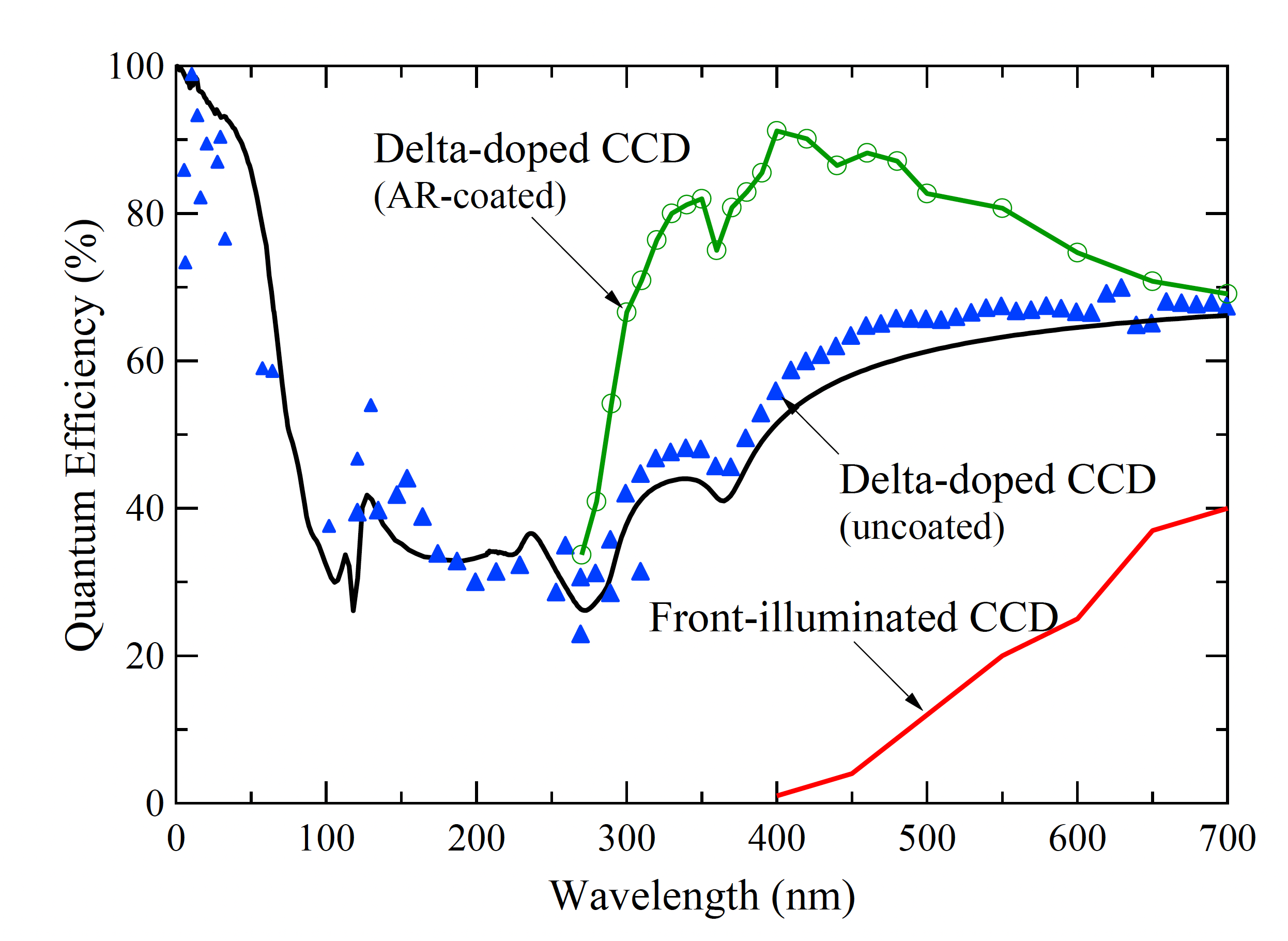} &
  \includegraphics[width=0.49\textwidth]{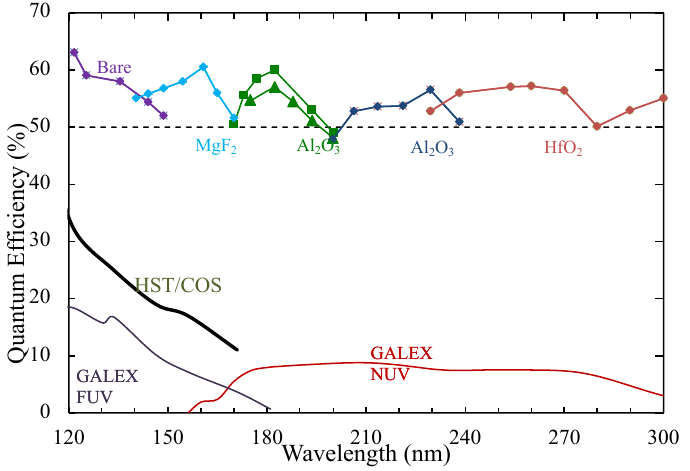}
  \end{tabular}
  \end{center}
  \caption{Delta-doping involves a doping layer to a back-illuminated silicon detector, which eliminates an electron well in the detector surface. The result is silicon detectors sensitive in the UV. The QE can be further increased by the addition of anti-reflection coatings. Left: solid lines are models, markers are measurements \cite{Hoenk2022}, showing early results of delta-doped and AR Coated devices. Right: single layer AR Coatings using atomic layer deposition (ALD) and thermal deposition (MgF$_2$ only) are applied for higher throughput in the NUV and FUV \cite{Nikzad2012}.}
  \label{fig:DeltaDoped} 
  \end{figure}

In the UV, the short photon absorption distance coupled with the formation of carrier traps (electrons in n-channel and holes in p-channel devices) in the Si-SiO\textsubscript{2} result in negligible QE below 400 nm (Figure \ref{fig:DeltaDoped}) in the front illuminated silicon devices or even unpassivated back illuminated silicon devices.

The naturally-occurring traps due to positive charge trapping in the Si-SiO2 interface, can be countered by the addition of charges of opposite polarity, e.g., a thin crystalline layer of silicon with embedded high boron density in a single atomic sheet. This delta-doping process \cite{Hoenk2022} results in silicon detectors that have reflection-limited response, i.e., near 100 \% internal QE from very soft x rays through near infrared, $\sim$ 1 $\mu$m, at which point silicon becomes transparent to photons. Because delta doping is a back surface process, it is agnostic to the readout structure and silicon detector architecture and essentially any silicon detector can become UV sensitive, allowing the mission to benefit from the very large commercial pool and latest designs in silicon detectors.

With photo-electrons efficiently collected in silicon detectors through delta doping, the QE can be further tailored by the addition of anti-reflection coatings or filters for out of band rejection. QEs as large as 50\% to 60\% are possible, with nearly 3-4 orders of magnitude out of band rejection with bandpasses whose location can be arbitrarily chosen.   

\begin{figure}
  \begin{center}

  \includegraphics[width=\textwidth]{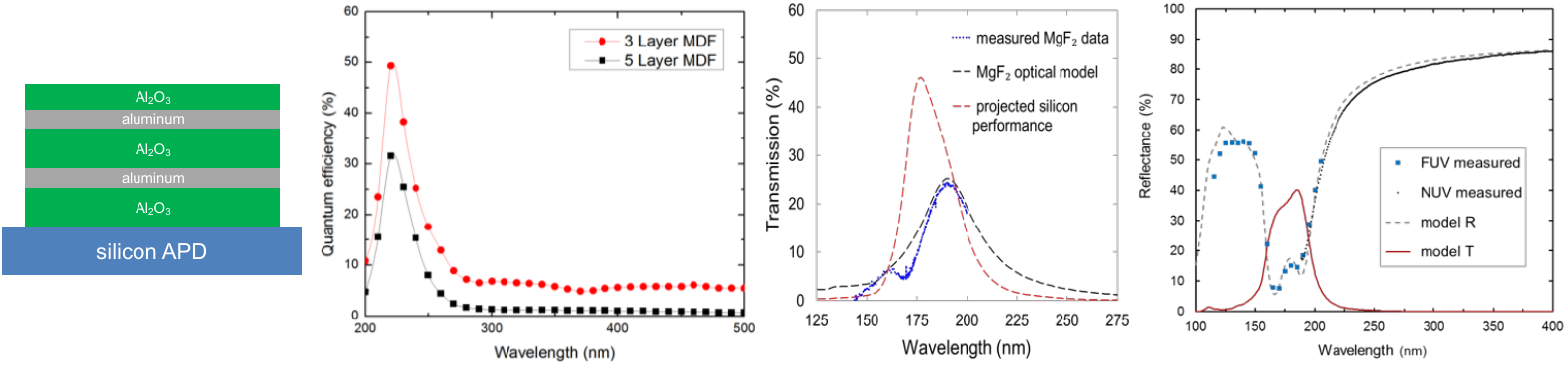}
  \end{center}
  \caption{Left: Schematic of the implementation of a metal dielectric filter (MDF). Left center: Linear scale QE of two example MDFs optimized for 225nm. Right Center: Alternative MDF example, with peak performance at 190 nm. Right: MDF optimized for FUV Bandpass, showing both reflectance from detector with integrated filter and transmission into detector.}
  \label{fig:mdf} 
  \end{figure} 

Further tailorability can be achieved by detector-integrated filters.Three to four orders of magnitude rejection has been obtained by incorporating metal dielectric filters directly on delta doped detectors. The think aluminum layer provides blocking of the visible light while allowing UV photons to reach the detector surface. Depending on the spectral range of interest, the dielectric material of choice would be an oxide or fluorides. Adding each metal-dielectric bilayer will provide additional out of band rejection but because the metal layer is not entirely transparent, there will be tradeoff between the out of band rejection and in band peak quantum efficiency.

\paragraph{Architecture variants} 

\begin{figure}
  \begin{center}

  \includegraphics[width=\textwidth]{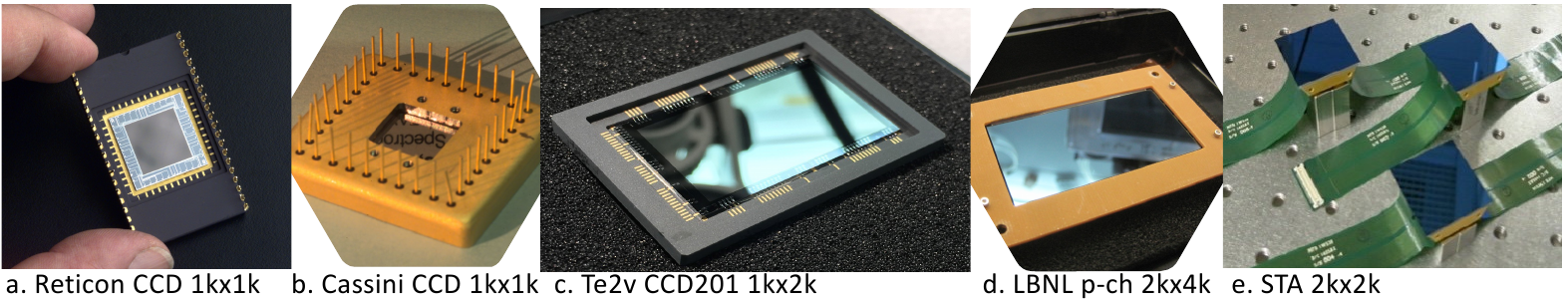}
  \end{center}
  \caption{A range of Delta doped and AR coated devices spanning various formats (1kx1k, 2kx1k, 2kx2k, 2kx4k, etc) and architecture (EMMCD, CCD, Full depletion CCD, etc) and polarity (n-channel and p-channel CCDs) from JPL which have been used for suborbital and ground-based missions. Because the processes developed are independent of the frontside architecture and format, the end-to-end post fabrication processing including delta doping and special coatings }
  \label{fig:detector_types} 
  \end{figure}

Solid-state Silicon devices all rely on similar architectures- a pixel for photo-electron collections, either directly connected to a readout amplifier in the case of CMOS devices or coupled to neighboring pixels in the case of CCDs. For a CCD, at least one and up to tens of readout amplifiers are used for reading out the charge in all pixels in a single device. The CCD architecture has been the dominant detector of choice for visible wavelength and ground based astronomy. Due to their serial readout scheme, CCDs have longer readout times than CMOS devices but have historically had lower noise in the amplifier, and fewer amplifiers per device make image processing more straightforward. In recent years, CMOS devices have advanced such that the read noise in each amplifier is low enough for scientific use and variations between pixels are no longer a significant disadvantage. Depending on how they are designed and operated, CMOS devices can have low fill factors (area of a pixel that is photo-sensitive) and are sometimes matched with microlenslet arrays to improve throughput. In recent years, low noise and back illuminated scientific CMOS imaging arrays have been developed. The recently selected Explorer UVEX is planning the use of 4kx4k delta doped and low noise CMOS arrays. The FUV version of the UVEX detectors plan to use detector-integrated filters. 

\paragraph{Solid state photon counting detectors}

When using a normal silicon CCD, it is impossible to tell the exact number of photo-generated electrons in a pixel because the added detector noise is usually too large; with conventional CCD read noise of a few electrons, a single sample will be unable to distinguish between individual electrons in a well. Several different technologies enable photon counting in a solid state detector, which we describe briefly below.

A variant of CCD is the Electron Multiplying CCD (EMCCD), which implements an additional set of pixels between the normal CCD serial register and the readout amplifier \cite{2001Mackay,2001Jerram,2004Gach}. They consist of a normal CCD image area with an additional serial register added after the normal serial register. This addition to the serial register of an EMCCD contains serial pixels that replace one clock of the pixel with a DC level pixel and a high voltage (HV) clock. These multiplication pixels have a high voltage applied to them, creating deep wells for electrons to impact and ionize as they enter the pixel. This impact ionization generates more electrons, yielding a significant gain over the course of several hundred multiplication pixels. This allows for photon counting with a CCD since the multiplication gain will increase the number of electrons in a pixel to well above the read noise. EMCCDs need to be operated carefully, however, to limit only a single photon event per pixel and have other noise sources that are added by the multiplication process.

\begin{figure}
  \begin{center}
  \begin{tabular}{cc} 
  \includegraphics[width=.25\textwidth]{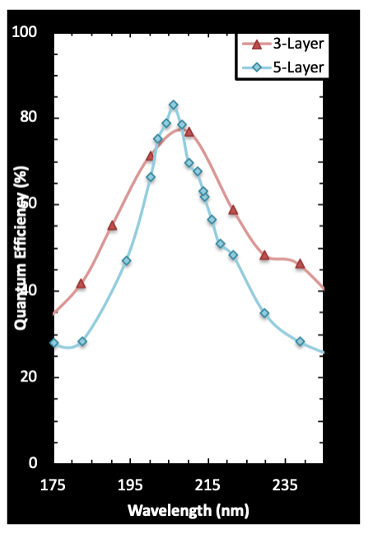} &
  \includegraphics[width=.7\textwidth]{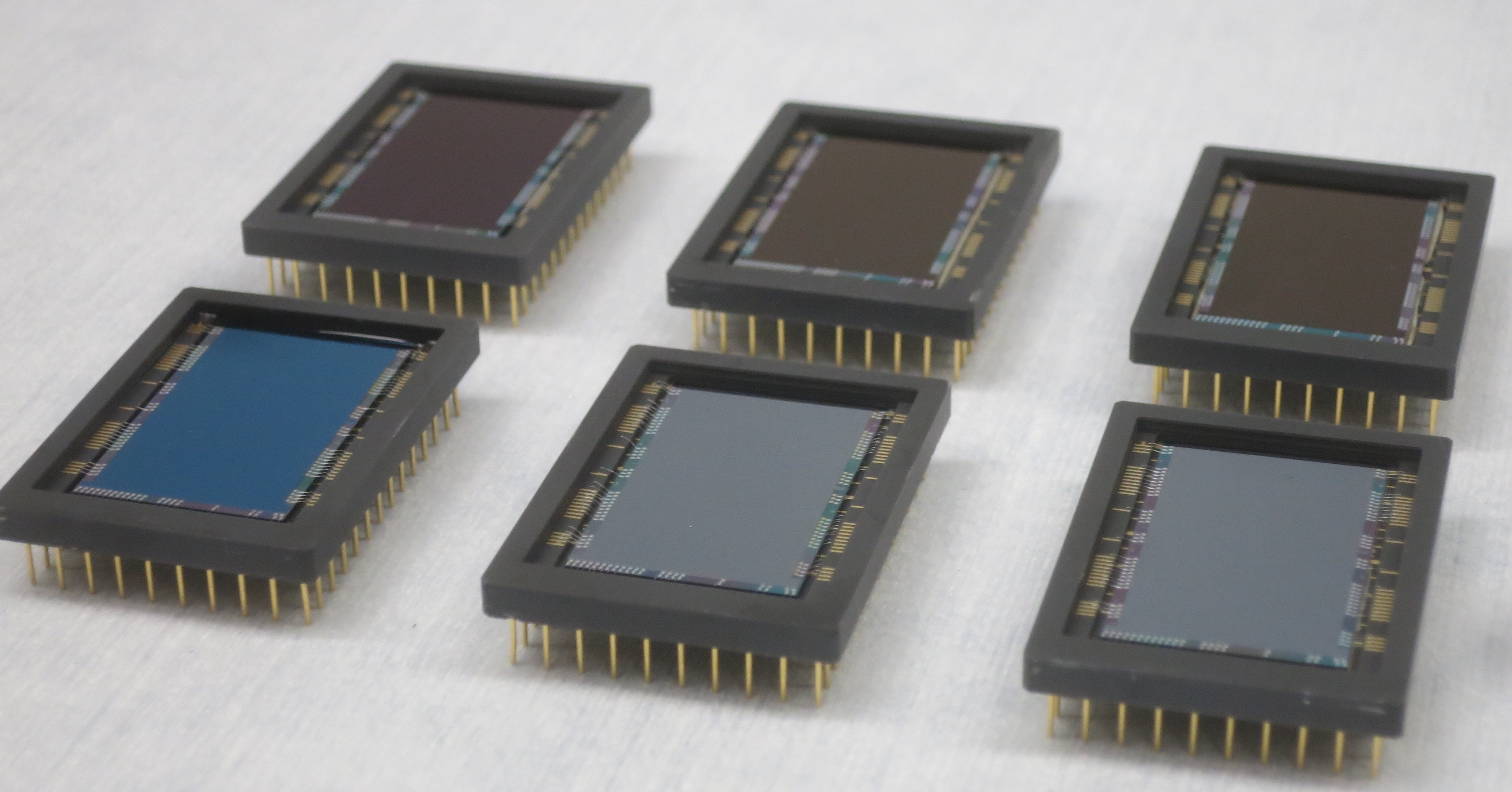}
  \end{tabular}
  \end{center}
  \caption{Delta doped and AR coated 1kx2K EMCCD prepared for FIREBall-2. LEFT: quantum efficiency of a three-layer AR coated device (red) and five layer AR coated device (blue). RIGHT: photographs of delta-doped 2KX1K EMCCDS. Different colored detectors have different AR coatings applied.}
  \label{fig:emccd} 
  \end{figure}

The advantage of the EM gain process is that it increases the signal from a single photo-electron to a value much larger than the on-chip amplifier read-noise. This process means single events can be detected by a threshold process \cite{2011Tulloch}. When operated in photon counting mode, pixels with counts greater than 5 times the read-noise are considered to have had 1 event. Pixels with counts less than this threshold are considered to have zero events. Some complications come along with this process. Not all electrons in the device will be amplified above the 5 sigma threshold. The multiplication process is stochastic and depends on the gain applied \cite{2008Daigle}. Additionally, noise comes from clock induced charge (CIC) and dark current. For an EMCCD, both of these noise sources will be amplified in the same manner as photo-electrons and need to be addressed in analysis. EMCCDs are currently used on several missions at a range of mission sizes: {\it FIREBall-2} (a balloon borne UV multi-object spectrograph), the {\it Roman Space Telescope}, {\it SHIMCO} (an astrophysics UV Rocket). EMCCDs are also baselined for several missions in planning stages: a New Frontier's {\it SILENUS} \cite{Balachandran2020}, {\it PRISM} (a lunar mission), a lunar DALI (Development and Advancement of Lunar Instrumentation) instrument, and was the baseline detector for {\it HabEx}.

\begin{figure}
  \begin{center}
  \begin{tabular}{c}  
  \includegraphics[width=0.56\textwidth]{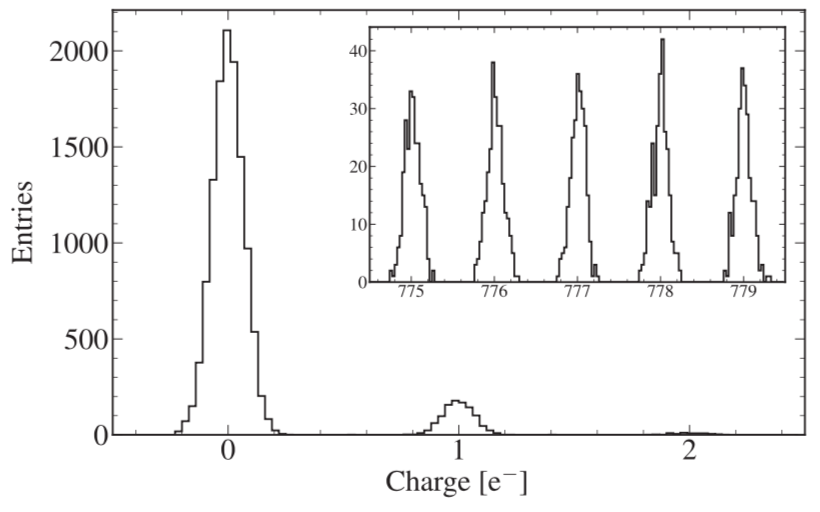}
  \end{tabular}
  \end{center}
  \caption{Histogram obtained in the visible showing the photon counting and photon number resolving capability of
Skipper CCD. Each peak represents an integer number of electrons in a bin. The main figure shows the clear delin-
eation between pixels with zero electrons and pixels with one electron. The inset figure shows that this clear separation is present at higher electron counts, 775 to 779 electrons per pixel. Figure from \cite{2017Tiffenberg}}
  \label{fig:skipper} 
  \end{figure}

\begin{figure}
  \centering
  \includegraphics[width=0.55\textwidth]{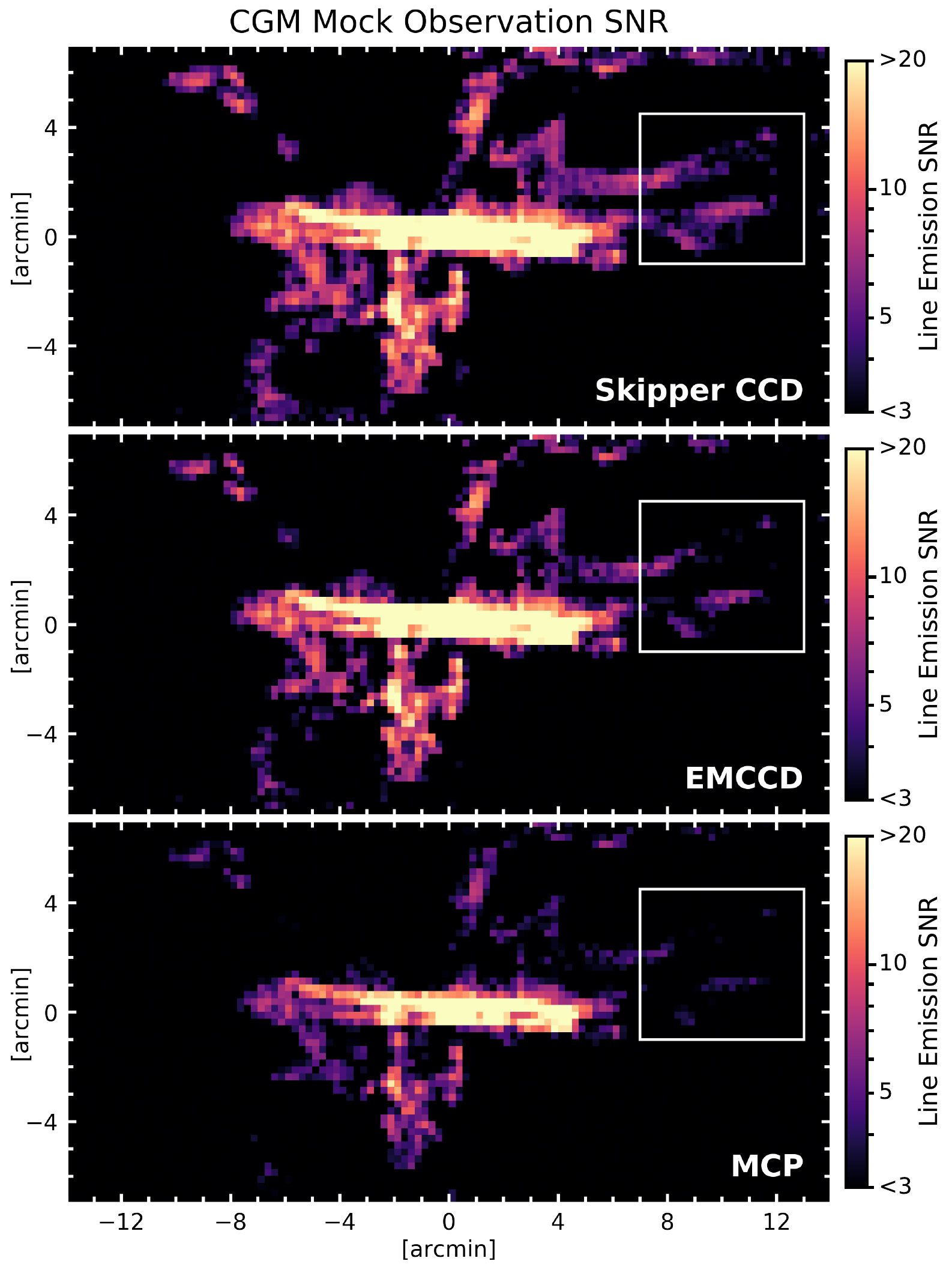}
  \caption{Simulated CGM around a Milky Way type galaxy shows faint, extended, filamentary structure. The panels show mock observations and the resulting SNR distribution (random-noise effect included) from an IFU with the simulated performance of a (Top) Skipper CCD, (Middle) EMCCD, and (Bottom) MCP. The white rectangle shows that the Skipper CCD can detect a very faint diffuse gas signal with greater fidelity compared to an EMCCD or MCP for otherwise the same conditions.}
  \label{fig:SNR}
\end{figure}

One of the emerging technologies that may provide promise for astrophysical photon counting is the skipper CCD architecture. A skipper CCD uses a non-destructive read-out amplifier and is able to sample a single pixel multiple times. This effectively reduces the read noise by several orders of magnitude depending on the number of samples. The reduction in noise is equal to the square root of the number of samples. This type of device had been described previously by \cite{Janesick2001} but has finally been made workable (e.g. \cite{Marrufo2022}). As an example, using a 3.5e- read noise amplifier and 4000 samples per pixel, the effective read noise is reduced to 0.068, well below the typical value of 0.16 usually required to clearly distinguish between numbers of electrons. Thus a skipper CCD can distinguish between 0 and 1 electrons in a pixel, which is the typical expectation for photon counting, and matches what an EMCCD or MCP can do. But it can also distinguish between 1 and 2, 2 and 3, 3 and 4, and so on up to the saturation limit of the well (see Figure \ref{fig:skipper}). This means a skipper can perform photon counting and is not limited by the typical rate requirements of an EMCCD (which can only distinguish between 0 and 1 electrons in a pixel, if the frame rate is correct for the observation) or an MCP. An example of the ability to detect faint sources is shown in Figure \ref{fig:SNR}, where various detector types are contrasted for an observation of faint emission from the CGM of a nearby galaxy.

\section{Gratings}\label{sec:gratings}

UV spectroscopy has historically suffered from low-performance hardware compared to other wavebands, in part due to the inherent challenges in material properties at UV wavelengths (absorption in common VIS/IR substrates, degradation over time, etc). 
In particular, UV high-resolution spectroscopy is limited by low efficiency gratings (typical efficiencies of $<$40\%; e.g., \cite{Content+1996,Hoadley+2016}) and scattered diffracted light that severely increases background levels (e.g., \cite{Landsman+1997,Hoadley+2016,Kruczek+2018}). 
The biggest limitation to building high-sensitivity, high-resolution UV spectrographs is the low performance of UV blazed gratings. Advances must be made to achieve the high-efficiency blazed gratings that will be necessary to enable a high-resolution (R$>$30,000) UV spectrograph on HWO. 

Historically, UV gratings used in astronomical instruments have been ruled either mechanically or holographically. Mechanically ruled gratings are produced using a diamond stylus to cut and shape grooves, while holographic gratings are produced using interference lithography, resulting in a sinusoidal groove profile at the specified density. Mechanically ruled gratings are capable of good diffraction efficiency as a result of their sharp, blazed facets, but typically suffer from groove period errors that induce scatter and ghosting. Conversely, holographic gratings are capable of precise groove placement, which minimizes stray light, but generally lack the diffraction efficiency achievable with a mechanical ruling process. While it is possible to improve upon holographic grating efficiency by directly recording a blazed facet in the photoresist using sophisticated holographic recording techniques, these techniques are resource intensive. In addition, for both ruling techniques, fabrication processes limit the achievable groove spacing; mechanical ruling is not suitable for very high groove densities, while holographic ruling is not well suited for low groove densities (providing desireable higher spectral resolution). \cite{Kruczek22}. Further, each process offers limited customizability in terms of variable line spacing and aberration control in the groove pattern, resulting in little flexibility in grating design and performance for UV spectrographs \cite{Grise21}.

\subsection{UV gratings with electron-beam lithography}

In response to NASA’s Astrophysical Strategic Technology Priorities and Gaps program, which recognized the need to improve all aspects of UV hardware, enormous progress has been made on the crucial task of improving UV gratings. Electron-beam lithography (EBL), enabled by semiconductor nanotechnology, offers a flexible method for patterning customized, minute features over large areas and, is therefore well-suited to patterning UV gratings. EBL achieves isolated feature sizes of $\approx$10 nm \cite{Vieu_EBLResolution_2000}, an order of magnitude smaller than required for most UV gratings. 

Recent development in EBL gratings for UV applications has leveraged years of progress in X-ray grating fabrication, where EBL and complementary nanofabrication techniques are used to manufacture small-period gratings capable of high spectral resolution and high diffraction efficiency \cite{McEntaffer13, Miles18, McEntaffer19, DeRoo20}. These nanofabrication processes allow for customizable groove layouts, densities, and facet angles, all of which are defined prior to fabrication. The desired groove pattern is then exposed with an EBL tool in an electron-beam-sensitive resist, which coats the grating substrate. The pattern is then further developed and etched to embed the desired groove pattern into the substrate itself. The etches that transfer the pattern into the substrate retain the groove layout exposed with the EBL tool such that the overall groove placement is largely determined by the quality of the EBL exposure.

\paragraph{Current state of the art in EBL gratings}

The state of the art in UV gratings fabricated via EBL relies on precise groove placement on the ~nm scale (as shown above) from advanced EBL tooling, pattern transfers into the grating substrate via a series of dry etches, and blazing the grooves to a custom facet angle with potassium hydroxide (KOH) etching \cite{Miles18, Kruczek21, Grise21}. An EBL grating blazed with KOH etching was fabricated for the ESCAPE small-explorer mission concept \cite{France19}. The ESCAPE prototype was coated with Au and achieved $>$75\% peak absolute diffraction efficiency. \cite{Kruczek22} further compared the performance of a prototype, blazed EBL grating to a previously flown mechanically ruled grating, finding that the EBL grating achieved a 50\% increase in absolute diffraction efficiency and a factor of $\approx$5 improvement in scatter. \cite{DeRoo20} showed that a 50~mm $\times$ 50~mm grating with a 1-$\mu$m period written with EBL is expected to reach a spectral resolution $R \approx 35,000$, matching grating performance onboard HST/COS when taking into account the difference in groove density. UV gratings have also been fabricated at a range of groove densities and patterns, ranging from $\approx$10-$\mu$m \cite{Kruczek21} down to $\approx$110-nm \cite{Grise21} groove spacing and from parallel to aberration-correcting groove profiles. 

There have also been substantial advances in patterning large areas with University-grade EBL tools. For example, the fabrication of a 100-mm $\times$ 107-mm ($>$100 cm$^2$) X-ray grating patterned using EBL in just a few days has been reported \cite{Miles19}, and UV gratings as large as 36 cm$^2$ have been manufactured and tested \cite{Kruczek21}. EBL tools are automated, requiring no human input to complete a patterning run once initiated. EBL tools maintain focus over long duration runs by checking alignment patterns outside of the pattern boundaries, so are robust against changes in environmental conditions that might occur even inside the vibration-isolated, climate-controlled facilities where the tools operate. 

EBL tools can also allow the patterning of curved substrates using z-stages that can move the substrate up and down to allow for the electron beam to focus on a variable-height surface. Apart from the curved, blazed grating flying on the Compact Reconnaissance Imaging Spectrometer for Mars (CRISM) instrument onboard Mars Reconnaissance Orbiter (MRO), however, EBL has not been routinely used to fabricate curved gratings for astrophysical applications. Current APRA and SAT proposals are focused on developing and optimizing methods to pattern such challenging substrates, encompassing spheres, and more complex shapes. Recently, a 39~mm $\times$ 20~mm variable-line-space grating with periods of 800 -- 1400~nm was successfully placed on a cylindrical substrate \cite{Grise23}.

\paragraph{Ongoing EBL development}

With promising early results on UV echelles, flat substrates, and a range of groove densities and facet angles, ongoing development is focused on maximizing customizability, one of the primary advantages of EBL processing. Active focus areas for curved gratings include scaling current EBL techniques to larger areas and the placement of curved grooves on non-planar substrates, both of which are critical for a broad range of aberration correction capabilities. Additional development efforts currently funded under NASA APRA and SAT programs include strategies for consistent, repeatable groove placement over large write areas, characterizing achievable resolving power and diffraction efficiency over a broad range of groove profiles, and maturing additional techniques to produce blazed facets as alternatives to KOH etching.

\cite{Grise21} showed that, while KOH etching yields atomically smooth groove facets capable of high diffraction efficiency, etch fidelity begins to deteriorate for groove directions that deviate $\gtrapprox$2$^\circ$ from a nominal layout direction. This degradation is due to the etch's reliance on the crystal structure of the silicon substrate; the KOH etches to a specific family of crystal planes, the \{111\} planes, and sets of \{111\} planes are parallel to each other. As the relative groove curvature exceeds $\approx$2$^\circ$, the etch conditions evolve such that the quality of the groove size and surface degrades. For applications that require aberration-correcting profiles with a large amount of curvature and where diffraction efficiency is paramount, alternative blaze techniques may be capable of improved performance over the existing state of the art with KOH etching.

Two alternative techniques are being developed to produce blazed UV gratings capable of high diffraction efficiency and low scatter. In each case, the techniques rely on EBL to expose the overall groove layout and therefore retain the resolving power capabilities possible with EBL gratings. Thermally activated selective topography equilibration (TASTE) has been in development for EUV and X-ray grating applications and uses ``greyscale'' EBL and thermally activated EBL resist to sculpt blazed facets \cite{McCoy20}. Greyscale EBL uses a modulated electron dose across individual grating groves to generate a variable-height groove profile, which is then thermally reflown to create a continuous, blazed facet. TASTE has been demonstrated on gratings with densities as high as 2500 grooves/mm, with early results showing EUV diffraction efficiency of $\approx$60\% \cite{McCoy20}. At higher groove densities, however, greyscale EBL becomes more challenging; further development and process optimization is needed for such applications. 

Gratings blazed using ion-beam etching (IBE) were recently developed for X-ray applications and the processes are transferable to UV gratings \cite{Miles22}. IBE, which has also been used to blaze holographically ruled gratings (e.g. \cite{Kowalski02}), uses EBL to expose the desired groove layout, a series of reactive-ion etches to transfer the EBL pattern into a substrate, and then a controlled, directional ion-beam etch to sculpt the blaze facets. Gratings fabricated with EBL and blazed with IBE are applicable to a broad range of groove densities, with demonstrations as high as 6000+ grooves/mm \cite{Miles22}. Studies involving the use of IBE with EBL are still in their infancy and, as with TASTE, further optimization on precise, controlled IBE etches, in particular for complex aberration-correcting profiles, is needed. 

KOH etching has shown excellent results for groove layouts with minimal curvature, while both TASTE and IBE are applicable to more complex aberration-correcting layouts with a large amount of curvature. TASTE has been demonstrated on groove patterns with small to medium groove densities, while IBE is suitable for high groove densities. In each case, the nanofabrication processes rely on precision EBL exposures to generate the groove layout with nm-scale precision, then a series of complementary processes to sculpt the grating facets into the desired blaze angle. Though further development and standardization is necessary in each case, the three blaze techniques represent a promising and varied approach to enabling the future of custom UV gratings. 

\section{Spectral Multiplexing Technologies}\label{sec:mplex}

Spectral multiplexing offers the simultaneous acquisition of spectroscopic information from multiple objects or regions within a telescope's field of view (FOV) within a single pointing. Goundbased spectral multiplexing techniques rely on fiber optics, multilenslet arrays (MLAs) \cite{Tuttle2014b, Hill2021}, or reflective image slicers (KCWI, MUSE \cite{Bacon2010}). These devices are used to dissect a target region into non-overlapping spatial regions to obtain large numbers of spectra ideally at the resolution limit of the telescope and spectrograph. MOS technologies used in groundbased instruments deploy individual fiber optics to target locations, historically by hand plugging plates, and most recently by implementing robotic positioners (i.e. MOONS, SDSS-V, DESI). IFUs will often use close packed fiber optic bundles along with a lenslet array to dissect a contiguous target field. Fibers have not historically had high transmission in the ultraviolet, although sub-orbital testing was done during the first Faint Intergalactic Redshifted Emission Balloon flights \cite{MatMat2010, Milliard2010, Tuttle2010} and development is ongoing \cite{Dewitt23}. MLAs for astronomical applications are made of fused silica, but are limited in transmission by anti-reflective coatings of limited bandpass. An FUV reflective image slicer has flown on the Integral Field Ultraviolet Spectrograph (INFUSE), a sounding rocket-borne instrument and the first FUV IFU to fly with access to wavelengths below 1150 \AA\ \cite{WittSPIE2023}.

Here we review micro-electro mechanical (MEMs) technologies that employ either arrays of clear aperture micro-slits, reflective micro-mirror arrays, or reflective "slit-farms" to partition the field prior to spectral acquisition as well as reflective image slicers, as potential effective flight qualified multiplexing approaches that have some space-based heritage and potential in the ultraviolet.

\subsection{Microshutter Arrays}

\begin{figure}[h]
  \centering
  \includegraphics[width=\textwidth]{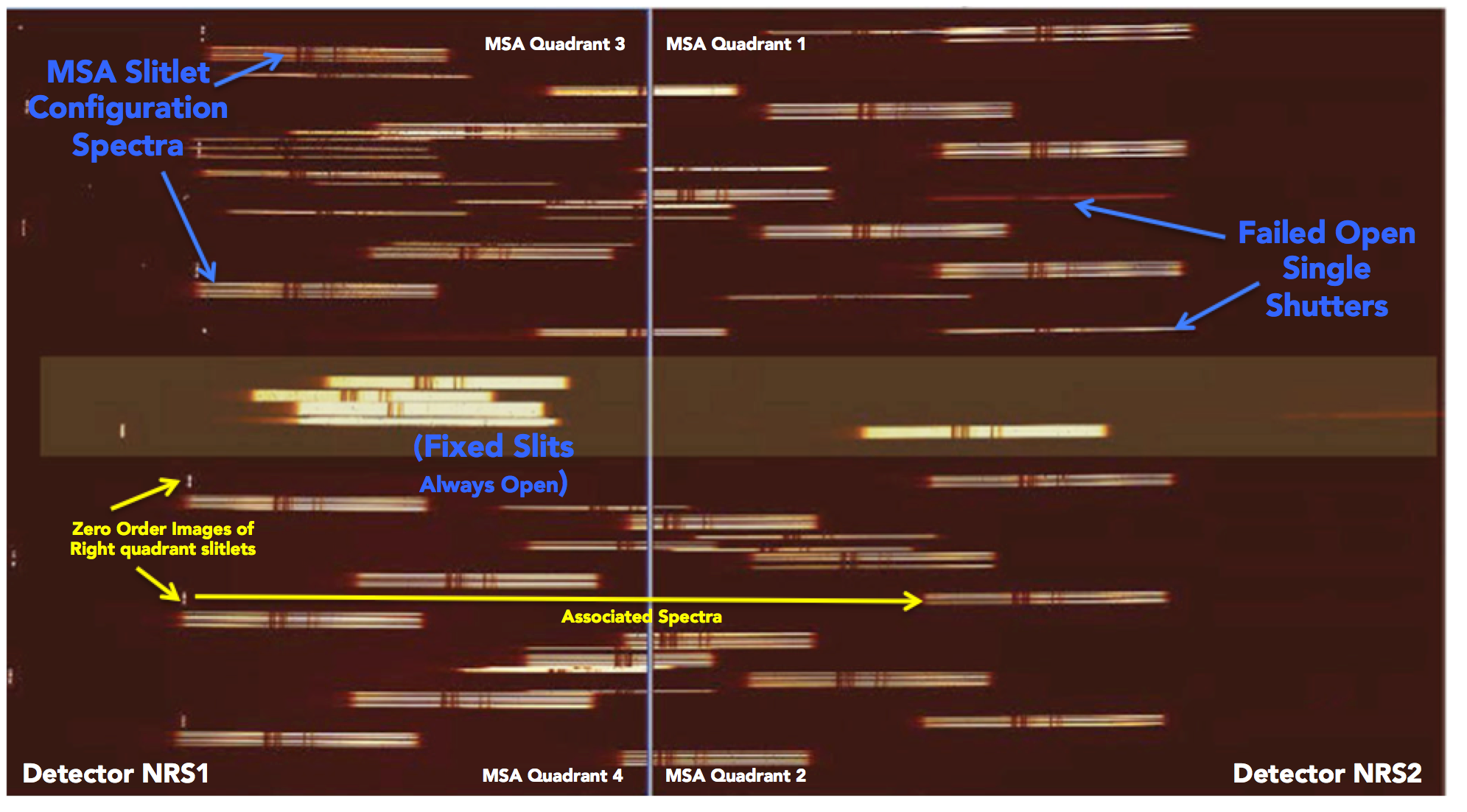}
  \caption{Example of NIRSpec MOS spectra acquired under illumination of a test slit pattern with a calibration lamp.}
  \label{fig:NIRSPEC}
\end{figure}

Microshutter arrays (MSA) are programmable multi-aperture micro-slit devices that were developed by GSFC for use on the {\it JWST} Near Infrared Spectrograph (NIRSpec). MSAs allow for the simultaneous acquisition of spectra from multiple objects in the FOV, enabling spectroscopic analysis on an industrial scale equivalent to that long enjoyed by multi-object fiber optic fed spectrographs in ground-based applications \cite{Smee:2013, Tamura:2016}. Their use on \jwst\ is ushering a revolution in space-based spectroscopic analysis \cite{Boker:2023, Jakobsen:2022}. These arrays can be programmed to provide any pattern of slits corresponding to sparsely distributed sources on the sky. It can also be programmed to provide shaped slits on extended sources. A NIRSpec MOS example is shown in Figure~\ref{fig:NIRSPEC}.

\begin{figure}
\centerline{
\includegraphics[width=.3\textwidth,viewport=.4in 9.5in 26in 29in,clip]{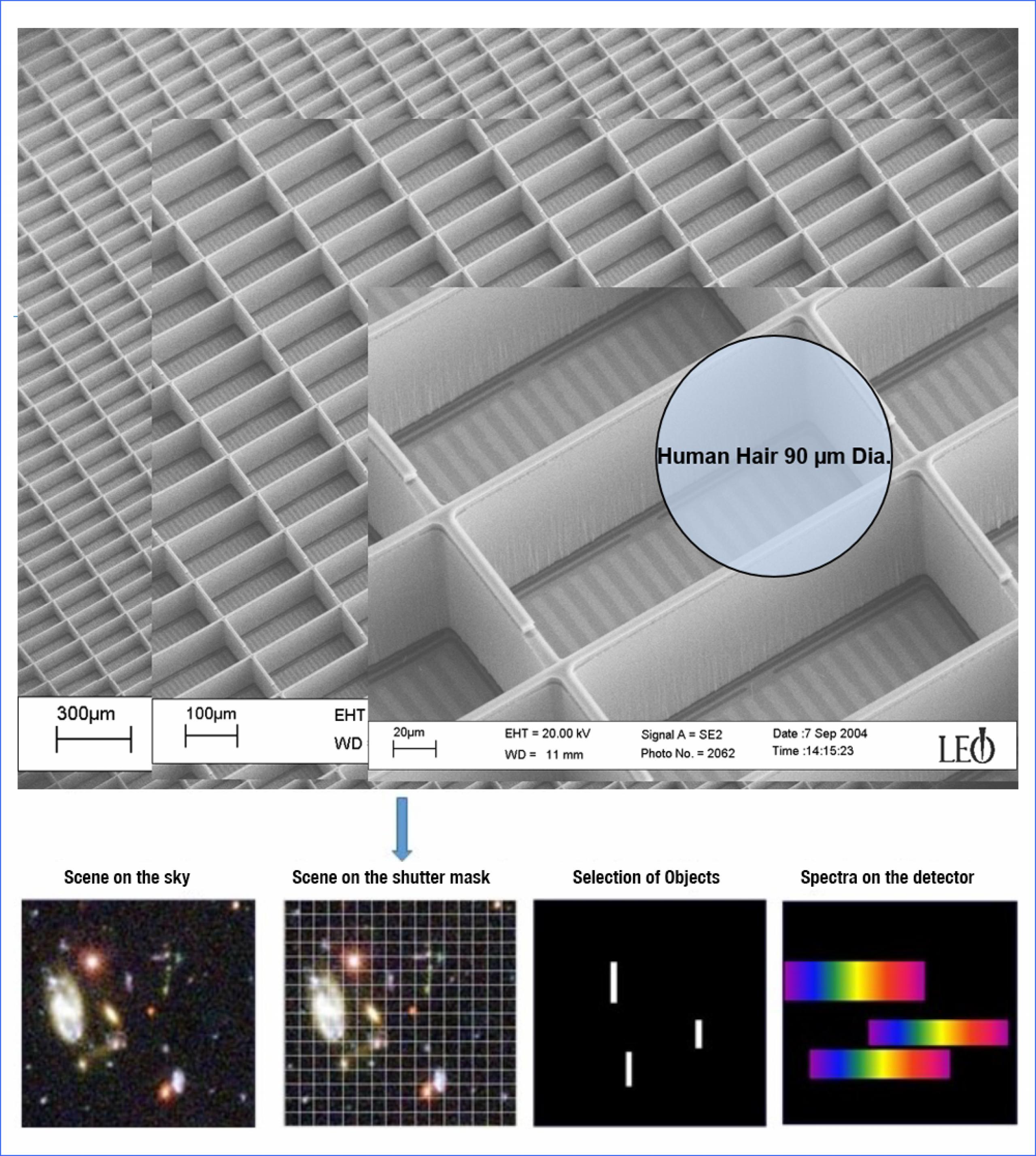}
\includegraphics[width=.21\textwidth,viewport=0in .05in 7in 7.7in,clip]
{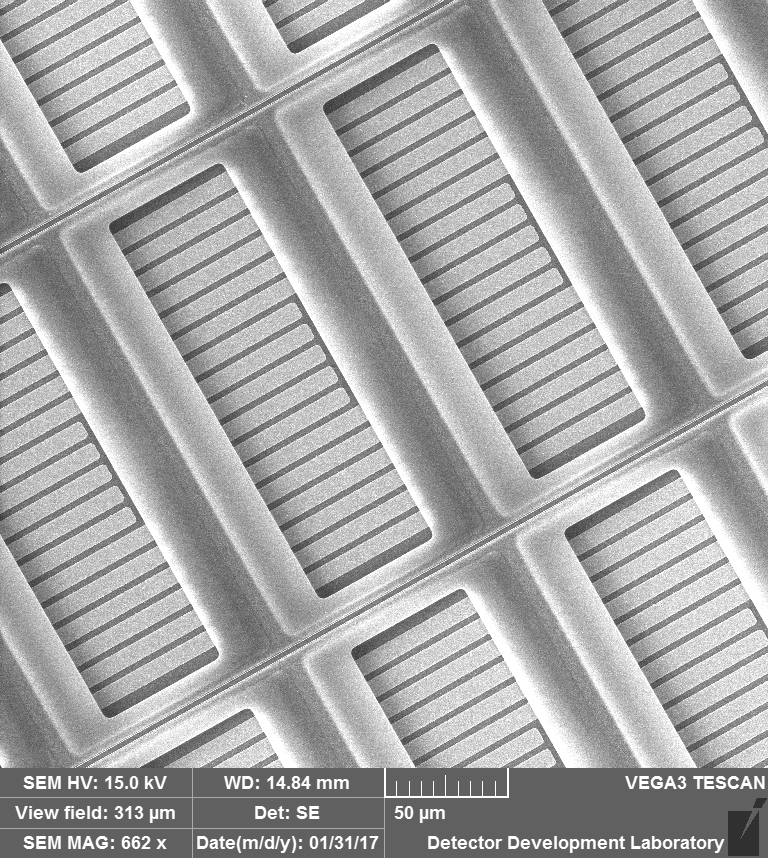}
\includegraphics[width=.5\textwidth,viewport=1.5in 1.5in 12.2in 7.5in,clip]{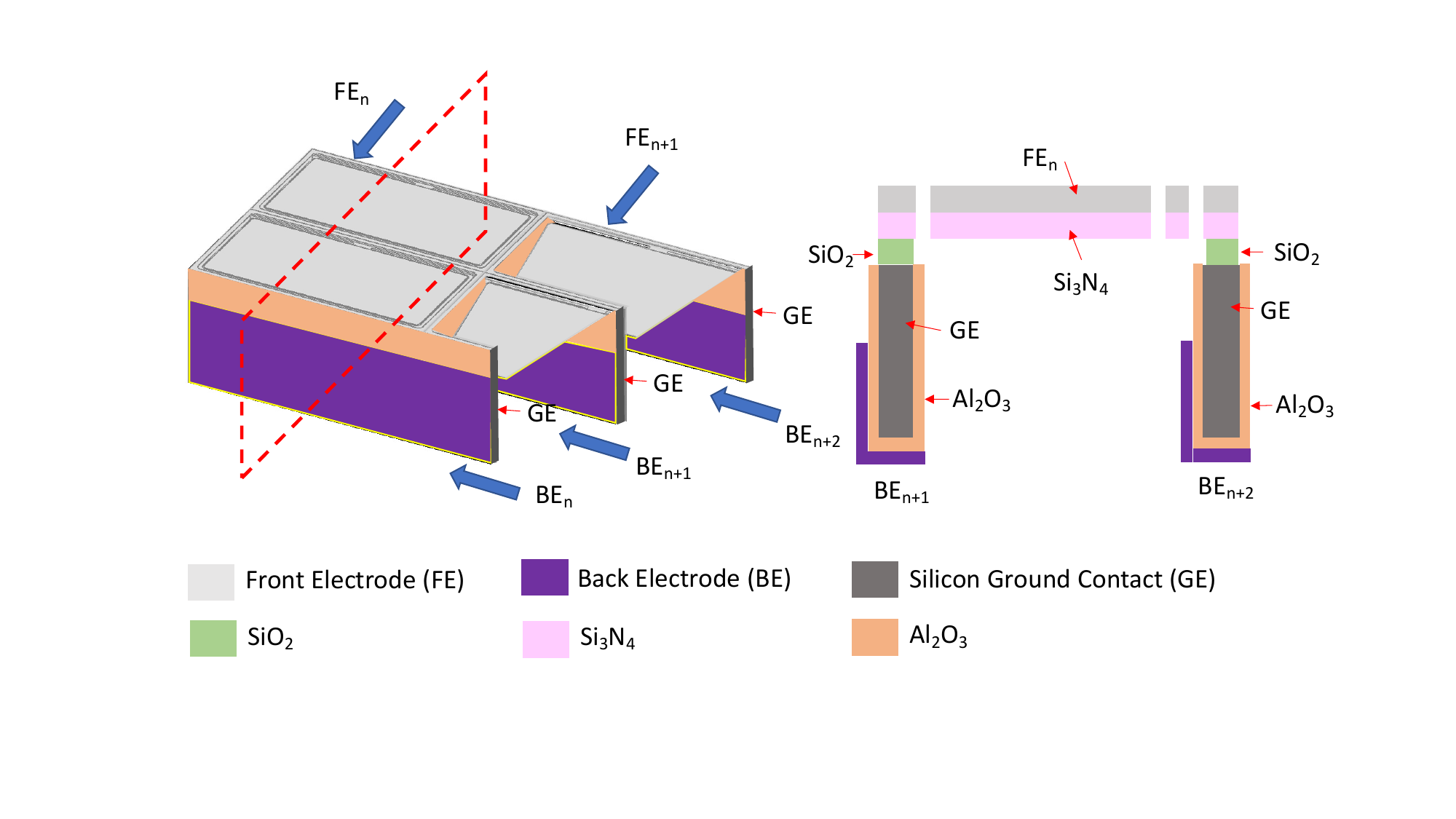}
}
\caption[]{Left -- MSA egg-crate detail. Middle -- Shutter blades viewed from light-shield side. Right -- Shutter unit cell with material layers (light-shield not shown). The egg-crate frame of Si is held at ground (GE) and coated with an insulating layer of ALD grown Al$_2$O$_3$. It is isolated from the Si$_3$N$_4$ shutter that is coated with Front Electrode (FE). A Back Electrode is deposited at an angle over the wall of the Al$_2$O$_3$. A potential difference between FE and BE attracts the shutter to the wall to latch it in the open position. Removal of the potential difference causes the shutter to relax to the closed position. }
\label{cross-section}
\end{figure}

The 1st generation MSAs flying on NIRSpec were optimized to provide high contrast in the cryogenic environment required for IR observation. The operation involved a combination of electrostatic and scanning magnet actuation that requires a heavy complex mechanical assembly. GSFC has recently developed Next Generation MSAs (NGMSAs) that retain much of the original architecture but have eliminated the need for a scanning magnet. This advance was enabled by the combination of thinner shutters and improvements in electrical isolation between shutters and ground. Together these improvements permit purely electrostatic actuation, greatly decreasing the time to open/close or address selected shutters from several seconds to a fraction of a second \cite{Kutyrev:2020, Ke:2022}.

\begin{figure}
\includegraphics[width=\textwidth]{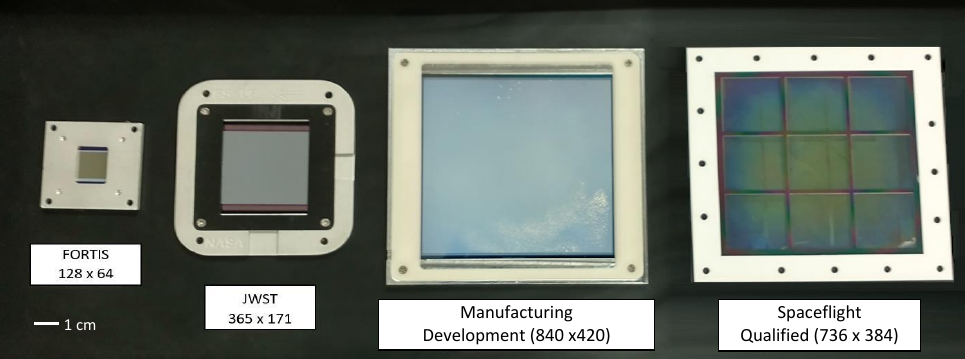}
\caption[]{MSA format development trend for Cosmic Origins is shown. FORTIS and JWST formats are shown on the left. A concept and manufacturing development array produced by GSFC fiscal year FY18 Strategic Astrophysics Technology (SAT) project is shown in the center. GSFC FY18 SAT project also produced a ruggedized design of an array mounted on a ceramic carrier with a supporting grid that is suitable for spaceflight is shown on the right. The design shown on the right has been demonstrated through tests to general environmental verification standard (GEVS) vibration levels for evolved expendable launch vehicle (EELV) launch.}\label{msa-formats}
\end{figure}

\begin{figure}
  \centerline{
  \includegraphics[width=.42\textwidth]{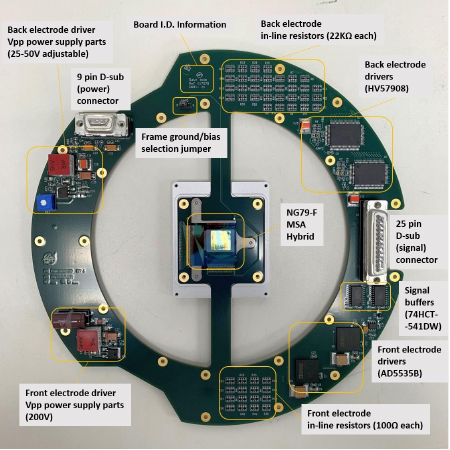}
\includegraphics[width=.42\textwidth,angle=90,viewport=1in 5in 8in 14in,clip]{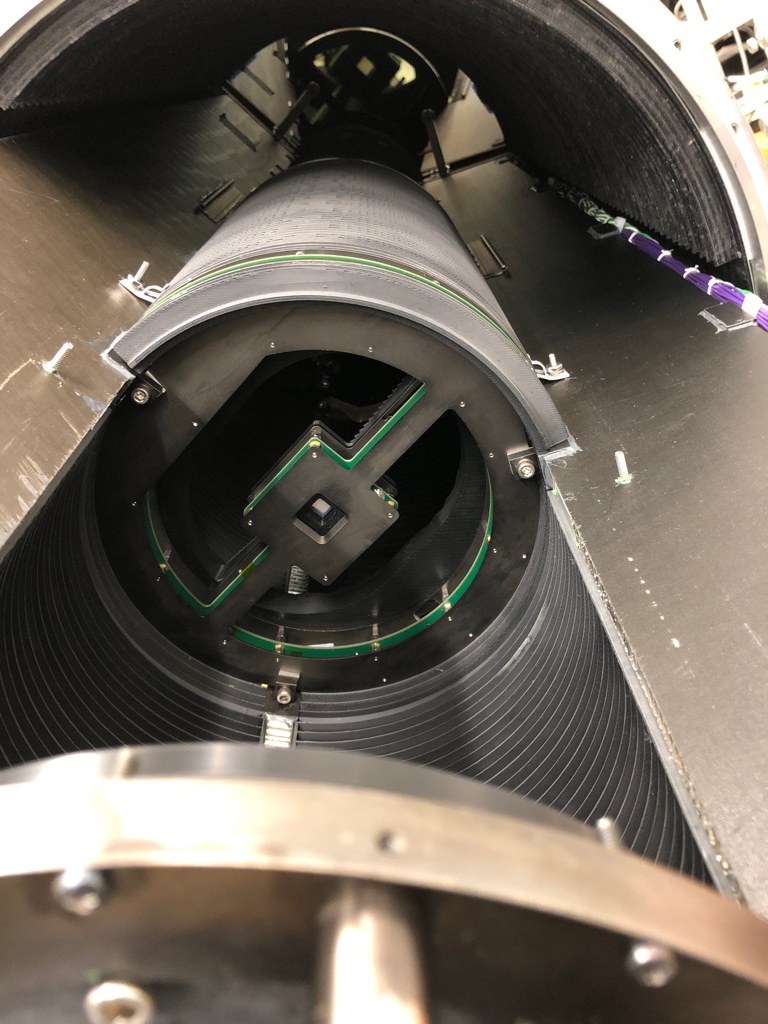}}
  \caption{Left - FORTIS 128x64 NGMSA on PCB carrier. Right - NGMAS assembly installed at the FORTIS prime focus.}
  \label{fig:fortisNGMSA}
\end{figure}

Shutters are etched into a silicon nitride (Si$_3$N$_4$) layer, forming an array with a pitch of 200 $\times$ 100 $\mu$m \cite{Kutyrev:2004}. Each shutter is connected to the array by a torsion hinge that is suspended above a 100 $\mu$m thick support grid called the ``egg-crate'' (Figure~\ref{cross-section} -- Left). Complete $90^{\circ}$ deflection and latching open of the shutter is provided by a potential difference between electrodes deposited on the interior walls of the egg-crate and top side of the shutters (Figure~\ref{cross-section} -- right), using a cross-point addressing scheme. Zeroing the potential difference allows the torsion spring to relax and return the shutter blade to its closed position. The shutter dimension is a few microns smaller ($\sim$ 3 $\mu$m) than the inner width and length egg-crate cell, so a light-shield fabricated in front of the shutter minimizes the amount of light that can pass between the gaps between the egg-crate walls and shutter perimeter. 

The 1st generation MSAs flying on NIRSpec were optimized to provide high contrast (open-to-close ratios) in the cryogenic environment required for IR observation. They exceeded the \jwst\ contrast requirement of $>$ 2000 by more than an order of magnitude \cite{Kutyrev:2008}.  NGMSA UV measurements are in progress with preliminary results showing contrast at 254 nm $>$ 50,000 for focal ratios slower than f/10 \cite{Mitchell:2023}. Work is underway to develop facilities to fully characterize the NGMSA contrast in the vacuum UV \cite{Kutyrev:2023}.

Figure~\ref{msa-formats} shows the MSA format development trends. The JHU sounding rocket program has partnered with GSFC in the development of Far-UV MOS technologies since 2011 with the Far-uv Off Rowland-circle Telescope for Imaging and Spectroscope (FORTIS) providing a platform for flight testing of next generation devices.

A flight-qualified NGMSA in the 128x64 format was integrated onto a carrier printed-circuit-board (PCB) with actuation circuitry, installed on FORTIS, and successfully operated on NASA sounding rocket flight 36.352 UG in 2019 raising this system TRL-9.

The 840x420 device shown in Figure~\ref{msa-formats} was subjected to finite element analysis (FEA), which suggested a mechanically robust structure would be achieved if it were divided into nine subarrays leading to the design of the 736x384 device. Current developments at GSFC are focused on raising the technical readiness of these larger devices from the current TRL-3 to TRL-5. Tasks to be performed under a recently awarded SAT22 include: finalizing the shape of the blade to increase actuation reliability; developing 2 sided buttable fanout board and driver circuity; life testing of full-scale breadboard in vacuum to 10,000 cycles per shutter; radiation testing to L2 total ionizing radiation dose; shake, shock and acoustic testing; and open/close contrast measurements.  

\subsection{Digital Micromirror Devices}
\begin{figure}
  \centering
  \includegraphics[width=1\textwidth]{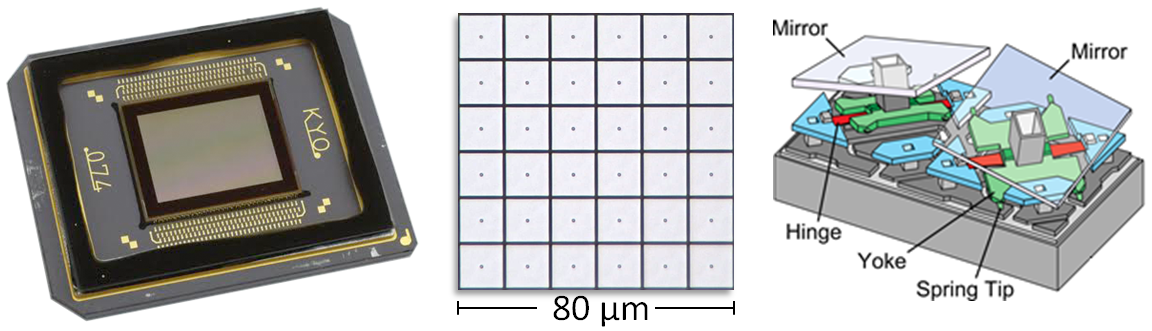}
  \caption{\textit{Left:} a $1024\times768$ mirror DMD with a custom NUV-transmissive window. \textit{Middle:} a close-up of several micromirrors, each $13 \mu m$ on a side, with a $\approx92\%$ fill-factor. \textit{Right:} A schematic of the micromirror hinge and tilt mechanism; note DMDs tilt about their diagonal.}
  \label{fig:DMDs}
\end{figure}

Multiobject and integral field spectroscopy can also be achieved using a digital micromirror device (DMD) - an array of many small mirrors ($\approx 10\mu m$ on a side) (Figure \ref{fig:DMDs}), developed by Texas Instruments and currently widely used in video projection and 3D printing applications. Each mirror of the DMD can be tilted into one of two orientations, usually $\pm12^{\circ}$ with respect to the device plane's normal orientation. In this way, light incident on the DMD can be directed into one of two directions, or ``channels''. In the most straightforward implementation, a DMD can be used to create hundreds of ``slitlets'' in the field of view and direct light from those locations to a spectrograph, while blocking the rest of the FoV. DMDs serve a similar function to MSAs in a spectrograph, but with a key fundamental difference: DMDs are reflective devices, whereas MSAs are transmissive. This allows DMD and MSA-based spectrographs unique auxiliary capabilities. For example, in a typical MOS observation, the slitlets only cover $\approx1\%$ of the field of view (by area). A DMD MOS can be designed with a parallel channel, to utilize the light not sent to the primary spectrograph (to do deep imaging, slitless spectroscopy, etc). However, a DMD-based spectrograph must contend with the reflectivity and scatter of the DMD mirrors, while MSA shutters are completely transmissive at all wavelengths.  

Several spectrographs based on DMDs have been built for ground-based telescopes, including: RITMOS \cite{Meyer2004}, IRMOS \cite{MacKenty2006}, BATMAN \cite{Zamkotsian2014}, and SAMOS \cite{Robberto2016}. To demonstrate the suitability of DMDs for deployment in space, a NASA SAT program was developed to perform flight qualification activities and to investigate the prospects for window replacement. Overall, it was determined that DMDs are not particularly sensitive to shock and vibration loads \cite{Vorobiev2016,Travinsky2017} or the radiation environment, except in the most extreme conditions \cite{Fourspring2013,Travinsky2016a,Travinsky2016b,Oram2019,Oram2020}.

Although commercially available DMDs are designed for use in the visible regime, their throughput can be extended into the NUV regime, to $\approx170$ nm by replacing the stock BK7 window with a suitable material (such as fused silica, sapphire, or MgF$_2$). For use in the shorter wavelengths, it may be possible to re-coat DMDs with the kinds of coatings that have been developed for monolithic optics in the FUV (see Section \ref{sec-coatings}). However, because only commercially available DMDs are sufficiently mature for use in space-based instruments, there are size limitations. In general, larger missions allow for telescopes with more fine spatial resolution and effective area; however, the fixed size of the DMD results in a decreasing field of view, with increased focal length. To enable a larger field of view, multiple groups are exploring the possibility of mosaicing DMDs.

\subsection{Integral Field Units}
IFUs powered by image slicers provide a way to obtain imaging spectroscopy in the FUV while avoiding the low transmission of fiber bundles. These systems consist of a slicer cube and a pupil array. The slicer cube is a stack of thin mirrors rotated at slightly different angles to reflect light onto a pupil array. Traditional image slicers use a pupil array of mirrors to realign the beams from each slicer element so that the image, when refocused, maintains a specific shape, generally an array of ``long-slit" fields of view (defined by the image slicer) arranged end-to-end to avoid spectral confusion. The system is then diffracted by a grating and re-imaged by a set of camera optics. An image slicer first flew on the \textit{JWST} Mid-Infrared Instrument (MIRI) \cite{Wells2015}. The first FUV IFU, an image slicer, flew on NASA sounding rocket flight 36.375 UG Fleming aboard INFUSE. The INFUSE slicer cube, provided by Canon, Inc., consisted of 26 slices micromachined out of invar and copper to 7.1 \AA\ RMS surface roughness and a $\lambda$/20 flat figure \cite{Sukegawa2023}. It was coated in conventional Al+LiF through a partnership with GSFC \cite{WittSPIE2023}. There are currently plans to coat the flight spare in Al+eLiF. 

\begin{figure}
  \centering
  \includegraphics[width=1\textwidth]{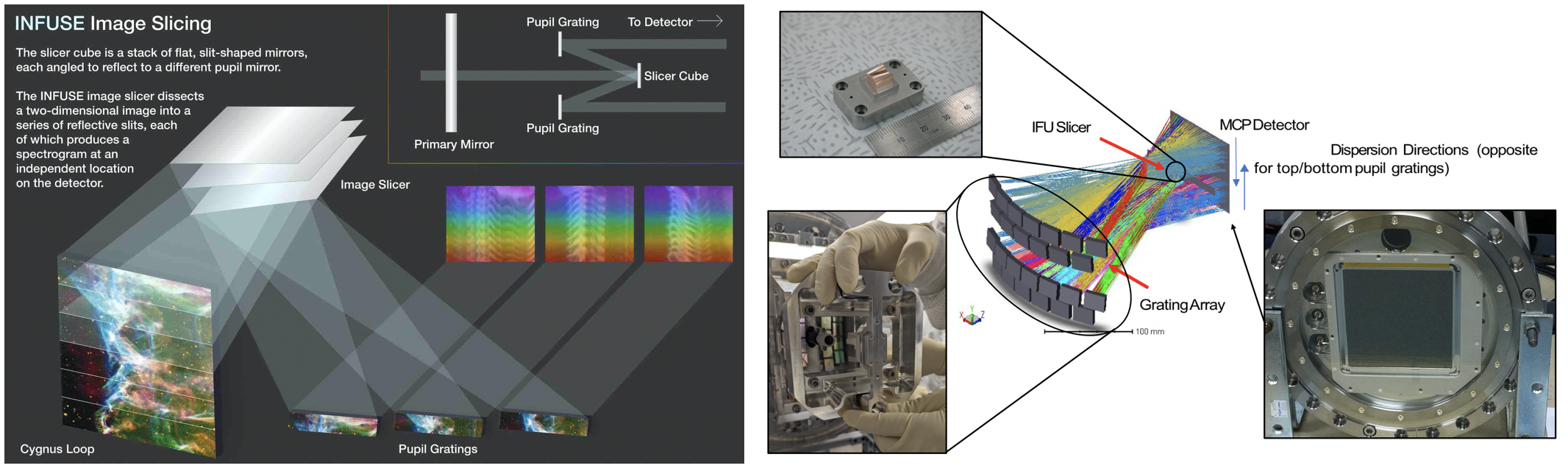}
  \caption{\textit{Left:} INFUSE image slicer concept. Cygnus Loop image credit: \cite{Danforth2001}. \textit{Right:} INFUSE raytrace with pictures of the slicer cube, grating array, and MCP detector inset.}
  \label{fig:INFUSEslicer}
\end{figure}

A traditional image slicer system is too inefficient for the FUV. In order to improve throughput, INFUSE replaced the pupil mirrors with aberration correcting, Type IV holographic gratings from Horiba JY (Figure \ref{fig:INFUSEslicer}). These gratings are identical replicas, blazed to 1300 \AA. Twenty-five are coated in conventional Al+LiF as the photoresist used to create replica gratings cannot tolerate the heat of a hot coating deposition. All of the gratings are arranged in two sets. Thirteen of the gratings are arranged so that the slicers are imaged end-to-end, with the dispersion going from the top of the large format MCP detector toward the bottom while the other thirteen are arranged with the dispersion going from the bottom to the top. These spectra were then reassembled into a 3D data cube after flight. This permitted moderate resolution integral field spectroscopy over a 2.48' $\times$ 2.57' FOV \cite{WittSPIE2023}.

\section{Contamination Reduction}\label{sec-contamination}

The UV is particularly sensitive to the presence of contaminants because of amplified scatter and the preferential absorption of this bandpass by molecular hydrocarbons. As such, strict practices of contamination control will need to be implemented to maintain an acceptable level of throughput through all critical surfaces of the observatory to ensure performance metrics are met at launch. Once attained, this performance can be maintained through the lifetime of the mission by measures such as baking, choice of low-outgassing materials, and vent path designs. This section will address the approaches that need to be taken to provide access to the far ultraviolet for HWO.

\subsection{Contamination Budgets}

The objectives of contamination control for space-borne optical instruments are a) to determine the maximum allowable contaminant quantities, or budgets, for the spacecraft or science instrument, and b) to select and implement the appropriate controls to prevent unacceptable contaminant-induced degradation.

A crucial early step in contamination control for highly contamination-sensitive systems is to establish the degradation allowables. In this step, some percentage of the performance margin of the system is reserved for contaminant-induced degradation. By analysis of degradation vs. contaminant quantity, ie., obscuration, absorption, and scatter, these degradation allowables form the basis to determine the maximum allowable lifetime contaminant quantities, called the contaminant budgets, for both particulates and molecular contaminants. Having accurate absorption coefficient data vs. wavelength for all non-metallics is necessary in order to develop accurate contaminant budgets.

In highly sensitive optical telescopes and instruments these contaminant budgets may differ for various surfaces within the instrument. For example, the particulate budget for a detector array may include a requirement disallowing particles or fibers greater than a specified percentage of the pixel dimensions.

Often it is necessary to assign larger portions of the contaminant budgets to more exposed and colder surfaces. These analyses also include allocation, or dividing, of the contaminant budgets over the several prelaunch environments and the space operations phase of the program. This requires extreme precision in cleaning operations, the correct choice of non-metallic materials and their preparation and cure, in-process contamination monitoring, and successful maintenance cleaning. 

Dependent on the results of contamination degradation analyses of the HWO telescope and instruments, the lifetime molecular contaminant budgets are expected to be far less than 50 ng/cm$^{2}$ per surface (roughly the equivalent of one monomolecular layer of water). 

\subsection{Contaminants and their Effects}

The contaminants that plague spacecraft and space science instruments are categorized as either particulate contaminants or molecular contaminants. Particulate contaminants are almost always microscopic and are relatively easy to describe. They may range from sub-micron size dust and smoke particles to visible garment fibers. Molecular contaminants cover the range from single atoms to complex molecules of several hundred atomic mass units (amu). Those particles which settle onto surfaces, and those molecular species which are adsorbed onto surfaces, and which change the properties of those surfaces, are the contaminants that degrade optical systems as well as other spacecraft surfaces and that require controls.

\begin{figure}
  \centerline{
  \includegraphics[width=.5\textwidth]{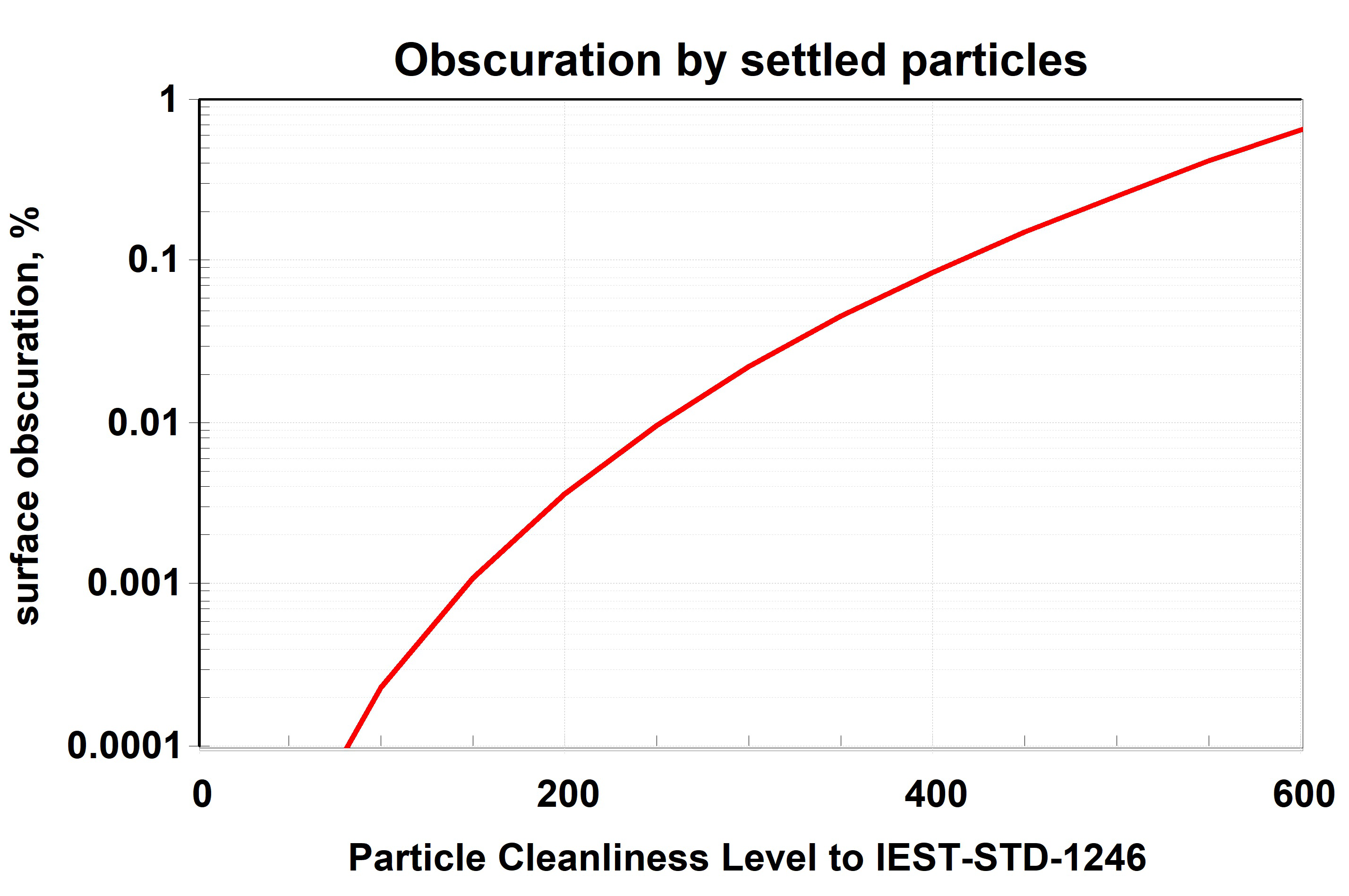}
\includegraphics[width=.5\textwidth]{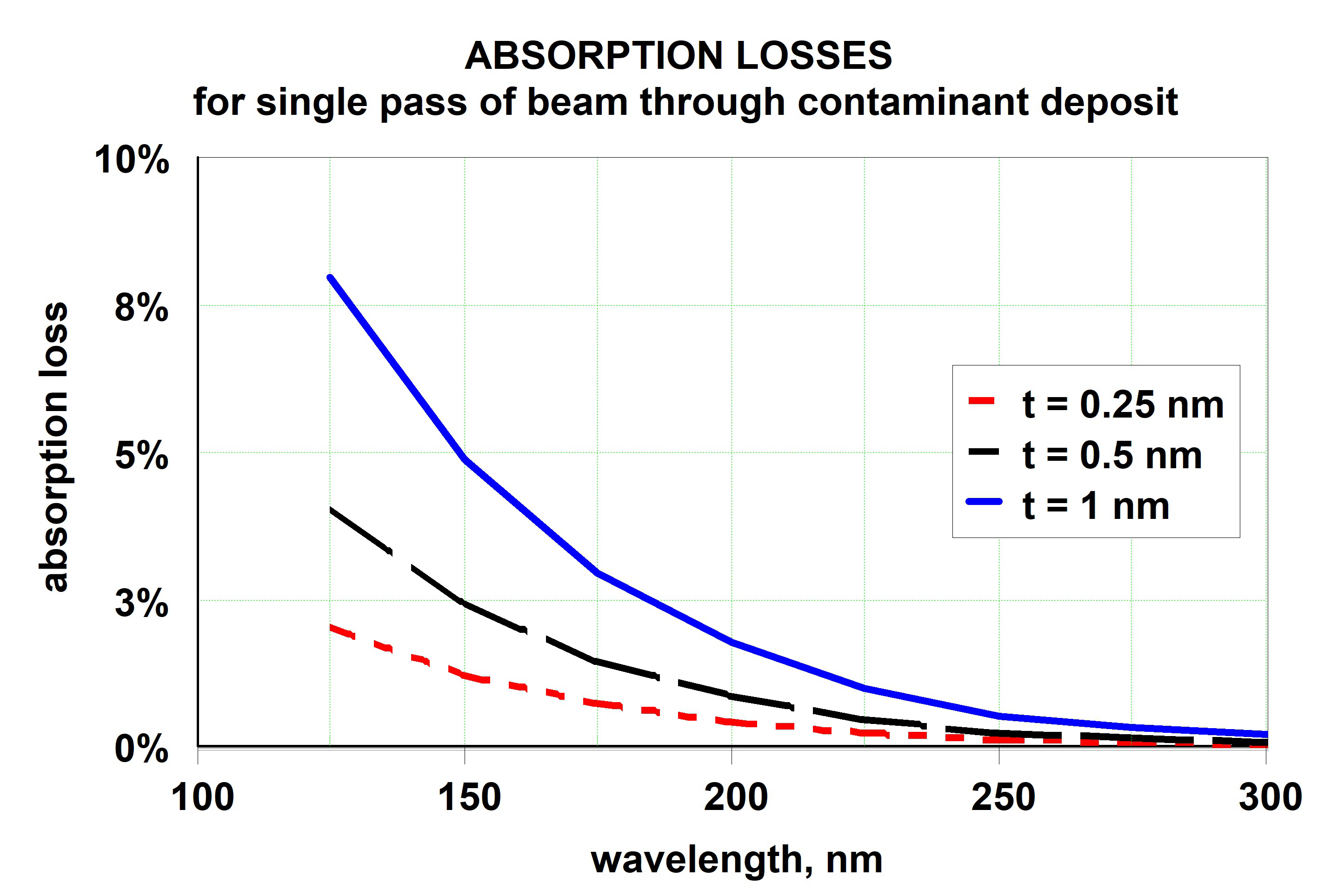}}
  \caption{Left - Change in surface obscuration as the particle Cleanliness Level increases (as set by IEST-STD-1246, referencing non-volatile molecular cleanliness levels). Right - Absorption loss of a UV light beam versus wavelength and contaminant deposit equivalent thickness. Combined, these plots motivate contamination control when working in the ultraviolet regime.}
  \label{fig:particles_losses}
\end{figure}

Settled particles obscure portions of optical beams and scatter light out of these beams as stray light (Figure ~\ref{fig:particles_losses}, left panel), while molecular deposits absorb as well as scatter light (Figure~\ref{fig:particles_losses}, right panel). Depending on the optical wavelengths and the contaminant quantities, a combination of particles and molecular species can reduce the optical efficiency of each optical surface by a few hundredths of a percent to several percent, the latter being unacceptable for optical systems such as the HWO ultraviolet instruments.

\subsection{Contamination Controls}

The primary factors of contamination control include initial and precision cleaning of physical components, cleanroom quality, personnel garmenting and operations, tool and work surface cleanliness, selection of low outgassing materials, cleanliness monitoring, maintenance cleaning, and specific controls during environmental testing. The contamination budgets derived for the HWO and the science instruments will lead to the specific contamination controls necessary for the observatory. 

Modern cleanrooms are quite capable of maintaining classes ISO 5 and ISO 6 \footnote{ISO 14644-1:2015 Cleanrooms and associated controlled environments, Part 1: Classification of air cleanliness by particle concentration.} airborne particle Cleanliness Levels, and even ISO 4 for periods when optical components are exposed even for lengthy time periods. The use of airborne particle counters and the collection of settled particles to determine particle settling rates during critical operations will be necessary for HWO optics. It is anticipated that, for most Assembly, Integration, and Test (AIT) operations with HWO optics, personnel garmenting will require full coverall suits with attached hood, booties, and face covers.

All flight and flight-spare optical elements, including detectors, must always be accompanied by optical witness samples (OWS) having surface coatings applied simultaneously with those of the optical elements. The only times the OWSs leave the optical elements would be during UV reflectance or other measurements to verify the optical element performance.

Throughout AIT operations the cleanliness of all hardware, tools, and work surfaces must be monitored either directly using high-precision Non-Volatile Residue (NVR) sampling and particle collection, including by tape-lift, either directly from hardware when possible, or from companion witness plates. Monitoring for molecular contaminant accretion must include real-time Quartz Crystal Microbalances (QCM) during all AIT operations. QCMs with 10 megahertz crystal frequency have sensitivity to detect equivalent contaminant equivalent thickness changes of 0.04 nm. Finally, the contamination controls must include maintenance cleaning procedures to be applied as necessary to maintain the hardware cleanliness within the derived budgets.

All work surfaces and tools should be maintained to the same cleanliness requirements as the hardware that will be in contact with the tools and surfaces. Unique controls must be developed for transfer of contamination-sensitive hardware between the cleanrooms and the vibration and thermal vacuum test facilities. And sufficient cleanliness of those facilities must be verified prior to their use. 

For example, prior to a vacuum test of the hardware, a chamber bakeout must be performed and an abbreviated ``dry run'' of the chamber, with contamination monitoring, must be made covering the temperature range of the pending test. The monitoring should include Residual Gas Analysis (RGA), QCM, and high precision NVR sampling to verify that the chamber and all internal test apparatus are not a contamination threat to the hardware.

It can be  shown by modeling that partial pressures of volatile-condensables at $<$ 10$^{-14}$ Torr at critical hardware temperatures do not present a significant contamination threat. The RGAs to be used should have at least this sensitivity. 

An extremely crucial control will be the selection of non-metallic materials; the paints, adhesives, pottings, wire insulation materials, and other plastics that must be used. Only materials that meet the low outgassing and molecular contamination requirements to be derived for the contamination budget can be used in HWO hardware. Traditionally the testing of materials for outgassing of molecular contaminants has included the basic test by ASTM-E595 \footnote{ASTM-E595, Standard Test Method for Total Mass loss and Collected Volatile Condensable Materials from Outgassing in a Vacuum Environment} which requires that Total Mass Loss (TML) during 24 hours in vacuum at 125 $^{\circ}$C not exceed 1 \%, nor shall Collected Volatile-Condensable Material (CVCM) exceed 0.1 \% on a collector at 25 $^{\circ}$C. However, this test alone certainly is insufficient for acceptance of materials for use on HWO hardware. A more appropriate test is MSFC-SPEC-1443 \footnote{MSFC-SPEC-1443, Outgassing Test for Nonmetallic Materials Associated with Sensitive Optical Surfaces in a Space Environment}. This test has the same requirements for TML and CVCM as ASTM-E595, plus it includes a MgF2/Al optical witness sample for ultraviolet reflectance measurements before and after the vacuum exposure. The basic UV requirement of MSFC-SPEC-1443 is that the reflectance of the mirror shall not change more than 3\% at 121.6 nm, 125 nm, 130 nm, and at 10 nm increments to 200 nm wavelengths. If this test proves sufficient for HWO, these reflectance criteria will require revision consistent with the wavelength and degradation requirements of the observatory.

Contaminant transport modeling will be a useful tool for predicting the molecular contamination accretion on critical surfaces during vacuum testing, helping to determine the necessary cleanliness of the vacuum test facilities, and on orbit to help with the allocation of the lifetime molecular contamination budget and specific materials choices.

Finally, on-board means of heating all optical elements should be provided to deplete molecular contaminants that may adsorb onto surfaces during space operations (e.g. \cite{DevaudUVContam2018}).

\subsection{Maintenance Cleaning}

During the expected months and years of assembly, integration and test operations cleanliness monitoring will be used to determine when maintenance cleaning is necessary. At the piece-part level maintenance cleaning procedures will essentially be the same as the final precision solvent cleaning and subsequent drying procedures, followed by cleanliness verification. However, at the subassembly and assembly levels solvent cleaning is most often not acceptable as solvents can redistribute both particulate and molecular contaminants into cavities and between contacting surfaces where they would be very difficult to remove. 

Typical procedures for maintenance cleaning to remove particles include vacuuming, tweezering of visible particles, and even tape lift. Generally, however, direct contact of surfaces should be avoided to prevent surface damage and additional contamination, and optical surfaces should never be contacted for cleaning purposes. Cleanliness control of molecular contaminants is significantly more difficult than it is for particles. The maintenance cleaning for molecular contaminants at the subassembly and assembly levels can only be done effectively by vacuum baking at temperatures typically between 60 and as high as 85 to 100 $^{\circ}$C. 

\subsection{Venting}

Adequate venting of all enclosures, particularly of enclosed optics, must be sufficient to allow rapid depletion of outgassing species during vacuum bakes and tests. The venting rates are functions of the heats of adsorption of the molecular contaminants, the temperatures, and the vent dimensions. The difference between large vents and launch-only sized vents can mean weeks rather than a few days to deplete molecular contaminants from an enclosed assembly. This is discussed in detail in Appendix B.  

\subsection{Future Testing}

A significant number of the non-metallic materials tests reported in the ASTM-E595 database are 30 to over 40 years old and many of the materials are no longer available. As noted above, acceptance of the ASTM-E595 criteria is certainly insufficient to select a material for HWO. Very few of the currently available materials used on spacecraft have been tested to MSFC-SPEC-1443. This is a very difficult test to pass, and at the same time, it remains to be seen if MSFC-SPEC-1443, even with extended wavelength range measurements, is the correct test method to be used for HWO materials, or if some modification of MSFC-SPEC-1443 or some other test method is more appropriate. In any case all non-metallic materials, and in some cases multiple lots of these materials, will have to be tested. 

Additionally, the UV absorption coefficients as well as the heats of adsorption, of the volatile-condensables of the materials expected to be used in the HWO must be determined. This should include determinations of multiple specimens of each material for statistical significance. 

\section{Development Priorities}

Previous UV technology development priorities can be found in Astrophysics Biennial Technology Report 2022 (\url{https://apd440.gsfc.nasa.gov/technology.html}). Here we offer the following updated listing of UV technology development priorities as a path towards raising the TRL of components and systems to enable FUV science objectives.

\begin{itemize}
\item Reflective coating development and characterization towards high reflectivity Broadband FUV to NIR Mirror Coatings.
\item Characterization of coating polarization and uniformity 
\item Large format, low dark count, high efficiency, high dynamic range, photon counting,
Solar blind FUV and NUV Detectors.
\item High throughput, large format technology for integral field and multi-object spectroscopy
\item High efficiency diffraction gratings for high and low spectral resolution from FUV to NIR wavelengths
\item FUV Imaging bandpass filters
\item Testbed development to raise systems level TRL for UV coronagraph, spectrograph and imaging systems. 

\end{itemize}

\section{Path to HWO}

New UV science and technology developments have been progressing at the speed that the Astrophysics Research and Analysis (APRA) \& Strategic Astrophysics Technology (SAT) programs allow. To accelerate the pace to match the timeline proposed by the GOMAP for HWO, we need tools that go beyond the individual/component-level technology developments offered by
APRA and SAT.

We require a combination of: 
\begin{enumerate}
\item Process level (``materials physics'') development. 
\item Scaling existing technologies to the sizes needed for HWO.
\item Shifting from individual component level proof-of-concept development to the production line development of optical elements required for HWO electro/optical systems.
\end{enumerate}

We also need to invest in system-level prototype/test instruments for both the laboratory and space. These instruments provide powerful inputs to Phase A-level trades and decisions (~2029), so we need to do this soon.
 We require a combination of: 
\begin{enumerate}
  \item Investment beyond APRA-levels in suborbital missions (balloons and rockets).
\item Development of laboratory prototype testbed instruments 
\item In-space technology demonstration missions, analogous to the Earth Science InVEST program, that can combine process development/scaling with systems-level testing and early-career training (Section \ref{sec:training})
\end{enumerate}





\subsection{Training}\label{sec:training}

Instrumental astrophysics is an unusual field, in the sense that it is populated by people with a broad range of backgrounds, and it is supported by a broad range of roles. We might train as engineers and then move to astrophysics, or train as astrophysicists and then move into labs. We might be faculty members or professional engineers. The breadth makes this subfield unique. The project scale is also very large, compared to many subfields in astrophysics (and on par with subfields in physics like particle physics) although this has shifted as mission and telescope size has grown. This makes training a particularly crucial aspect of our human infrastructure. A great deal of knowledge is passed person to person during hands-on lab experiences. A peer reviewed journal specifically for instrumentation (SPIE JATIS) has existed for less than a decade, with a great deal captured in local observatory or project documentation or in conference proceedings. It is crucial as we shape the future of UV hardware and space missions we recognize that we not only need the right hardware - we need pathways to train and support early career researchers so they can take leadership roles in the mission in Phase A $\longrightarrow$~E. As we move to increasingly large space-based missions and surveys on the ground and in space, many astronomy programs focus on the data handling skills necessary to navigate the enormous results. We must also keep pathways open for training builders to sustain projects for the upcoming decades \cite{Tuttle2014}. We present below one potential pathway to increasing the ranks of builders for the ultraviolet.

\subsection{Smallsats for Accelerated Technology and Workforce Maturation}

Many of the key technologies in the development queue for HWO require the combined activities of 1) facility and process development for validation of technologies at the scale required for HWO and 2) deployment in the ‘real world’ environment of mission Integration \& Test followed by on-orbit operations. Many of these goals could be accomplished simultaneously through the development of a dedicated program combining laboratory advancement and process scaling conducted in parallel with the deployment of these technologies to space.
Parallel and closely-linked facility, laboratory, and instrument prototype development program that can be applied to any of NASA’s Future Great Observatories (FGOs), and an initial framework for this program has been outlined as the Smallsat Technology Accelerated Maturation Platform (STAMP)~\cite{france2024}. Advanced broadband optical coatings, high-sensitivity ultraviolet detector systems, and multi-object selection technology could all be brought to TRL 6 and flight demonstrated through such a program. STAMP advances HWO technology on an accelerated timescale, building on current ROSES SAT+APRA programs, reducing cost and schedule risk for HWO.  while conducting a compelling program of preparatory science and workforce development with direct cost and schedule risk reduction for HWO mission implementation in the 2030s.

The STAMP science and instrument teams would be made up of $>$ 50\% early-career researchers; this builds the critical relationships between observers, theorists, and instrumentalists that enable open communication and a cohesive development environment for HWO. The organizational layout of the mission would feature ‘deputies’ in all key mission science roles, including deputy-PI (dPI), deputy-Project Scientist (dPS), deputy-Instrument Scientist (dIS), and in all the major engineering and program management positions (Program Manager, Project Systems Engineer, Instrument Systems Engineer, etc). These positions would be fully funded and provide a direct route for the training of the scientists and engineers with the expertise to serve in science, instrument, and mission leadership roles for HWO. 

\section{Summary}\label{sec-summary}

We have presented here a summary of the current status of UV technology, establishing a baseline for the development of the upcoming Habitable World Observatories instrumentation. This representative sample of the last several decades of development demonstrates not only the thriving UV technology but also illustrates the areas for advancement in preparation for HWO. Multiple pathways exist for successful instrumentation across the ultraviolet bandpass, and rapid investment will support the scaling up of technology that has been flight qualified on smaller scales. Coordination across the community will provide a path for rapid subsystems level flight qualification along with full integrated systems level tests. This is an exciting time to work in UV hardware, and we hope the knowledge gathered here provides a strong foundation for the exciting challenges posed by HWO.

\centerline{\bf{ACKNOWLEDGMENTS}}
This work has been supported by the following NASA grants:  NNX11AG54G, NNX14AI78G, NNG15PF22P, NNG16LI16P, NNX17AC26G, 80NSSC19K1040, 80NSSC22K0940, 80NSSC22K1698, NNH08ZDA001N, NNH12ZDA001N-RTF, NNX16AG28G, NNX12AG06G, 80NSSC20K0412, 80NSSC21K2016, 80NSSC21K1667, 80NSSC19K0661, NNX17AI84G, 80NSSC23K0754, 80NSSC22M0081. A portion of this research was done at Jet Propulsion Laboratory, California Institute of Technology, under a contract with the National Aeronautics and Space Administration  (80NM0018D0004). This work was supported in part by the NASA Cosmic Origins Program Office under the auspices of the Cosmic Origins UV Working Group.

\appendix

\section{UV Technology Frequently Asked Questions}\label{app-FAQ}

\begin{itemize}
\item \textbf{What are some of the challenges associated with combining ultra-high-contrast exoplanet observations and a UV-sensitive telescope?}

\begin{itemize}
  
\item Wavefront Control: Going to shorter wavelengths means that the same wavefront error in physical dimension becomes a larger fraction of wavelength. This places more stringent requirements on the deformable-mirror system. 
\item Polarization aberration: Although effects of polarization aberration are more pronounced in the UV, this degradation is concentrated at small angles in terms of $\lambda /D$. Habitable zones of the best UV target stars tend to be at larger angular separations and angular separation per $\lambda /D$ is also larger in the UV compared to longer wavelengths can mitigate the science impact of this change.
\item Coating uniformity: A recent study (Krist, priv. comm.) using the USORT pupil (which has 19 hexagonal segments in the primary-mirror assembly) found that reflectance differences at the segment-to-segment level of up to 3\% is tolerable in terms of contrast performance. 
\item Scattered light: Scattering, due to surface roughness and particle contaminants, is stronger in the UV. Accurate estimates of near-angle scattering (NAS) is currently lacking, mainly due to lack of measurement data \cite{pfisterer}. NASA Engineering and Safety Center is funding a study (PI: Gaskin) to better quantify the impact of NAS on ultra-high-contrast imaging, which can lead to improved mirror-PSD and contamination-control requirements. 

\end{itemize}

\item \textbf{Will UV coating surface roughness be capable of working with coronagraphy at the level needed for HWO?}

The current roughness requirements driven by the nominal coronagraphy design are applied to the full wavelength range. Work is in process to test and quantify roughness, but the requirement as currently understood is not met by any coating at any wavelength range.

\item \textbf{What contamination control requirements does UV sensitivity impose on the observatory?}

Contamination control is necessary for the reflective coatings in the ultraviolet. However, the contamination control is well understood and defined as shown in Section \ref{sec-contamination} and Appendix \ref{app:contamination}.

\item \textbf{What are some of the key science cases that need wavelength coverage down to 100~nm?}

We highlight key science cases in Section \ref{Science}, and summarize them below. 

\begin{itemize}
\item 102 — 115~nm : FUV inputs into exoplanetary atmospheres, including emission lines like SIV, OVI, NeV, and FeXIX
\item 100 — 115~nm : Direct measurement of absolute abundance and temperature of inner disk gas of protoplanetary disks.
\item 103nm : Provides access to CGM OVI at $z = 0$
\item 100 — 110~nm : CGM high ionization lines like Ne VIII (0.29 $< z <$  0.41) and Si XII ( 1 $< z <$ 1.2)
\item 100 — 120~nm: Lyman continuum escape fraction at very low redshift (0.1 $< z <$ 1)
\end{itemize}

\item \textbf{MKIDS and nanowire detectors have made a lot of progress - can we use those technologies?} 

Not for this. Currently, these detectors are both at a relatively low TRL and not yet achieving desired QE in the FUV for potential UV spectrographs. Detectors beyond those discussed here are being considered for other instrumentation on HWO. 

\item \textbf{Does pushing the observatory's wavelength coverage below the short wavelength cutoff of Hubble (115~nm) increase the technical challenges?}

No. In the decades since Hubble was built, technology development has pushed forward performance below 115~nm, including in optical coatings and detectors. Sections \ref{sec-coatings} and \ref{sec-detectors} demonstrate the significant progress that decreases the technical risk of extending UV coverage beyond the Hubble cutoff.

\item \textbf{Are MCP detectors limited in their ultimate count rate and susceptible to gain sag that limits their lifetime?}

High rate real time gain sag is due to the local recharge time of a pore. During instrument design, strategies can be used to improve the detector dynamic range, such as using lower resistance MCPs and low gain compatible readouts such as the cross strip to achieve 1000 events/sec for a resolution element. 
The MCP lifetime is related to total charge extraction. Traditional MCPs drop in gain substantially with extracted charge. The lifetime of new technology borosilicate MCPs with ALD functionalization have been shown to maintain stable high gain during extended life testing ($>$\qty{7}{\coulomb\per\square\centi\metre}) to more than 5 x 10 13 events per cm2 which is substantially more than most missions accumulate.

\item \textbf{Are MCP detectors at risk when exposed to high count rates?}

In practice, no. The rate would have to be extremely high (focused beam of sunlight) to physically heat the MCP. This is usually avoided (and proven at TRL 9) by multiple layers of hardware, software, and observation planning restrictions. Newer MCPs also tolerate much higher count rates (Siegmund 2023 NIM) as demonstrated by full recovery after exposure to laser illumination.

\end{itemize}

\section{Contamination Control}\label{app:contamination}

\paragraph{Particles}

Typical particles, including fibers, that contaminate space hardware range from approximately 1 µm to 1 mm in maximum dimension. Their sources include air-borne dust in the local environment and fibers and particles released from tools, work surfaces, garments, human hair and skin, etc.

The dominant effects of settled particles for optical systems are light scatter and obscuration (sometimes parametrized as percent area covered). IEST-STD-1246 \footnote{IEST-STD-CC1246, Product Cleanliness Levels and Contamination Control Program, Institute of Environmental Sciences and Technology} establishes particle Cleanliness Levels listing the maximum number of particles per 0.1 m$^{2}$ \footnote{The strange notation of \# / 0.1 m$^{2}$ is a result of the historical attempt to equate Imperial unit values which were given in \# / ft$^{2}$ to metric nomenclature. Of course, this leads to an error of about 9.3 \% in comparing Imperial data to metric data.} surface area for ranges of particle maximum linear dimensions.

\begin{figure}
\centering 
\includegraphics[width=0.7\textwidth]{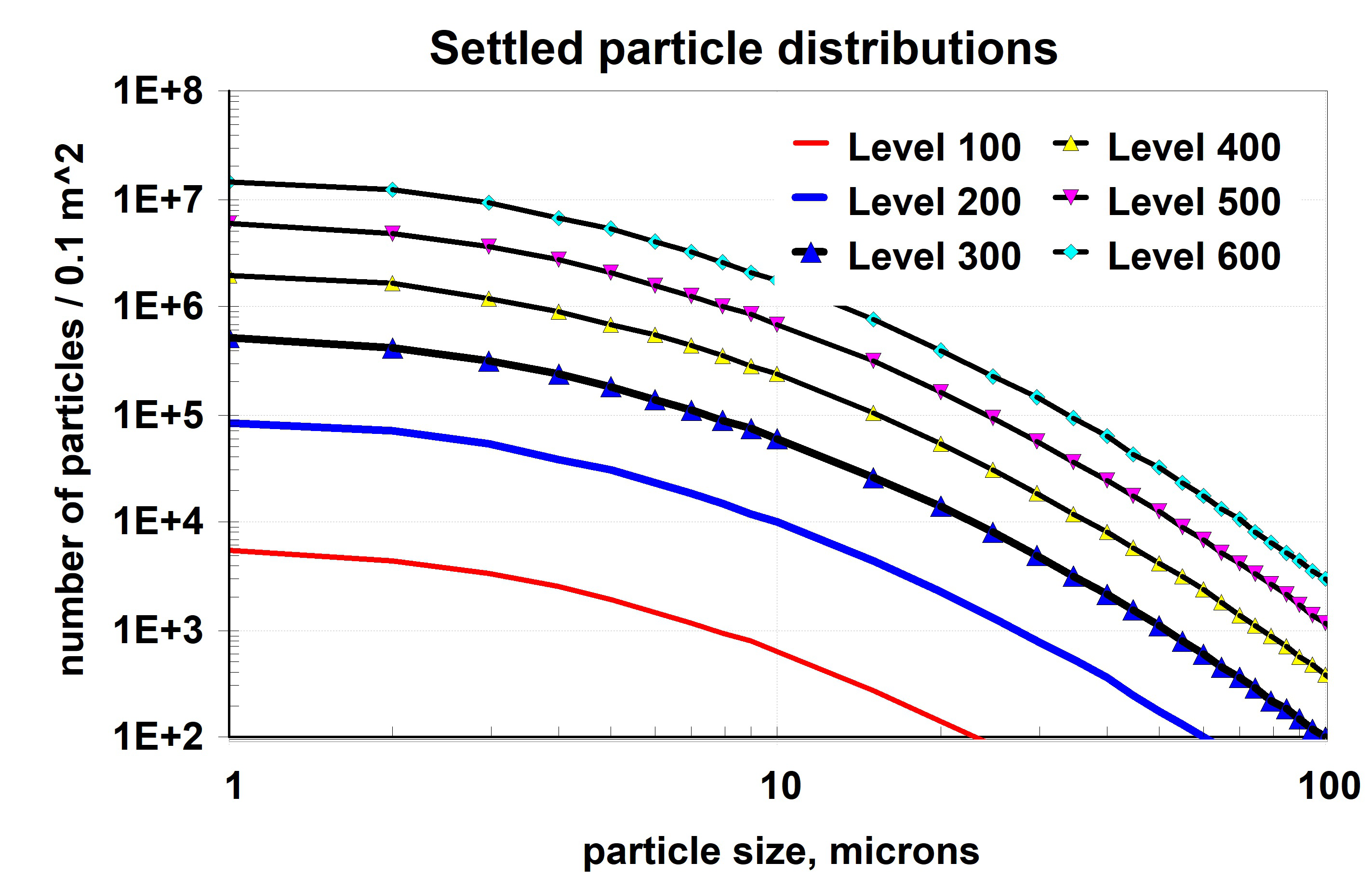}
  \caption{Cumulative particle size distributions for selected Cleanliness Levels}
\label{fig:psizes}
\end{figure}

Typical criteria for contamination-sensitive optical instruments are often in the range of particle Cleanliness Levels of 200 to 300 (Figure ~\ref{fig:psizes}) which are not unusually difficult to achieve, even for large optical systems. Even with 3 or 4 optical surfaces in series, this represents less than 0.1 \% optical obscuration for the optical system (Figure \ref{fig:particles_losses}). However, for scatter-sensitive optical instruments such as coronagraphs, complex scatter analyses, involving measurements of the actual size distribution, may be necessary to determine acceptable particle deposition.

\paragraph{Molecular Contaminants}

It is difficult to bound the sources and variety of molecular contaminants to which space hardware can be exposed. Even after thorough initial and precision cleaning, the molecular contaminants found on space hardware in cleanrooms include volatile-condensable molecular species of masses up to approximately 300 amu. These molecular contaminants may exist on surfaces as randomly spaced single molecules, coalesced nanometer sized ``deposits,'' monolayer to multilayer films less than 0.5 nm to a few nm in thickness, or even micro droplets, all possibly consisting of a mix of molecular species.

The dominant effect of these molecular contaminants for optical systems is light absorption, which can be quantified using Lambert’s Law when the absorption coefficients of the contaminants are known. In a worst-case of molecular contaminant deposits as micro-droplets, light scatter will also occur.

IEST-STD-1246 defines non-volatile molecular cleanliness levels over a very broad range from 10 ng/0.1m$^{2}$ to 25 mg/0.1m$^{2}$ (alternatively, 10 pg/cm$^{2}$ to 25 $\mu$g/cm$^{2}$). Molecular contaminant lifetime allowables for spacecraft and space science instruments have typically been in the range of 50 to 500 $\mu$g/0.1m$^{2}$	 (50 to 500 ng/cm$^{2}$; equivalent to thicknesses of 0.5 to 5 nm for contaminants with specific gravity = 1.0). However, dependent on the results of contamination degradation analyses of the HWO telescope and instruments, the molecular contaminant allowables for HWO may be significantly lower\footnote{For example, with a design consisting of 5 surfaces (primary, secondary, pick-off mirror, instrument reflecting surface, detector surface) there will be 9 passes through contaminant deposits. For deposits 0.5 nm thick, the total degradation between 100 nm and 150 nm will be 30\% to 50\%.}.

\paragraph{Molecular Contaminant Transfer}

In controlled environments such as cleanrooms and vacuum chambers troublesome molecular contaminants are transferred to and accrete onto surfaces primarily by desorption from a source surface and subsequent adsorption onto another surface. These are called volatile-condensable species. The standard term for this desorption is outgassing and is quantified as mass loss per unit area, typically in micrograms per cm$^{2}$ (µg/cm$^{2}$) or nanograms per cm$^{2}$ (ng/cm$^{2}$). These outgassing species can be both molecules of, or molecular fragments of the base material, foreign contaminants from the surface of the base material, or contaminants diffused from the interior of the base material.

Another source of contaminant transfer between surfaces is contact transfer. As the name implies, contact transfer is contacting one surface with another and the transfer of contaminants from one to the other. An example is to touch a clean surface with a contaminated glove and leave some of the contaminants on the clean surface.

Molecular contaminant transfer, or transport, is discussed further in Appendix \ref{sec-molecularContaminantEffect}.

\paragraph{Molecular Contaminant Effect}
\label{sec-molecularContaminantEffect}
The magnitude of the optical absorption losses are predictable using Lambert’s Law: the intensity of a light beam inside an absorbing medium decreases exponentially as a function of the distance the beam has traveled into the medium and as a function of the absorption coefficient, $\alpha$, of the absorbing material.

The absorption coefficients, $\alpha$, given in units of inverse thickness (1/t), or inverse mass per areal density (i.e., cm$^{2}$/ng), have widely ranging values over several orders of magnitude for wavelengths from the far UV through the IR for the various molecular contaminants typically found.

The available data on absorption coefficients for molecular contaminant deposits is very limited, as these values are exceptionally difficult and expensive to measure, particularly for UV wavelengths. In support of the Wide Field Planetary Camera (UV wavelength range: 200-1000 nm) and beneficial to the Faint Object Spectrograph (UV wavelength range: 115-850 nm) which were flown on Hubble, $\alpha$ values of contaminant deposits were measured at several UV wavelengths and for multiple thicknesses by Dr. Joseph Muscari, Martin Marietta Corp., for several adsorbed contaminants. \footnote{Nonmetallic Materials Contamination Studies, Final Technical Report, J A Muscari, Martin Marietta Corp., JPL contract 955426, December 16, 1980}  Other limited compilations of $\alpha$ values have been produced showing a broad range of values vs. wavelength for the few contaminant source materials which have been tested. Detailed information about the test methods and test sample variations are unknown and may contribute to the data variations. 

\begin{figure}
\centering 
\includegraphics[width=0.7\textwidth]{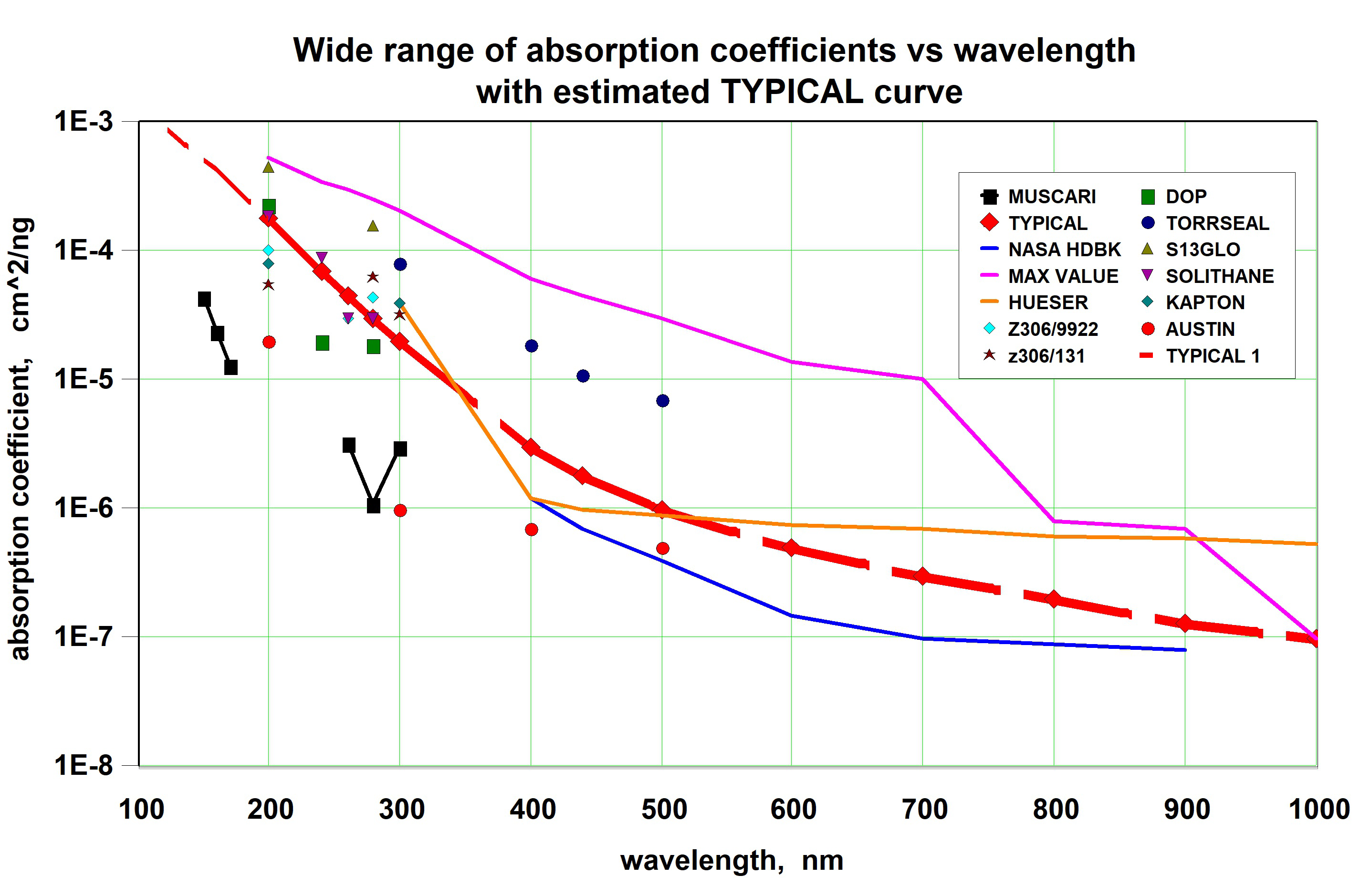}
  \caption{Range of Absorption Coefficients vs. Wavelength with estimated TYPICAL Curve. }
\label{fig:abscoeff}
\end{figure}

Data currently available can be used to bracket the probable range of UV $\alpha$ values. Figure~\ref{fig:abscoeff} shows a summary of the available absorption coefficient data compiled mostly by Joseph Hueser, Ball Aerospace (unpublished), over the wavelength range of 150 to 1000 nm. A ``typical value'' curve has been included in the Hueser plot. Many of the data points shown in the Hueser plot were derived from Muscari measurements. Great caution should be used in applying this data as in most cases the precision of unknown, was typically measured on relatively thick contaminant deposits, and may not be representative of currently available materials.

Cautiously using the estimated ``typical value'' curve given in Figure~\ref{fig:abscoeff} for molecular contaminant absorption values, and for want of more and better data, the predicted absorption over the range of 122 nm to 300 nm for contaminant equivalent thicknesses of 0.25, 0.5 and 1 nm was shown in Figure~\ref{fig:particles_losses}. This illustrates how very important it is to minimize molecular contamination in the vacuum UV and the need for precise adsorption coefficient data on materials to be used on the HWO. For a telescope and UV science instrument in series, there will be a minimum of 4 optical surfaces and 7 passes of the optical beam through contaminant deposits, which can severely degrade instrument performance.  $\alpha$ values are orders of magnitude higher for most molecular contaminants in the UV vs. the near IR and Mid-IR, making HWO far more contaminant-sensitive than JWST.

\paragraph{Molecular Contaminant Cleanliness Levels}

Table~\ref{tab:cleanlevels} is a listing of selected molecular Cleanliness Levels from IEST-STD-CC1246E with level names shown in column 1. Column 2 is the defined maximum allowable Non-Volatile Residue (NVR) per 0.1 m$^{2}$ for each level. When collecting contaminant samples from surfaces for quantifying, the measurement results are often given in nanograms per cm$^{2}$ (ng/cm$^{2}$). 
Column 3 shows this equivalent areal density, while the Column 4 shows the approximate equivalent thickness were the molecular contaminant to be uniform in thickness and to have a density of 1 g/cm$^{3}$. \footnote{The density of most contaminants is expected to be in the range of 0.8 to 1.2 g/cm$^{3}$ based on the densities of the typical source plastics.}

\begin{table}
  \centering
\begin{tabular}{c|c|c|c}
\hline
Molecular & Maximum Allowable & Equivalent & Approx. \\
Contaminant & NVR Mass / 0.1 m$^{2}$ & Areal Density & Equivalent \\
Level & Surface Area & (ng/cm$^{2}$) & Thickness (nm) \\
\hline
	R5E-3	&	5 µg	&	5	&	0.05 \\
	R1E-2	&	10 µg	&	10	&	0.1 \\
	R2E-2	&	20 µg	&	20	&	0.2 \\
	R5E-2	&	50 µg	&	50	&	0.5 \\
	R1E-1	&	100 µg	&	100	&	1 \\
	R2E-1	&	200 µg	&	200	&	2 \\
	R5E-1	&	500 µg	&	500	&	5 \\
 \hline
\end{tabular}
\caption{Non-volatile molecular cleanliness levels from IEST-STD-1246}
  \label{tab:cleanlevels}
\end{table}

\paragraph{Molecular Mass Transport}

Outgassing is not just a vacuum phenomenon, it occurs in all environments. As molecular contaminants transfer between surfaces at room temperature, such as during exposed periods in cleanrooms, they desorb from source surfaces, adsorb onto exposed surfaces, and come to rest for predictable time periods, with the mean molecular rest times dependent on their heats-of-adsorption \footnote{Heats of adsorption can differ for a specific molecular species on different surfaces, dependent on other contaminants present and the unique Van der Waals forces for the contaminant-surface pair. } and the local temperature. Often the total adsorbed mass can be reduced or essentially depleted by elevating the temperature. However, this must be done in high vacuum. The species with lower heats of adsorption, less than about 23 kc/mol deplete readily in vacuum even at near room temperature. The molecular species which are the most difficult to deplete are those with heats of adsorption greater than about 24 kc/mol. And the species with heats of absorption greater than about 27 kc/mol are so stable that they are affected very little at spacecraft bakeout temperatures, and may more likely be chemisorbed rather than physisorbed as is the norm for volatile-condensable molecular contaminants.

Cleanrooms are far cleaner than laboratories and workshops, but still, they are not perfect. They will have measurable quantities of airborne particles and volatile-condensable molecular species leading to settling of particles and adsorption of molecular contaminants. Historical evidence from measurements on witness plates of volatile-condensables has shown that molecular contaminants do adsorb onto these surfaces with adsorption rates and maximum quantities being dependent on the cleanroom quality and assembly and integration operations that are occurring. Such evidence has shown that the time for the adsorbed species to reach some apparent equilibrium with the gaseous phase species to be on the order of a few weeks to a couple of months for the few cleanrooms that were monitored.

\begin{figure}
\centering 
\includegraphics[width=0.7\textwidth]{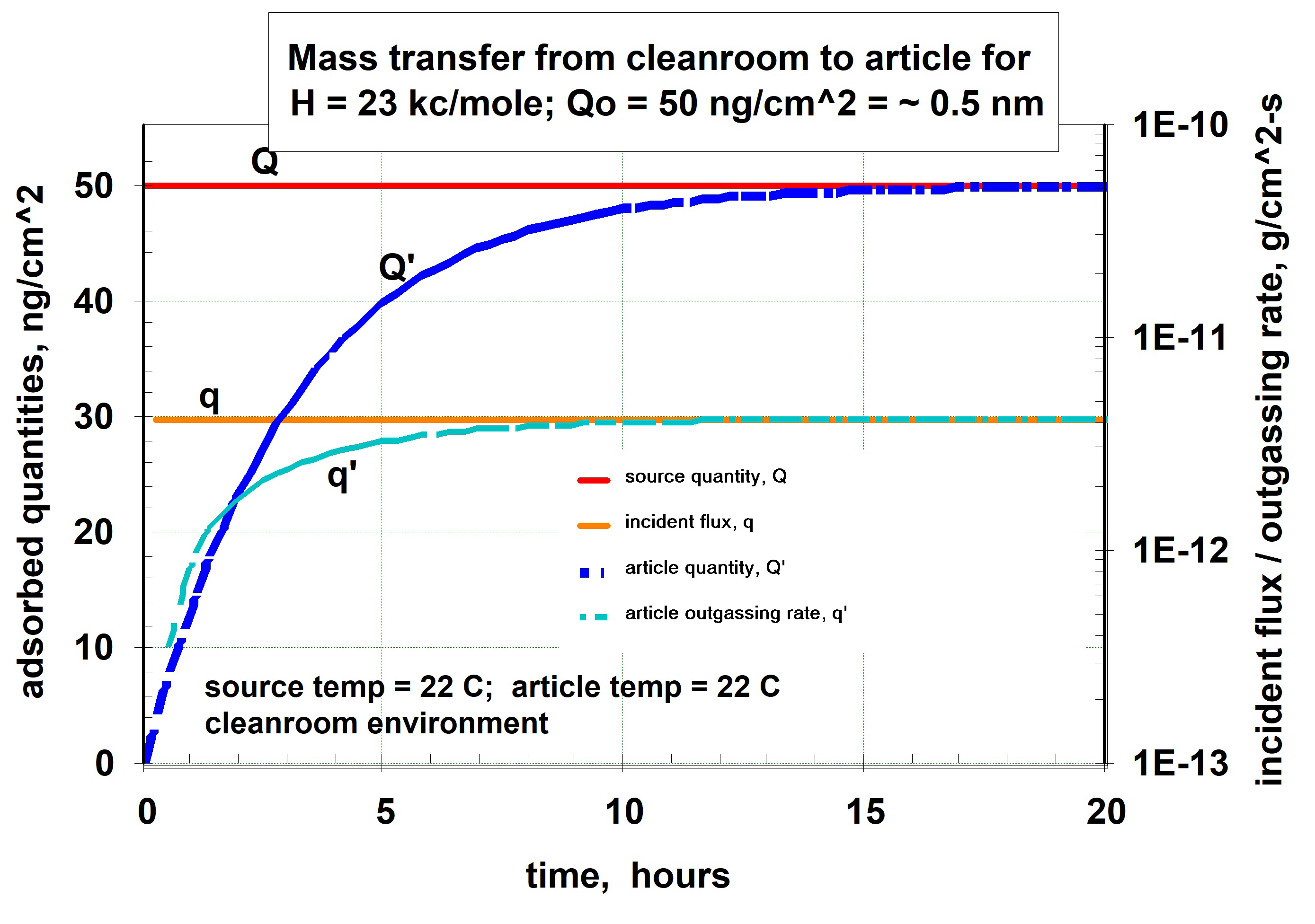}
  \caption{Mass transfer contaminant accretion onto clean surfaces at an H of 23 kc/mol over 15-20 hrs.}
\label{fig:masstrans1}
\end{figure}

\begin{figure}
\centering 
\includegraphics[width=0.7\textwidth]{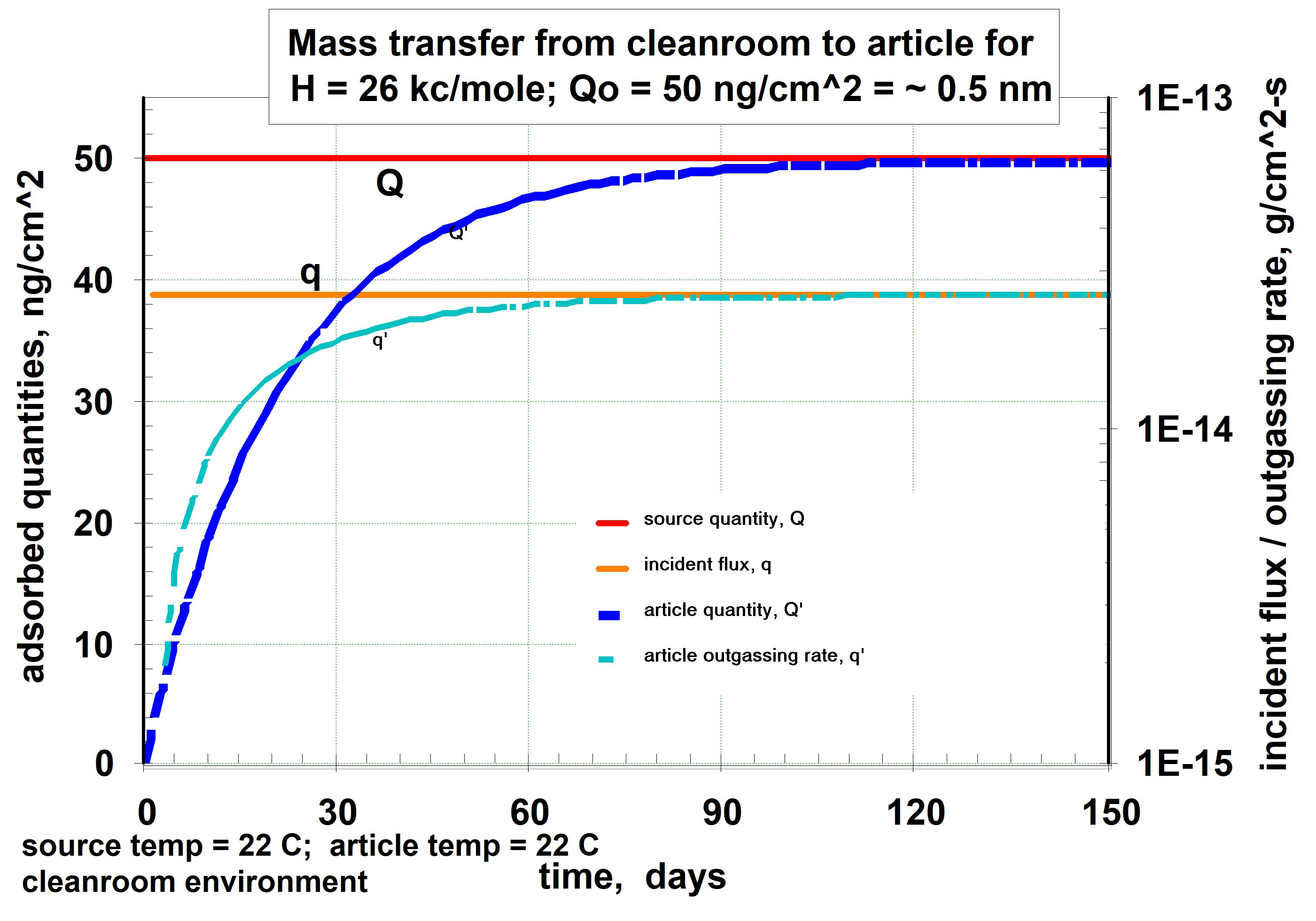}
  \caption{Mass transfer contaminant accretion onto clean surfaces at an H of 26 kc/mol over 120 hrs.}
\label{fig:masstrans2}
\end{figure}

\begin{figure}
\centering 
\includegraphics[width=0.7\textwidth]{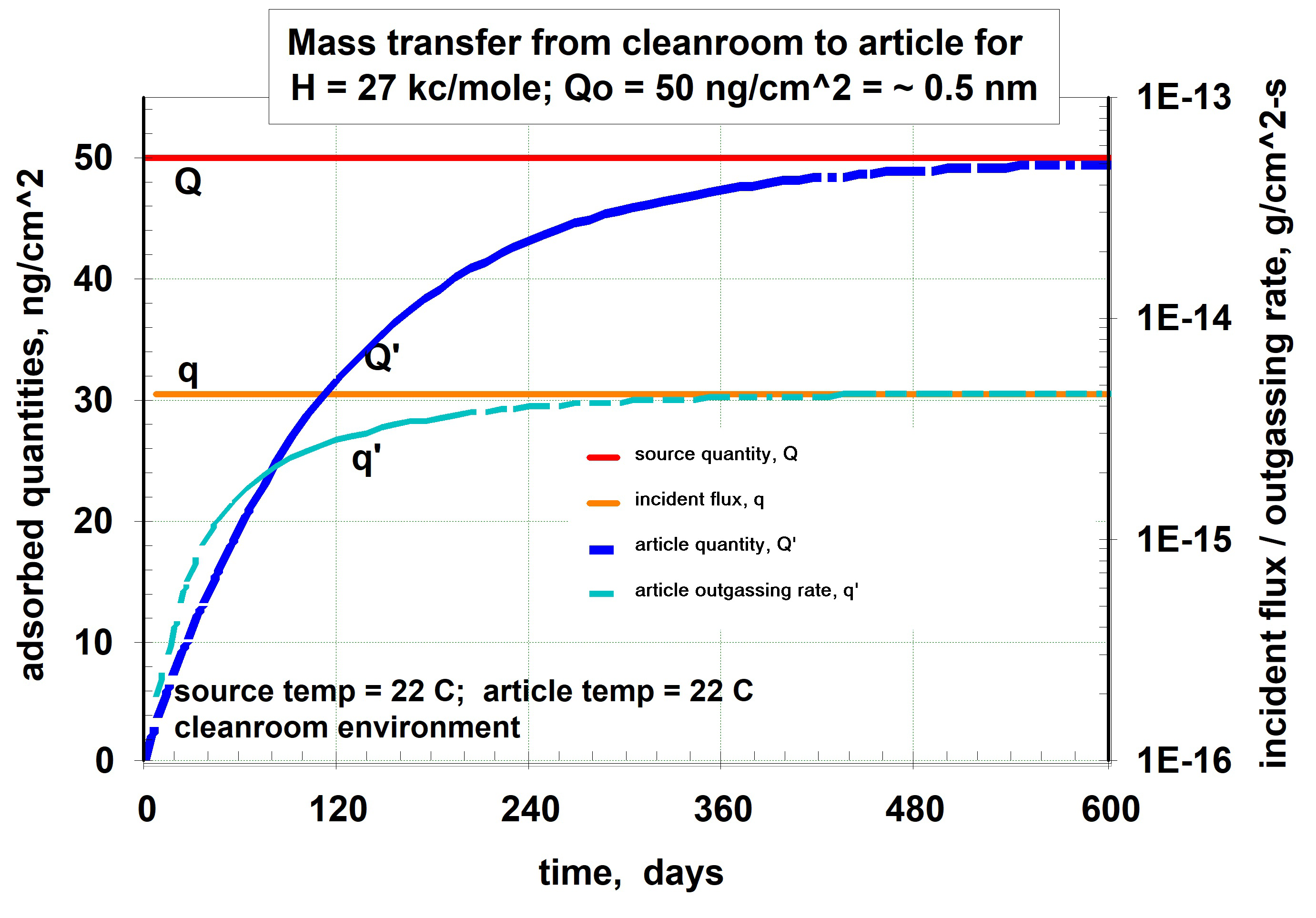}
  \caption{Mass transfer contaminant accretion onto clean surfaces at an H of 27 kc/mol indicating low levels of contamination.}
\label{fig:masstrans3}
\end{figure}

\begin{figure}
\centering 
\includegraphics[width=0.7\textwidth]{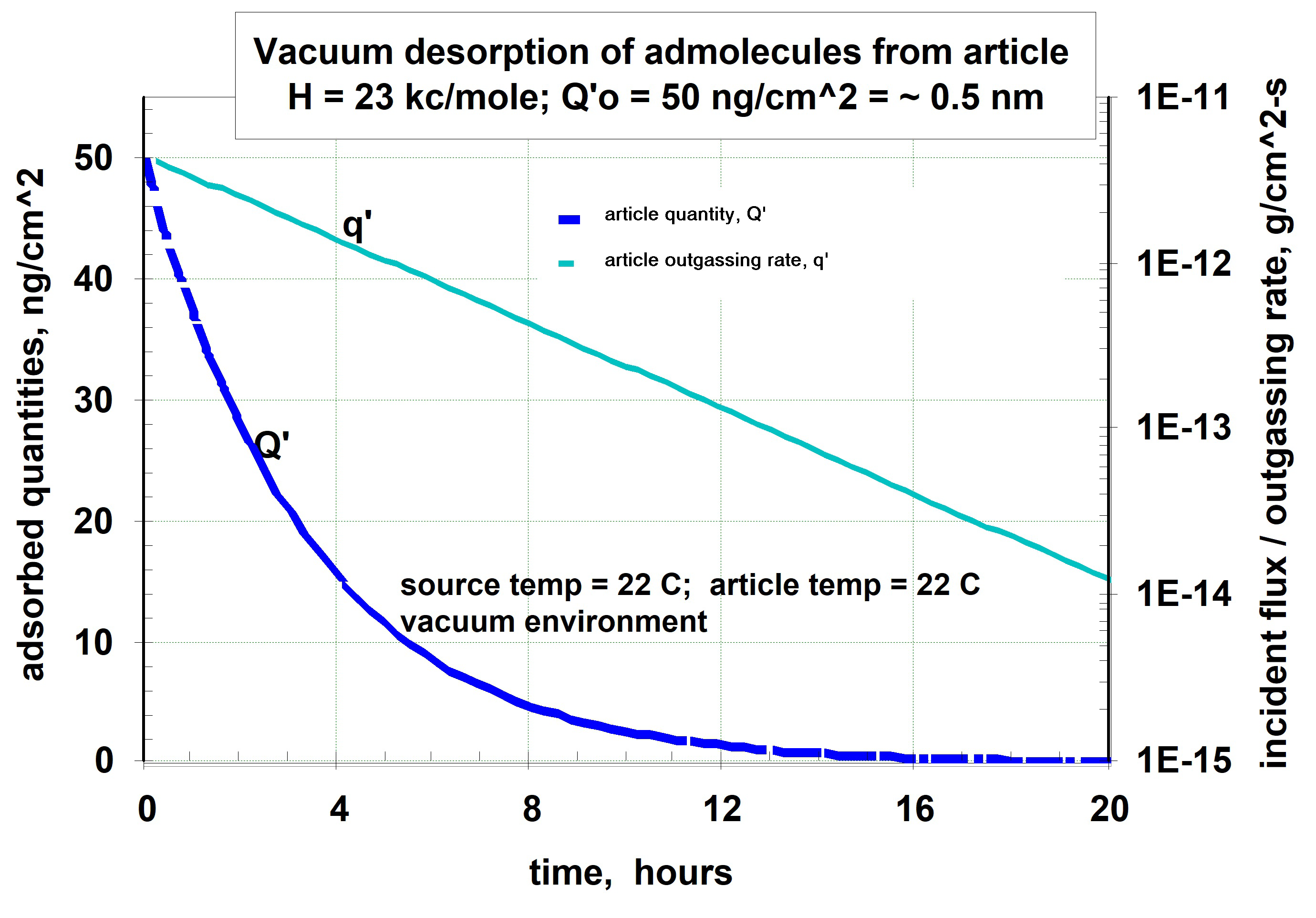}
  \caption{Analysis showing that when a plane contaminated surface is placed in vacuum the molecular contaminants desorb relatively quickly as a function of temperature and heats of adsorption. }
\label{fig:masstrans4}
\end{figure}

Figures ~\ref{fig:masstrans1} and ~\ref{fig:masstrans2} show the contaminant accretion rates onto clean surfaces for an important range of volatile-condensables. For the analysis, it was assumed that the areal density of each adsorbed contaminant at equilibrium was 50 ng/cm$^{2}$ (roughly equivalent to a single monomolecular layer of water) and does not represent a particular cleanroom. Figure ~\ref{fig:masstrans3} shows that the predicted time for a clean, exposed surface to equilibrate to the cleanroom quantity of species with heats of adsorption of 23 kc/mol is on the order of about 15-20 hours, whereas Figure ~\ref{fig:masstrans4} shows it could take up to 120 hours for a 26 kc/mol contaminant to reach cleanroom equilibrium.  Additional analyses show that the volatile-condensibles having heats of adsorption less than about 23 kc/mol move very quickly through the cleanroom, while volatile-condensables having heats of adsorption greater than 26 kc/mole travel so slowly from source surfaces at room temperature that they are not a significant problem in the cleanroom. This is shown in Figure ~\ref{fig:masstrans3}.

The message is that molecular contaminants do transfer throughout cleanrooms with those with low heats-of-adsorption reaching some equilibrium on exposed surfaces in hours to days, while those with higher heats-of-adsorption may take so long to transfer that they do not present a significant contaminant risk to hardware that is sealed or bagged and purged for the majority of the time in AIT.

Figure ~\ref{fig:masstrans4} shows that when a plane, fully exposed contaminated surface is placed in vacuum the molecular contaminants desorb relatively quickly in a predictable manner as a function of temperature, heats of adsorption, and elimination of the original contaminant sources. However, consider as an example, a short telescope tube with dimensions of length/diameter = 4 and open on the aperture end. The probability of a molecule desorbing from a random spot inside this open-ended tube venting out the aperture in each desorb-adsorb cycle is only 0.06, based on view factors. In this case, the time to depletion of the contaminant is increased from about 12 hours to roughly 200 hours. This has important implications for vacuum baking cleanup of complex assemblies.

Venting is an extremely important factor for efficient vacuum bake maintenance cleaning of assemblies. Vents of enclosures such as electronics boxes are normally designed to prevent pressure differentials during vacuum pumpdown and during launch from mechanically damaging the enclosure. A normal launch vent design criterion is given by:

\begin{equation}
  A > \frac{V}{2000 mm}
\end{equation}

where A = cross sectional area of the vents, and V = volume of the vented enclosure.

This is sufficient for launch venting, but not for vacuum bake cleaning. The internal surface area of even an empty enclosure with vents to this criterion is approximately 1000 greater than the area of the vents only, leading to a ``vent factor" (defined as the ratio of the area of the vent aperture to the total surface area inside vented cavity) of ~0.001. Vacuum baking of such articles for even weeks would be very ineffective, leaving these articles as potential contaminant sources when in assemblies adjacent to optical elements, particularly during vacuum tests.

\begin{figure}
\centering 
\includegraphics[width=0.7\textwidth]{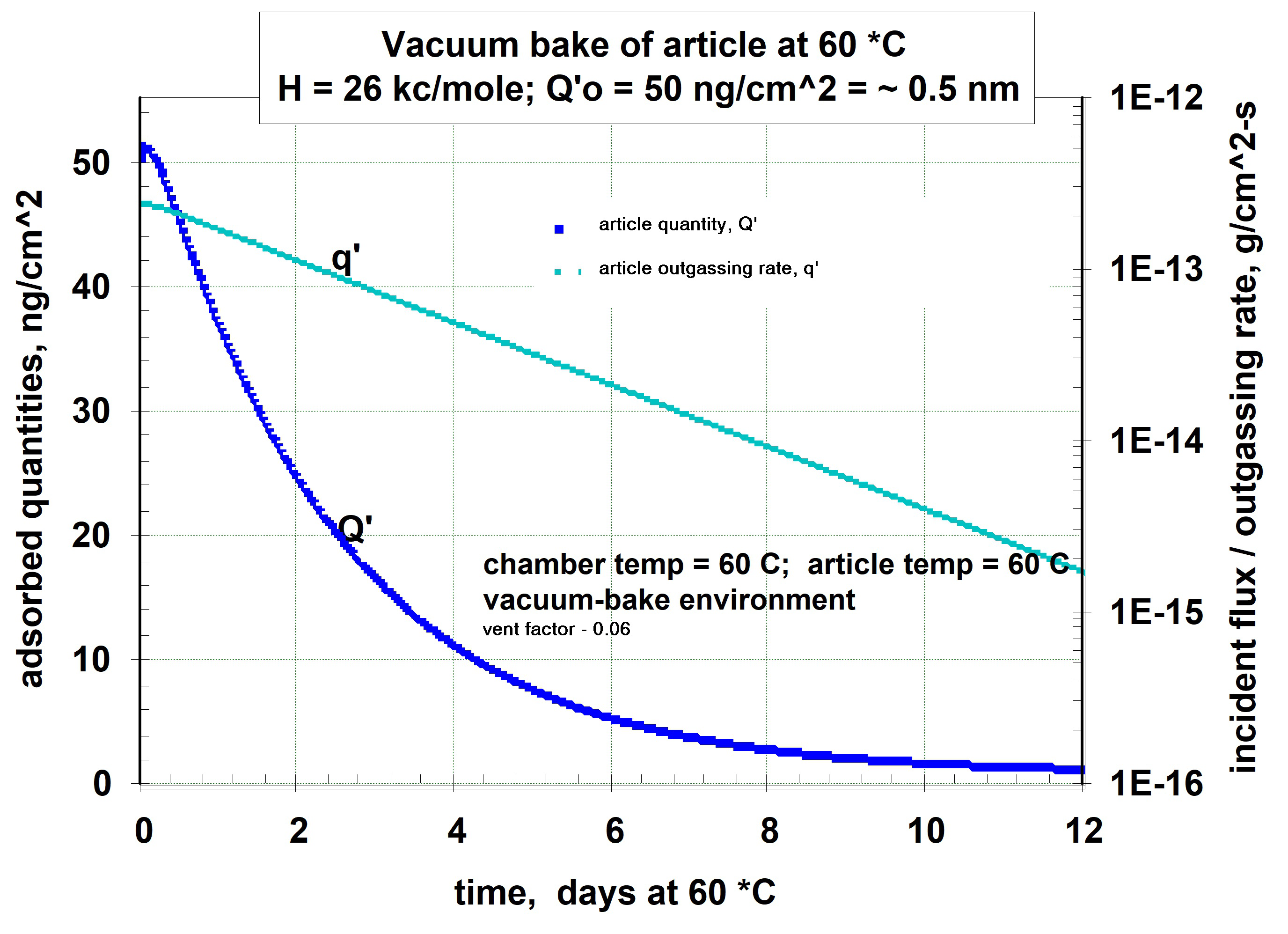}
  \caption{Vacuum bake efficacy: the predicted time to essentially deplete a heat of adsorption 26 kc/mol contaminant is on the order of 10 to 12 days at 60 $^{\circ}$C for surfaces with a vent factor of 0.06.}
\label{fig:vacbake}
\end{figure}

The maximum temperature for vacuum bakes of spacecraft assemblies is often limited by the batteries to about 60 $^{\circ}$C. Figure ~\ref{fig:vacbake} shows that the predicted time to essentially deplete a heat of adsorption 26 kc/mol contaminant is on the order of 10 to 12 days at 60 $^{\circ}$C for surfaces with a vent factor of 0.06. Increasing the vacuum bake temperature to 82 $^{\circ}$C will reduce the time to deplete the 26 kc/mol species to about 1 day. This can result in significant savings when vacuum baking subassemblies. And the contaminants with lower heats of adsorption will be depleted much quicker. However, regardless of the heat of adsorption, the times to deplete will be far longer for surfaces inside boxes with only small launch vents. 

It is shown by analysis that the most troublesome volatile-condensable contaminants to deplete with vacuum baking are those with heats of absorption in the range of 23 to 26 kcal/mol. Molecular contaminants with lower heats of absorption are readily depleted in vacuum at room temperature and above. However, the contaminants with high heats of absorption require higher temperatures and perhaps long times in vacuum bake to deplete. 

The above data show that volatile-condensable molecular contaminants can be depleted by vacuum baking, however adequate venting of enclosed volumes is necessary. Enclosures of articles other than optical components should be opened for vacuum bake cleanup and should be vacuum baked separate from the optics. The troublesome contaminants must be depleted to acceptable levels prior to top assemblies of optics. Mechanical and electronic assemblies must be exposed in vacuum together, apart from optical elements, to avoid the optics being contaminated. 

To take scheduling advantage of this knowledge requires monitoring all vacuum bakes with high-sensitivity, broad-range residual gas analyzers (RGA), and determining the heats of adsorption of the specific higher mass (100 – 300 amu) species. In some cases, a temperature-controlled quartz crystal microbalance (TQCM) can be used to determine approximate heats of adsorption simultaneously while monitoring the RGA-identified mass peaks.

\newpage


\bibliography{uvtech}   

\begin{thebibliography}{100}

\bibitem{Astro2020}
NASEM, {\em {Pathways to Discovery in Astronomy and Astrophysics for the 2020s}}  (2021).

\bibitem{bryson20}
S.~{Bryson}, J.~{Coughlin}, N.~M. {Batalha}, {\em et~al.}, ``{A Probabilistic Approach to Kepler Completeness and Reliability for Exoplanet Occurrence Rates},'' {\em \aj} {\bf 159}, 279  (2020).

\bibitem{dressing15}
C.~D. {Dressing} and D.~{Charbonneau}, ``{The Occurrence of Potentially Habitable Planets Orbiting M Dwarfs Estimated from the Full Kepler Dataset and an Empirical Measurement of the Detection Sensitivity},'' {\em \apj} {\bf 807}, 45  (2015).

\bibitem{Harman2015}
C.~E. {Harman}, E.~W. {Schwieterman}, J.~C. {Schottelkotte}, {\em et~al.}, ``{Abiotic O$_{2}$ Levels on Planets around F, G, K, and M Stars: Possible False Positives for Life?},'' {\em \apj} {\bf 812}, 137  (2015).

\bibitem{Zahnle2017}
K.~J. {Zahnle} and D.~C. {Catling}, ``{The Cosmic Shoreline: The Evidence that Escape Determines which Planets Have Atmospheres, and what this May Mean for Proxima Centauri B},'' {\em \apj} {\bf 843}, 122  (2017).

\bibitem{segura05}
A.~{Segura}, J.~F. {Kasting}, V.~{Meadows}, {\em et~al.}, ``{Biosignatures from Earth-Like Planets Around M Dwarfs},'' {\em Astrobiology} {\bf 5}, 706--725  (2005).

\bibitem{hu12}
R.~{Hu}, S.~{Seager}, and W.~{Bains}, ``{Photochemistry in Terrestrial Exoplanet Atmospheres. I. Photochemistry Model and Benchmark Cases},'' {\em \apj} {\bf 761}, 166  (2012).

\bibitem{tian14}
F.~{Tian}, K.~{France}, J.~L. {Linsky}, {\em et~al.}, ``{High stellar FUV/NUV ratio and oxygen contents in the atmospheres of potentially habitable planets},'' {\em Earth and Planetary Science Letters} {\bf 385}, 22--27  (2014).

\bibitem{meadows18}
V.~S. {Meadows}, G.~N. {Arney}, E.~W. {Schwieterman}, {\em et~al.}, ``{The Habitability of Proxima Centauri b: Environmental States and Observational Discriminants},'' {\em Astrobiology} {\bf 18}, 133--189  (2018).

\bibitem{fulton18}
B.~J. {Fulton} and E.~A. {Petigura}, ``{The California-Kepler Survey. VII. Precise Planet Radii Leveraging Gaia DR2 Reveal the Stellar Mass Dependence of the Planet Radius Gap},'' {\em \aj} {\bf 156}, 264  (2018).

\bibitem{owen18}
J.~E. {Owen} and D.~{Lai}, ``{Photoevaporation and high-eccentricity migration created the sub-Jovian desert},'' {\em \mnras} {\bf 479}, 5012--5021  (2018).

\bibitem{tsai23}
S.-M. {Tsai}, E.~K.~H. {Lee}, D.~{Powell}, {\em et~al.}, ``{Photochemically produced SO$_{2}$ in the atmosphere of WASP-39b},'' {\em \nat} {\bf 617}, 483--487  (2023).

\bibitem{france14}
K.~{France}, G.~J. {Herczeg}, M.~{McJunkin}, {\em et~al.}, ``{CO/H$_{2}$ Abundance Ratio {\ensuremath{\approx}} 10$^{-4}$ in a Protoplanetary Disk},'' {\em \apj} {\bf 794}, 160  (2014).

\bibitem{cauley21}
P.~W. {Cauley}, K.~{France}, G.~J. {Herzceg}, {\em et~al.}, ``{A CO-to-H$_{2}$ Ratio of {\ensuremath{\approx}}10$^{-5}$ toward the Herbig Ae Star HK Ori},'' {\em \aj} {\bf 161}, 217  (2021).

\bibitem{duvvuri21}
G.~M. {Duvvuri}, J.~{Sebastian Pineda}, Z.~K. {Berta-Thompson}, {\em et~al.}, ``{Reconstructing the Extreme Ultraviolet Emission of Cool Dwarfs Using Differential Emission Measure Polynomials},'' {\em \apj} {\bf 913}, 40  (2021).

\bibitem{Feinstein2022}
A.~D. {Feinstein}, K.~{France}, A.~{Youngblood}, {\em et~al.}, ``{AU Microscopii in the Far-UV: Observations in Quiescence, during Flares, and Implications for AU Mic b and c},'' {\em \aj} {\bf 164}, 110  (2022).

\bibitem{herczeg02}
G.~J. {Herczeg}, J.~L. {Linsky}, J.~A. {Valenti}, {\em et~al.}, ``{The Far-Ultraviolet Spectrum of TW Hydrae. I. Observations of H$_{2}$ Fluorescence},'' {\em \apj} {\bf 572}, 310--325  (2002).

\bibitem{france11}
K.~{France}, R.~{Schindhelm}, E.~B. {Burgh}, {\em et~al.}, ``{The Far-ultraviolet ``Continuum'' in Protoplanetary Disk Systems. II. Carbon Monoxide Fourth Positive Emission and Absorption},'' {\em \apj} {\bf 734}, 31  (2011).

\bibitem{ingleby11}
L.~{Ingleby}, N.~{Calvet}, E.~{Bergin}, {\em et~al.}, ``{Near-ultraviolet Excess in Slowly Accreting T Tauri Stars: Limits Imposed by Chromospheric Emission},'' {\em \apj} {\bf 743}, 105  (2011).

\bibitem{france12}
K.~{France}, R.~{Schindhelm}, G.~J. {Herczeg}, {\em et~al.}, ``{A Hubble Space Telescope Survey of H$_{2}$ Emission in the Circumstellar Environments of Young Stars},'' {\em \apj} {\bf 756}, 171  (2012).

\bibitem{arulanantham18}
N.~{Arulanantham}, K.~{France}, K.~{Hoadley}, {\em et~al.}, ``{A UV-to-NIR Study of Molecular Gas in the Dust Cavity around RY Lupi},'' {\em \apj} {\bf 855}, 98  (2018).

\bibitem{alcala19}
J.~M. {Alcal{\'a}}, C.~F. {Manara}, K.~{France}, {\em et~al.}, ``{HST spectra reveal accretion in MY Lupi},'' {\em \aap} {\bf 629}, A108  (2019).

\bibitem{roberge00}
A.~{Roberge}, P.~D. {Feldman}, A.~M. {Lagrange}, {\em et~al.}, ``{High-Resolution Hubble Space Telescope STIS Spectra of C I and CO in the{\ensuremath{\beta}} Pictoris Circumstellar Disk},'' {\em \apj} {\bf 538}, 904--910  (2000).

\bibitem{roberge01}
A.~{Roberge}, A.~{Lecavelier des Etangs}, C.~A. {Grady}, {\em et~al.}, ``{FUSE and Hubble Space Telescope/STIS Observations of Hot and Cold Gas in the AB Aurigae System},'' {\em \apjl} {\bf 551}, L97--L100  (2001).

\bibitem{Keeney2017}
B.~A. {Keeney}, S.~A. {Stern}, M.~F. {A'Hearn}, {\em et~al.}, ``{H$_{2}$O and O$_{2}$ absorption in the coma of comet 67P/Churyumov-Gerasimenko measured by the Alice far-ultraviolet spectrograph on Rosetta},'' {\em \mnras} {\bf 469}, S158--S177  (2017).

\bibitem{Hendrix2020}
A.~R. {Hendrix}, T.~M. {Becker}, D.~{Bodewits}, {\em et~al.}, ``{Ultraviolet-Based Science in the Solar System: Advances and Next Steps},'' {\em arXiv e-prints} , arXiv:2007.14993  (2020).

\bibitem{tumlinson17}
J.~{Tumlinson}, M.~S. {Peeples}, and J.~K. {Werk}, ``{The Circumgalactic Medium},'' {\em \araa} {\bf 55}, 389--432  (2017).

\bibitem{Tripp13}
T.~{Tripp}, ``{Quasar Absorption Lines in the Far Ultraviolet: An Untapped Gold Mine for Galaxy Evolution Studies},'' {\em arXiv e-prints} , arXiv:1303.0043  (2013).

\bibitem{peeples19}
M.~S. {Peeples}, L.~{Corlies}, J.~{Tumlinson}, {\em et~al.}, ``{Figuring Out Gas \& Galaxies in Enzo (FOGGIE). I. Resolving Simulated Circumgalactic Absorption at 2 {\ensuremath{\leq}} z {\ensuremath{\leq}} 2.5},'' {\em \apj} {\bf 873}, 129  (2019).

\bibitem{oppenheimer16}
B.~D. {Oppenheimer}, R.~A. {Crain}, J.~{Schaye}, {\em et~al.}, ``{Bimodality of low-redshift circumgalactic O VI in non-equilibrium EAGLE zoom simulations},'' {\em \mnras} {\bf 460}, 2157--2179  (2016).

\bibitem{bertone13}
S.~{Bertone}, A.~{Aguirre}, and J.~{Schaye}, ``{How the diffuse Universe cools},'' {\em \mnras} {\bf 430}, 3292--3313  (2013).

\bibitem{werketal16}
J.~K. {Werk}, J.~X. {Prochaska}, S.~{Cantalupo}, {\em et~al.}, ``{The COS-Halos Survey: Origins of the Highly Ionized Circumgalactic Medium of Star-Forming Galaxies},'' {\em \apj} {\bf 833}, 54  (2016).

\bibitem{nicastroetal18}
F.~{Nicastro}, J.~{Kaastra}, Y.~{Krongold}, {\em et~al.}, ``{Observations of the missing baryons in the warm-hot intergalactic medium},'' {\em \nat} {\bf 558}, 406--409  (2018).

\bibitem{McCandliss2017}
S.~R. {McCandliss} and J.~M. {O'Meara}, ``{Flux Sensitivity Requirements for the Detection of Lyman Continuum Radiation Drop-ins from Star-forming Galaxies below Redshifts of 3},'' {\em \apj} {\bf 845}, 111  (2017).

\bibitem{Inoue2008}
A.~K. {Inoue} and I.~{Iwata}, ``{A Monte Carlo simulation of the intergalactic absorption and the detectability of the Lyman continuum from distant galaxies},'' {\em \mnras} {\bf 387}, 1681--1692  (2008).

\bibitem{Inoue2014}
A.~K. {Inoue}, I.~{Shimizu}, I.~{Iwata}, {\em et~al.}, ``{An updated analytic model for attenuation by the intergalactic medium},'' {\em \mnras} {\bf 442}, 1805--1820  (2014).

\bibitem{Angel1961}
{D.W. Angel, W.R. Hunter, and R. Tousey}, ``{Extreme Ultraviolet Reflectance of LiF-Coated Aluminum Mirrors},'' {\em JOSA} {\bf 51(8)}, 913--914  (1961).

\bibitem{Cox1968}
{J.T. Cox, G. Hass, and J. E. Waylonis}, ``{Further Studies on LiF-Overcoated Aluminum Mirrors with Highest Reflectance in the Vacuum Ultraviolet},'' {\em Appl. Opt.} {\bf 7}, 1535--1540  (1968).

\bibitem{Hutcheson1972}
{E. T. Hutcheson, G. Hass, and J. T. Cox}, ``{Effect of Deposition Rate and Substrate Temperature on the Vacuum Ultraviolet Reflectance of MgF$_2$- and LiF-Overcoated Aluminum Mirrors},'' {\em Appl. Opt.} {\bf 11}, 2245--2248  (1972).

\bibitem{Quijada2014}
M.~A. {Quijada}, J.~{Del Hoyo}, and S.~{Rice}, ``{Enhanced far-ultraviolet reflectance of MgF$_{2}$ and LiF over-coated Al mirrors},'' in {\em Space Telescopes and Instrumentation 2014: Ultraviolet to Gamma Ray},  T.~{Takahashi}, J.-W.~A. {den Herder}, and M.~{Bautz}, Eds., {\em Society of Photo-Optical Instrumentation Engineers (SPIE) Conference Series} {\bf 9144}, 91444G  (2014).

\bibitem{Quijada2012}
M.~A. {Quijada}, S.~{Rice}, and E.~{Mentzell}, ``{Enhanced MgF$_{2}$ and LiF over-coated Al mirrors for FUV space astronomy},'' in {\em Modern Technologies in Space- and Ground-based Telescopes and Instrumentation II},  R.~{Navarro}, C.~R. {Cunningham}, and E.~{Prieto}, Eds., {\em Society of Photo-Optical Instrumentation Engineers (SPIE) Conference Series} {\bf 8450}, 84502H  (2012).

\bibitem{Rodriguez2022}
L.~V. Rodr{\'\i}guez-de Marcos, B.~T. Fleming, J.~Hennessy, {\em et~al.}, ``Advanced al/elif mirrors for the sprite cubesat,'' in {\em Advances in Optical and Mechanical Technologies for Telescopes and Instrumentation V},   {\bf 12188}, 709--720, SPIE  (2022).

\bibitem{Bowen2023}
M.~{Bowen}, B.~{Fleming}, B.~{Indahl}, {\em et~al.}, ``{Preliminary optical performance of the SPRITE CubeSat instrument},'' in {\em UV, X-Ray, and Gamma-Ray Space Instrumentation for Astronomy XXIII},  O.~H. {Siegmund} and K.~{Hoadley}, Eds., {\em Society of Photo-Optical Instrumentation Engineers (SPIE) Conference Series} {\bf 12678}, 126780A  (2023).

\bibitem{Fleming2017}
B.~Fleming, M.~Quijada, J.~Hennessy, {\em et~al.}, ``Advanced environmentally resistant lithium fluoride mirror coatings for the next generation of broadband space observatories,'' {\em Applied Optics} {\bf 56}(36), 9941--9950  (2017).

\bibitem{Hennessy2021}
J.~{Hennessy}, A.~D. {Jewell}, M.~E. {Hoenk}, {\em et~al.}, ``{Advances in detector-integrated filter coatings for the far ultraviolet},'' in {\em UV, X-Ray, and Gamma-Ray Space Instrumentation for Astronomy XXII},  O.~H. {Siegmund}, Ed., {\em Society of Photo-Optical Instrumentation Engineers (SPIE) Conference Series} {\bf 11821}, 118211A  (2021).

\bibitem{Quijada2022}
M.~A. {Quijada}, L.~V. {Rodriguez de Marcos}, J.~G. {Del Hoyo}, {\em et~al.}, ``{Advanced Al mirrors protected with LiF overcoat to realize stable mirror coatings for astronomical telescopes},'' in {\em Society of Photo-Optical Instrumentation Engineers (SPIE) Conference Series},  {\em Society of Photo-Optical Instrumentation Engineers (SPIE) Conference Series} {\bf 12188}, 121881V  (2022).

\bibitem{Nell2020}
N.~{Nell}, K.~{France}, N.~{Kruczek}, {\em et~al.}, ``{The assembly, calibration, and laboratory performance of the SISTINE rocket payload: demonstrating ultraviolet hardware for large UV/optical observatories},'' in {\em American Astronomical Society Meeting Abstracts \#235},  {\em American Astronomical Society Meeting Abstracts} {\bf 235}, 373.18  (2020).

\bibitem{Nell23}
N.~{Nell}, K.~{France}, N.~{Kruczek}, {\em et~al.}, ``{The third flight of SISTINE: far-ultraviolet spectroscopic observations of {\ensuremath{\alpha}} Centauri A and B},'' in {\em American Astronomical Society Meeting Abstracts},  {\em American Astronomical Society Meeting Abstracts} {\bf 55}, 461.11  (2023).

\bibitem{Indahl23}
B.~Indahl, B.~Fleming, D.~Vorobiev, {\em et~al.}, ``{Status and mission operations of the SPRITE 12U CubeSat: a probe of star formation feedback from stellar to galactic scales with far-UV imaging spectroscopy},'' in {\em UV, X-Ray, and Gamma-Ray Space Instrumentation for Astronomy XXIII},  O.~H. Siegmund and K.~Hoadley, Eds.,  {\bf 12678}, 1267806, International Society for Optics and Photonics, SPIE  (2023).

\bibitem{Rodriguez2021}
L.~V. {Rodriguez de Marcos}, D.~R. {Boris}, E.~{Gray}, {\em et~al.}, ``{Room temperature plasma-etching and surface passivation of far-ultraviolet Al mirrors using electron beam generated plasmas},'' {\em Optical Materials Express} {\bf 11}, 740  (2021).

\bibitem{Quijada2018}
M.~A. {Quijada}, D.~R. {Boris}, J.~{del Hoyo}, {\em et~al.}, ``{E-beam generated plasma etching for developing high-reflectance mirrors for far-ultraviolet astronomical instrument applications},'' in {\em Space Telescopes and Instrumentation 2018: Ultraviolet to Gamma Ray},  J.-W.~A. {den Herder}, S.~{Nikzad}, and K.~{Nakazawa}, Eds., {\em Society of Photo-Optical Instrumentation Engineers (SPIE) Conference Series} {\bf 10699}, 106992X  (2018).

\bibitem{Rodriguez2018}
L.~V.~R. {De Marcos}, J.~I. {Larruquert}, J.~A. {M{\'e}ndez}, {\em et~al.}, ``{Optimization of MgF2-deposition temperature for far UV Al mirrors},'' {\em Optics Express} {\bf 26}, 9363  (2018).

\bibitem{Oliveira1999}
C.~M. Oliveira, K.~Retherford, S.~J. Conard, {\em et~al.}, ``Aging studies of lif-coated optics for use in the far ultraviolet,'' in {\em EUV, X-Ray, and Gamma-Ray Instrumentation for Astronomy X},   {\bf 3765}, 52--60, SPIE  (1999).

\bibitem{Stempfhuber2020}
S.~{Stempfhuber}, N.~{Felde}, S.~{Schwinde}, {\em et~al.}, ``{Influence of seed layers on optical properties of aluminum in the UV range},'' {\em Optics Express} {\bf 28}, 20324  (2020).

\bibitem{Larruquert2021}
J.~I. {Larruquert}, C.~{Honrado-Ben{\'\i}tez}, N.~{Guti{\'e}rrez-Luna}, {\em et~al.}, ``{Far UV-enhanced Al mirrors with a Ti seed film},'' {\em Optics Express} {\bf 29}, 7706  (2021).

\bibitem{Quijada2021}
M.~A. {Quijada}, J.~G. {Del Hoyo}, E.~{Gray}, {\em et~al.}, ``{Influence of evaporation rate and chamber pressure on the FUV reflectance and physical characteristics of aluminum films},'' in {\em UV/Optical/IR Space Telescopes and Instruments: Innovative Technologies and Concepts X},  A.~A. {Barto}, J.~B. {Breckinridge}, and H.~P. {Stahl}, Eds., {\em Society of Photo-Optical Instrumentation Engineers (SPIE) Conference Series} {\bf 11819}, 118190G  (2021).

\bibitem{wilbrandt2014protected}
S.~Wilbrandt, O.~Stenzel, H.~Nakamura, {\em et~al.}, ``Protected and enhanced aluminum mirrors for the vuv,'' {\em Applied Optics} {\bf 53}(4), A125--A130  (2014).

\bibitem{balasubramanian2015}
K.~Balasubramanian, J.~Hennessy, N.~Raouf, {\em et~al.}, ``Coatings for uvoir telescope mirrors,'' in {\em UV/Optical/IR Space Telescopes and Instruments: Innovative Technologies and Concepts VII},   {\bf 9602}, 195--203, SPIE  (2015).

\bibitem{aguirre2021assembly}
F.~C. Aguirre, N.~Nell, N.~Kruczek, {\em et~al.}, ``The assembly, calibration, and predicted performance of the sistine-2 sounding rocket payload,'' in {\em UV, X-Ray, and Gamma-Ray Space Instrumentation for Astronomy XXII},   {\bf 11821}, 178--187, SPIE  (2021).

\bibitem{Hennessy2018}
{J. Hennessy and S. Nikzad}, ``{Atomic Layer Deposition of Lithium Fluoride Optical Coatings for the Ultraviolet},'' {\em Inorganics} {\bf 6(2)}, 46  (2018).

\bibitem{putkonen2009ald}
M.~I. Putkonen, ``Ald applications beyond outside ic technology-existing and emerging possibilities,'' {\em ECS transactions} {\bf 25}(4), 143  (2009).

\bibitem{fryauf2018scaling}
D.~M. Fryauf, A.~C. Phillips, M.~J. Bolte, {\em et~al.}, ``Scaling atomic layer deposition to astronomical optic sizes: low-temperature aluminum oxide in a meter-sized chamber,'' {\em ACS applied materials \& interfaces} {\bf 10}(48), 41678--41689  (2018).

\bibitem{Ohl2000}
R.~G. {Ohl}, R.~H. {Barkhouser}, S.~J. {Conard}, {\em et~al.}, ``{Performance of the Far Ultraviolet Spectroscopic Explorer mirror assemblies},'' in {\em Instrumentation for UV/EUV Astronomy and Solar Missions},  S.~{Fineschi}, C.~M. {Korendyke}, O.~H. {Siegmund}, {\em et~al.}, Eds., {\em Society of Photo-Optical Instrumentation Engineers (SPIE) Conference Series} {\bf 4139}, 137--148  (2000).

\bibitem{Moos2000}
H.~W. {Moos}, W.~C. {Cash}, L.~L. {Cowie}, {\em et~al.}, ``{Overview of the Far Ultraviolet Spectroscopic Explorer Mission},'' {\em \apjl} {\bf 538}, L1--L6  (2000).

\bibitem{France2016}
K.~{France}, K.~{Hoadley}, B.~T. {Fleming}, {\em et~al.}, ``{The SLICE, CHESS, and SISTINE Ultraviolet Spectrographs: Rocket-Borne Instrumentation Supporting Future Astrophysics Missions},'' {\em Journal of Astronomical Instrumentation} {\bf 5}, 1640001  (2016).

\bibitem{DelHoyo2023}
J.~{Del Hoyo}, L.~{Rodriguez-de-Marcos}, J.~{Hennessy}, {\em et~al.}, ``{Cryolite overcoated aluminum reflectors for far-ultraviolet spectroscopy},'' in {\em UV/Optical/IR Space Telescopes and Instruments: Innovative Technologies and Concepts XI},  A.~A. {Barto}, F.~{Keller}, and H.~P. {Stahl}, Eds., {\em Society of Photo-Optical Instrumentation Engineers (SPIE) Conference Series} {\bf 12676}, 126760I  (2023).

\bibitem{de2021room}
L.~V.~R. de~Marcos, D.~R. Boris, E.~Gray, {\em et~al.}, ``Room temperature plasma-etching and surface passivation of far-ultraviolet al mirrors using electron beam generated plasmas,'' {\em Optical Materials Express} {\bf 11}(3), 740--756  (2021).

\bibitem{LUVOIR}
{The LUVOIR Team}, ``{The LUVOIR Mission Concept Study Final Report},'' {\em arXiv e-prints} , arXiv:1912.06219  (2019).

\bibitem{HabEx}
B.~S. {Gaudi}, S.~{Seager}, B.~{Mennesson}, {\em et~al.}, ``{The Habitable Exoplanet Observatory (HabEx) Mission Concept Study Final Report},'' {\em arXiv e-prints} , arXiv:2001.06683  (2020).

\bibitem{wizaMicrochannelPlateDetectors1979}
J.~L. Wiza, ``Microchannel plate detectors,'' {\em Nuclear Instruments and Methods} {\bf 162}, 587--601  (1979).

\bibitem{siegmundMicrochannelPlateEUV1984}
O.~H.~W. Siegmund, R.~F. Malina, K.~Coburn, {\em et~al.}, ``Microchannel {{Plate EUV Detectors}} for the {{Extreme Ultraviolet Explorer}},'' {\em IEEE Trans. Nucl. Sci.} {\bf 31}(1), 776--779  (1984).

\bibitem{siegmundPerformanceDoubleDelay1997}
O.~H.~W. Siegmund, M.~A. Gummin, J.~M. Stock, {\em et~al.}, ``Performance of the double delay line microchannel plate detectors for the {{Far-Ultraviolet Spectroscopic Explorer}},'' in {\em Optical {{Science}}, {{Engineering}} and {{Instrumentation}} '97},  O.~H.~W. Siegmund and M.~A. Gummin, Eds., 283--294, ({San Diego, CA})  (1997).

\bibitem{jelinskyPerformanceResultsGALEX2003}
P.~N. Jelinsky, P.~F. Morrissey, J.~M. Malloy, {\em et~al.}, ``Performance results of the {{GALEX}} cross delay line detectors,'' in {\em Proc. {{SPIE}}},  J.~C. Blades and O.~H.~W. Siegmund, Eds.,  {\bf 4854}, 233, ({Waikoloa, Hawai'i, United States})  (2003).

\bibitem{vallergaHSTCOSFarultravioletDetector2001}
J.~V. Vallerga, J.~B. McPhate, A.~P. Martin, {\em et~al.}, ``{{HST-COS}} far-ultraviolet detector: Final ground calibration,'' in {\em Proc. {{SPIE}}},  O.~H.~W. Siegmund, S.~Fineschi, and M.~A. Gummin, Eds., 141--151, ({San Diego, CA})  (2001).

\bibitem{korpelaInFlightPerformanceICON2023}
E.~J. Korpela, M.~M. Sirk, J.~Edelstein, {\em et~al.}, ``In-{{Flight Performance}} of the {{ICON EUV Spectrograph}},'' {\em Space Sci Rev} {\bf 219}, 24  (2023).

\bibitem{mcclintockGlobalScaleObservations2020b}
W.~E. McClintock, R.~W. Eastes, A.~C. Hoskins, {\em et~al.}, ``Global-{{Scale Observations}} of the {{Limb}} and {{Disk Mission Implementation}}: 1. {{Instrument Design}} and {{Early Flight Performance}},'' {\em JGR Space Physics} {\bf 125}  (2020).

\bibitem{darlingMicrochannelPlateLife2017a}
N.~Darling, T.~Curtis, J.~Tedesco, {\em et~al.}, ``Microchannel plate life testing for {{UV}} spectroscopy instruments,'' in {\em {{UV}}, {{X-Ray}}, and {{Gamma-Ray Space Instrumentation}} for {{Astronomy XX}}},  O.~H. Siegmund, Ed., 38, {SPIE}, ({San Diego, United States})  (2017).

\bibitem{mcphateLifeTestingConventional2019}
J.~B. McPhate, N.~T. Darling, T.~L. Curtis, {\em et~al.}, ``Life testing of conventional and atomic layer deposition functionalized microchannel plates,'' in {\em Proc. {{SPIE}}},  O.~H. Siegmund, Ed.,  {\bf 11118}, 111180P, {SPIE}, ({San Diego, United States})  (2019).

\bibitem{martinFurtherScrubbingQuantum2003}
A.~P. Martin, J.~V. Vallerga, J.~B. McPhate, {\em et~al.}, ``Further scrubbing and quantum efficiency results of the {{HST-COS}} far-ultraviolet detector,'' in {\em Proc. {{SPIE}}},  J.~C. Blades and O.~H.~W. Siegmund, Eds.,  {\bf 4854}, 526, ({Waikoloa, Hawai'i, United States})  (2003).

\bibitem{schindhelmMicrochannelPlateDetector2017}
E.~R. Schindhelm, J.~C. Green, O.~H.~W. Siegmund, {\em et~al.}, ``Microchannel plate detector technology potential for {{LUVOIR}} and {{HabEx}},'' in {\em Proc. {{SPIE}}},  O.~H. Siegmund, Ed.,  {\bf 10397}, 1039711, {SPIE}, ({San Diego, United States})  (2017).

\bibitem{jokelaCharacterizationSecondaryElectron2011}
S.~J. Jokela, I.~V. Veryovkin, A.~V. Zinovev, {\em et~al.}, ``The characterization of secondary electron emitters for use in large area photo-detectors,'' in {\em {{AIP Conference Proceedings}}},   {\bf 1336}, 208--212, ({Fort Worth, Texas, (USA)})  (2011).

\bibitem{davisJUICEUltravioletSpectrograph2021}
M.~W. Davis, G.~R. Gladstone, K.~D. Retherford, {\em et~al.}, ``The {{JUICE}} ultraviolet spectrograph: A next-generation compact {{UVS}} for {{ESA}}'s {{JUICE}} mission,'' in {\em International {{Conference}} on {{Space Optics}} \textemdash{} {{ICSO}} 2020},  Z.~Sodnik, B.~Cugny, and N.~Karafolas, Eds., 77, {SPIE}, ({Online Only, France})  (2021).

\bibitem{siegmundPerformanceCharacteristicsAtomic2013}
O.~H.~W. Siegmund, N.~Richner, G.~Gunjala, {\em et~al.}, ``Performance characteristics of atomic layer functionalized microchannel plates,'' in {\em {{SPIE Optical Engineering}} + {{Applications}}},  O.~H. Siegmund, Ed., 88590Y, ({San Diego, California, United States})  (2013).

\bibitem{ertleyMicrochannelPlateDetectors2018}
C.~Ertley, O.~Siegmund, J.~Vallerga, {\em et~al.}, ``Microchannel plate detectors for future {{NASA UV}} observatories,'' in {\em Proc. {{SPIE}}},  J.-W.~A. {den Herder}, K.~Nakazawa, and S.~Nikzad, Eds.,  {\bf 10699}, 106993H, {SPIE}, ({Austin, United States})  (2018).

\bibitem{siegmundDevelopmentUVImaging2020}
O.~H.~W. Siegmund, J.~McPhate, T.~Curtis, {\em et~al.}, ``Development of {{UV}} imaging detectors with atomic layer deposited microchannel plates and cross strip readouts,'' in {\em Proc. {{SPIE}}},  A.~D. Holland and J.~Beletic, Eds.,  {\bf 11454}, 114541H, {SPIE}, ({Online Only, United States})  (2020).

\bibitem{siegmundNovelLargeFormat2011}
O.~Siegmund, J.~McPhate, J.~Vallerga, {\em et~al.}, ``Novel large format sealed tube microchannel plate detectors for {{Cherenkov}} timing and imaging,'' {\em Nucl Instrum Methods Phys Res A} {\bf 639}, 165--168  (2011).

\bibitem{jelinskyProgressSoftXray1996}
S.~R. Jelinsky, O.~H.~W. Siegmund, and J.~A. Mir, ``Progress in soft x-ray and {{UV}} photocathodes,'' in {\em Proc. {{SPIE}}},  O.~H.~W. Siegmund and M.~A. Gummin, Eds.,  {\bf 2808}, 617--625, ({Denver, CO})  (1996).

\bibitem{tremsinUVRadiationResistance2001}
A.~Tremsin and O.~Siegmund, ``{{UV}} radiation resistance and solar blindness of {{CsI}} and {{KBr}} photocathodes,'' {\em IEEE Trans. Nucl. Sci.} {\bf 48}, 421--425  (2001).

\bibitem{larruquertOpticalPropertiesQuantum2002}
J.~I. Larruquert, J.~A. M{\'e}ndez, J.~A. Azn{\'a}rez, {\em et~al.}, ``Optical properties and quantum efficiency of thin-film alkali halides in the far ultraviolet,'' {\em Appl. Opt.} {\bf 41}, 2532  (2002).

\bibitem{tremsinQuantumEfficiencyStability2005}
A.~S. Tremsin and O.~H.~W. Siegmund, ``The quantum efficiency and stability of {{UV}} and soft x-ray photocathodes,'' in {\em Proc. {{SPIE}}},   {\bf 5920}, 59200I, ({San Diego, California, USA})  (2005).

\bibitem{ertleyDevelopmentOpaquePhotocathodes2017}
C.~Ertley, O.~Siegmund, A.~Tremsin, {\em et~al.}, ``Development of {{Opaque Photocathodes Deposited}} onto {{Microchannel Plates}},'' in {\em {{IEEE Nuclear Science Symposium}} and {{Medical Imaging Conference Proceedings}} ({{NSS}}/{{MIC}})},  1--5, {IEEE}, ({Atlanta, GA})  (2017).

\bibitem{seljakASICsReadoutSystem2018}
A.~Seljak, G.~Varner, J.~Vallerga, {\em et~al.}, ``{{ASICs}} and readout system for a multi {{Mpixel}} single photon {{UV}} imaging detector capable of space applications,'' in {\em Proc. {{TWEPP}}},  025, {Sissa Medialab}, ({Santa Cruz, California})  (2018).

\bibitem{Nikzad2012}
S.~{Nikzad}, M.~E. {Hoenk}, F.~{Greer}, {\em et~al.}, ``{Delta-doped electron-multiplied CCD with absolute quantum efficiency over 50\% in the near to far ultraviolet range for single photon counting applications},'' {\em \ao} {\bf 51}, 365  (2012).

\bibitem{Janesick2001}
J.~Janesick, {\em {Scientific Charge-Coupled Devices}}  (2001).

\bibitem{NikzadChapter2015}
S.~Nikzad, A.~D. Jewell, A.~Carver, {\em et~al.}, ``Digital imaging for planetary exploration,'' in {\em Handbook of Digital Imaging},  {\em Remote Imaging}, John Wiley \& Sons, Ltd  (2015).

\bibitem{Hoenk2022}
M.~E. {Hoenk}, A.~D. {Jewell}, G.~{Kyne}, {\em et~al.}, ``{2D-doped silicon detectors for UV/optical/NIR and x-ray astronomy},'' in {\em X-Ray, Optical, and Infrared Detectors for Astronomy X},  A.~D. {Holland} and J.~{Beletic}, Eds., {\em Society of Photo-Optical Instrumentation Engineers (SPIE) Conference Series} {\bf 12191}, 1219113  (2022).

\bibitem{2001Mackay}
C.~D. {Mackay}, R.~N. {Tubbs}, R.~{Bell}, {\em et~al.}, ``{Subelectron read noise at MHz pixel rates},'' in {\em Sensors and Camera Systems for Scientific, Industrial, and Digital Photography Applications II},  M.~M. {Blouke}, J.~{Canosa}, and N.~{Sampat}, Eds., {\em Society of Photo-Optical Instrumentation Engineers (SPIE) Conference Series} {\bf 4306}, 289--298  (2001).

\bibitem{2001Jerram}
P.~{Jerram}, P.~J. {Pool}, R.~{Bell}, {\em et~al.}, ``{The LLCCD: low-light imaging without the need for an intensifier},'' in {\em Sensors and Camera Systems for Scientific, Industrial, and Digital Photography Applications II},  M.~M. {Blouke}, J.~{Canosa}, and N.~{Sampat}, Eds., {\em Society of Photo-Optical Instrumentation Engineers (SPIE) Conference Series} {\bf 4306}, 178--186  (2001).

\bibitem{2004Gach}
J.~L. {Gach}, C.~{Guillaume}, O.~{Boissin}, {\em et~al.}, ``{First Results of aN L3CCD in Photon Counting Mode},'' in {\em Scientific Detectors for Astronomy, The Beginning of a New Era},  P.~{Amico}, J.~W. {Beletic}, and J.~E. {Belectic}, Eds., {\em Astrophysics and Space Science Library} {\bf 300}, 611--614  (2004).

\bibitem{2011Tulloch}
S.~M. {Tulloch} and V.~S. {Dhillon}, ``{On the use of electron-multiplying CCDs for astronomical spectroscopy},'' {\em \mnras} {\bf 411}, 211--225  (2011).

\bibitem{2008Daigle}
O.~{Daigle}, J.-L. {Gach}, C.~{Guillaume}, {\em et~al.}, ``{CCCP: a CCD controller for counting photons},'' in {\em Ground-based and Airborne Instrumentation for Astronomy II},  I.~S. {McLean} and M.~M. {Casali}, Eds., {\em Society of Photo-Optical Instrumentation Engineers (SPIE) Conference Series} {\bf 7014}, 70146L  (2008).

\bibitem{Balachandran2020}
K.~{Balachandran}, E.~{Nathan}, M.~{Rovira-Navarro}, {\em et~al.}, ``{A Mission Concept Investigating the Habitability of Enceladus: S.I.L.E.N.U.S.},'' in {\em 51st Annual Lunar and Planetary Science Conference},  {\em Lunar and Planetary Science Conference}, 1339  (2020).

\bibitem{Marrufo2022}
E.~{Marrufo Villalpando}, A.~{Drlica-Wagner}, M.~{Bonati}, {\em et~al.}, ``{Design of a Skipper CCD Focal Plane for the SOAR Integral Field Spectrograph},'' {\em arXiv e-prints} , arXiv:2210.03665  (2022).

\bibitem{2017Tiffenberg}
J.~{Tiffenberg}, M.~{Sofo-Haro}, A.~{Drlica-Wagner}, {\em et~al.}, ``{Single-Electron and Single-Photon Sensitivity with a Silicon Skipper CCD},'' {\em \prl} {\bf 119}, 131802  (2017).

\bibitem{Content+1996}
D.~A. {Content}, R.~A. {Boucarut}, C.~{Bowers}, {\em et~al.}, ``{Development and testing of diffraction gratings for the Space Telescope Imaging Spectrograph},'' in {\em Space Telescopes and Instruments IV},  P.~Y. {Bely} and J.~B. {Breckinridge}, Eds., {\em Society of Photo-Optical Instrumentation Engineers (SPIE) Conference Series} {\bf 2807}, 267--278  (1996).

\bibitem{Hoadley+2016}
K.~{Hoadley}, K.~{France}, N.~{Kruczek}, {\em et~al.}, ``{The re-flight of the Colorado high-resolution Echelle stellar spectrograph (CHESS): improvements, calibrations, and post-flight results},'' in {\em Space Telescopes and Instrumentation 2016: Ultraviolet to Gamma Ray},  J.-W.~A. {den Herder}, T.~{Takahashi}, and M.~{Bautz}, Eds., {\em Society of Photo-Optical Instrumentation Engineers (SPIE) Conference Series} {\bf 9905}, 99052V  (2016).

\bibitem{Landsman+1997}
W.~{Landsman} and C.~{Bowers}, ``{Scattered Light in the STIS Echelle Modes},'' in {\em The 1997 HST Calibration Workshop with a New Generation of Instruments},  S.~{Casertano}, R.~{Jedrzejewski}, T.~{Keyes}, {\em et~al.}, Eds., 132  (1997).

\bibitem{Kruczek+2018}
N.~{Kruczek}, N.~{Nell}, K.~{France}, {\em et~al.}, ``{The fourth flight of CHESS: spectral resolution enhancements for high-resolution FUV spectroscopy},'' in {\em Space Telescopes and Instrumentation 2018: Ultraviolet to Gamma Ray},  J.-W.~A. {den Herder}, S.~{Nikzad}, and K.~{Nakazawa}, Eds., {\em Society of Photo-Optical Instrumentation Engineers (SPIE) Conference Series} {\bf 10699}, 106990K  (2018).

\bibitem{Kruczek22}
N.~{Kruczek}, D.~M. {Miles}, B.~{Fleming}, {\em et~al.}, ``{High efficiency echelle gratings for the far ultraviolet},'' {\em \ao} {\bf 61}, 6430  (2022).

\bibitem{Grise21}
F.~{Gris{\'e}}, N.~{Kruczek}, B.~{Fleming}, {\em et~al.}, ``{Fabrication of custom astronomical gratings for the extreme and far ultraviolet bandpasses},'' in {\em UV, X-Ray, and Gamma-Ray Space Instrumentation for Astronomy XXII},  O.~H. {Siegmund}, Ed., {\em Society of Photo-Optical Instrumentation Engineers (SPIE) Conference Series} {\bf 11821}, 1182112  (2021).

\bibitem{Vieu_EBLResolution_2000}
C.~{Vieu}, F.~{Carcenac}, A.~{P{\'e}pin}, {\em et~al.}, ``{Electron beam lithography: resolution limits and applications},'' {\em Applied Surface Science} {\bf 164}, 111--117  (2000).

\bibitem{McEntaffer13}
R.~{McEntaffer}, C.~{DeRoo}, T.~{Schultz}, {\em et~al.}, ``{First results from a next-generation off-plane X-ray diffraction grating},'' {\em Experimental Astronomy} {\bf 36}, 389--405  (2013).

\bibitem{Miles18}
D.~M. {Miles}, J.~A. {McCoy}, R.~L. {McEntaffer}, {\em et~al.}, ``{Fabrication and Diffraction Efficiency of a Large-format, Replicated X-Ray Reflection Grating},'' {\em \apj} {\bf 869}, 95  (2018).

\bibitem{McEntaffer19}
R.~L. {McEntaffer}, ``{Reflection grating concept for the Lynx X-Ray Grating Spectrograph},'' {\em Journal of Astronomical Telescopes, Instruments, and Systems} {\bf 5}, 021002  (2019).

\bibitem{DeRoo20}
C.~T. {DeRoo}, J.~{Termini}, F.~{Gris{\'e}}, {\em et~al.}, ``{Limiting Spectral Resolution of a Reflection Grating Made via Electron-beam Lithography},'' {\em \apj} {\bf 904}, 142  (2020).

\bibitem{Kruczek21}
N.~{Kruczek}, F.~{Gris{\'e}}, D.~M. {Miles}, {\em et~al.}, ``{Performance of anisotropically-etched gratings in the extreme and far ultraviolet bandpasses},'' in {\em UV, X-Ray, and Gamma-Ray Space Instrumentation for Astronomy XXII},  O.~H. {Siegmund}, Ed., {\em Society of Photo-Optical Instrumentation Engineers (SPIE) Conference Series} {\bf 11821}, 118210X  (2021).

\bibitem{France19}
K.~{France}, B.~T. {Fleming}, J.~J. {Drake}, {\em et~al.}, ``{The extreme-ultraviolet stellar characterization for atmospheric physics and evolution (ESCAPE) mission concept},'' in {\em UV, X-Ray, and Gamma-Ray Space Instrumentation for Astronomy XXI},  O.~H. {Siegmund}, Ed., {\em Society of Photo-Optical Instrumentation Engineers (SPIE) Conference Series} {\bf 11118}, 1111808  (2019).

\bibitem{Miles19}
D.~M. {Miles}, S.~V. {Hull}, T.~B. {Schultz}, {\em et~al.}, ``{Water Recovery X-Ray Rocket grating spectrometer},'' {\em Journal of Astronomical Telescopes, Instruments, and Systems} {\bf 5}, 044006  (2019).

\bibitem{Grise23}
F.~Gris{\'e}, C.~DeRoo, J.~McCoy, {\em et~al.}, ``{Fabrication of gratings on curved substrates using electron-beam lithography},'' in {\em Optics for EUV, X-Ray, and Gamma-Ray Astronomy XI},  S.~L. O'Dell, J.~A. Gaskin, G.~Pareschi, {\em et~al.}, Eds.,  {\bf 12679}, 126790K, International Society for Optics and Photonics, SPIE  (2023).

\bibitem{McCoy20}
J.~A. {McCoy}, R.~L. {McEntaffer}, and D.~M. {Miles}, ``{Extreme Ultraviolet and Soft X-Ray Diffraction Efficiency of a Blazed Reflection Grating Fabricated by Thermally Activated Selective Topography Equilibration},'' {\em \apj} {\bf 891}, 114  (2020).

\bibitem{Miles22}
D.~M. {Miles}, R.~L. {McEntaffer}, and F.~{Gris{\'e}}, ``{Blazed reflection gratings with electron-beam lithography and ion-beam etching},'' in {\em Society of Photo-Optical Instrumentation Engineers (SPIE) Conference Series},  J.-W.~A. {den Herder}, S.~{Nikzad}, and K.~{Nakazawa}, Eds., {\em Society of Photo-Optical Instrumentation Engineers (SPIE) Conference Series} {\bf 12181}, 1218153  (2022).

\bibitem{Kowalski02}
M.~P. {Kowalski}, R.~G. {Cruddace}, J.~C. {Rife}, {\em et~al.}, ``{Record High EUV Efficiency from Multilayer-Coated Liquid-Overcoated Blazed Ion-Etched Gratings},'' in {\em American Astronomical Society Meeting Abstracts},  {\em American Astronomical Society Meeting Abstracts} {\bf 201}, 82.10  (2002).

\bibitem{Tuttle2014b}
S.~E. {Tuttle}, G.~J. {Hill}, H.~{Lee}, {\em et~al.}, ``{The construction, alignment, and installation of the VIRUS spectrograph},'' in {\em Ground-based and Airborne Instrumentation for Astronomy V},  S.~K. {Ramsay}, I.~S. {McLean}, and H.~{Takami}, Eds., {\em Society of Photo-Optical Instrumentation Engineers (SPIE) Conference Series} {\bf 9147}, 91470R  (2014).

\bibitem{Hill2021}
G.~J. {Hill}, H.~{Lee}, P.~J. {MacQueen}, {\em et~al.}, ``{The HETDEX Instrumentation: Hobby-Eberly Telescope Wide-field Upgrade and VIRUS},'' {\em \aj} {\bf 162}, 298  (2021).

\bibitem{Bacon2010}
R.~{Bacon}, M.~{Accardo}, L.~{Adjali}, {\em et~al.}, ``{The MUSE second-generation VLT instrument},'' in {\em Ground-based and Airborne Instrumentation for Astronomy III},  I.~S. {McLean}, S.~K. {Ramsay}, and H.~{Takami}, Eds., {\em Society of Photo-Optical Instrumentation Engineers (SPIE) Conference Series} {\bf 7735}, 773508  (2010).

\bibitem{MatMat2010}
M.~{Matuszewski}, J.~{Evrard}, F.~{Mirc}, {\em et~al.}, ``{FIREBALL: instrument pointing and aspect reconstruction},'' in {\em Space Telescopes and Instrumentation 2010: Ultraviolet to Gamma Ray},  M.~{Arnaud}, S.~S. {Murray}, and T.~{Takahashi}, Eds., {\em Society of Photo-Optical Instrumentation Engineers (SPIE) Conference Series} {\bf 7732}, 773229  (2010).

\bibitem{Milliard2010}
B.~{Milliard}, D.~C. {Martin}, D.~{Schiminovich}, {\em et~al.}, ``{FIREBALL: the Faint Intergalactic medium Redshifted Emission Balloon: overview and first science flight results},'' in {\em Space Telescopes and Instrumentation 2010: Ultraviolet to Gamma Ray},  M.~{Arnaud}, S.~S. {Murray}, and T.~{Takahashi}, Eds., {\em Society of Photo-Optical Instrumentation Engineers (SPIE) Conference Series} {\bf 7732}, 773205  (2010).

\bibitem{Tuttle2010}
S.~E. {Tuttle}, D.~{Schiminovich}, R.~{Grange}, {\em et~al.}, ``{FIREBALL: the first ultraviolet fiber fed spectrograph},'' in {\em Space Telescopes and Instrumentation 2010: Ultraviolet to Gamma Ray},  M.~{Arnaud}, S.~S. {Murray}, and T.~{Takahashi}, Eds., {\em Society of Photo-Optical Instrumentation Engineers (SPIE) Conference Series} {\bf 7732}, 773227  (2010).

\bibitem{Dewitt23}
D.~{Dewitt}, D.~{Vorobiev}, T.~{Livingston}, {\em et~al.}, ``{Transmission and bend loss in far ultraviolet hollow-core fibers for compact fiber-fed, multi-object spectrographs and reflectometers},'' in {\em Society of Photo-Optical Instrumentation Engineers (SPIE) Conference Series},  O.~H. {Siegmund} and K.~{Hoadley}, Eds., {\em Society of Photo-Optical Instrumentation Engineers (SPIE) Conference Series} {\bf 12678}, 126780U  (2023).

\bibitem{WittSPIE2023}
E.~M. {Witt}, B.~T. {Fleming}, J.~C. {Green}, {\em et~al.}, ``{INFUSE: preflight performance of a rocket-borne FUV integral field spectrograph},'' in {\em Society of Photo-Optical Instrumentation Engineers (SPIE) Conference Series},  O.~H. {Siegmund} and K.~{Hoadley}, Eds., {\em Society of Photo-Optical Instrumentation Engineers (SPIE) Conference Series} {\bf 12678}, 1267808  (2023).

\bibitem{Smee:2013}
S.~A. {Smee}, J.~E. {Gunn}, A.~{Uomoto}, {\em et~al.}, ``{The Multi-object, Fiber-fed Spectrographs for the Sloan Digital Sky Survey and the Baryon Oscillation Spectroscopic Survey},'' {\em \aj} {\bf 146}, 32  (2013).

\bibitem{Tamura:2016}
N.~{Tamura}, N.~{Takato}, A.~{Shimono}, {\em et~al.}, ``{Prime Focus Spectrograph (PFS) for the Subaru telescope: overview, recent progress, and future perspectives},'' in {\em Ground-based and Airborne Instrumentation for Astronomy VI},  C.~J. {Evans}, L.~{Simard}, and H.~{Takami}, Eds., {\em Society of Photo-Optical Instrumentation Engineers (SPIE) Conference Series} {\bf 9908}, 99081M  (2016).

\bibitem{Boker:2023}
T.~{B{\"o}ker}, T.~L. {Beck}, S.~M. {Birkmann}, {\em et~al.}, ``{In-orbit Performance of the Near-infrared Spectrograph NIRSpec on the James Webb Space Telescope},'' {\em \pasp} {\bf 135}, 038001  (2023).

\bibitem{Jakobsen:2022}
P.~{Jakobsen}, P.~{Ferruit}, C.~{Alves de Oliveira}, {\em et~al.}, ``{The Near-Infrared Spectrograph (NIRSpec) on the James Webb Space Telescope. I. Overview of the instrument and its capabilities},'' {\em \aap} {\bf 661}, A80  (2022).

\bibitem{Kutyrev:2020}
A.~S. Kutyrev, M.~Greenhouse, M.~J. Li, {\em et~al.}, ``{Programmable microshutter selection masks in application to UV spectroscopy},'' in {\em Space Telescopes and Instrumentation 2020: Optical, Infrared, and Millimeter Wave},  M.~Lystrup, M.~D. Perrin, N.~Batalha, {\em et~al.}, Eds.,  {\bf 11443}, 114431D, International Society for Optics and Photonics, SPIE  (2020).

\bibitem{Ke:2022}
M.~Ke, A.~S. Kutyrev, C.~A. Kotecki, {\em et~al.}, ``Modelling analysis for design of microshutters with new electrode configuration,'' {\em Sensors and Actuators A: Physical} {\bf 343}, 113661  (2022).

\bibitem{Kutyrev:2004}
A.~S. {Kutyrev}, R.~{Arendt}, S.~H. {Moseley}, {\em et~al.}, ``{Programmable Microshutter Arrays for the JWST NIRSpec: Optical Performance},'' {\em IEEE Journal of Selected Topics in Quantum Electronics} {\bf 10}, 652--661  (2004).

\bibitem{Kutyrev:2008}
A.~S. Kutyrev, N.~Collins, J.~Chambers, {\em et~al.}, ``{Microshutter arrays: high contrast programmable field masks for JWST NIRSpec},'' in {\em Space Telescopes and Instrumentation 2008: Optical, Infrared, and Millimeter},  J.~M.~O. Jr., M.~W.~M. de~Graauw, and H.~A. MacEwen, Eds.,  {\bf 7010}, 70103D, International Society for Optics and Photonics, SPIE  (2008).

\bibitem{Mitchell:2023}
O.~H. Mitchell, S.~R. McCandliss, R.~Pelton, {\em et~al.}, ``{An enhanced contrast evaluation testbed for next-generation microshutter arrays},'' in {\em UV, X-Ray, and Gamma-Ray Space Instrumentation for Astronomy XXIII},  O.~H. Siegmund and K.~Hoadley, Eds.,  {\bf 12678}, 126780V, International Society for Optics and Photonics, SPIE  (2023).

\bibitem{Kutyrev:2023}
A.~Kutyrev, M.~Greenhouse, M.-P. Chang, {\em et~al.}, ``{Scalable microshutter focal plane masks for UV, visible, and infrared spectroscopy},'' in {\em UV, X-Ray, and Gamma-Ray Space Instrumentation for Astronomy XXIII},  O.~H. Siegmund and K.~Hoadley, Eds.,  {\bf 12678}, 126780Q, International Society for Optics and Photonics, SPIE  (2023).

\bibitem{Meyer2004}
R.~D. Meyer, K.~J. Kearney, Z.~Ninkov, {\em et~al.}, ``Ritmos- a micromirror-based multi-object spectrometer,'' {\em Proc. SPIE} {\bf 5492}, 200--219  (2004).

\bibitem{MacKenty2006}
J.~W. {MacKenty}, I.~{Ohl}, Raymond~G., M.~A. {Greenhouse}, {\em et~al.}, ``{Commissioning of the IRMOS MEMS spectrometer},'' in {\em Ground-based and Airborne Instrumentation for Astronomy},  I.~S. {McLean} and M.~{Iye}, Eds., {\em Society of Photo-Optical Instrumentation Engineers (SPIE) Conference Series} {\bf 6269}, 626915  (2006).

\bibitem{Zamkotsian2014}
F.~Zamkotsian, P.~Spano, P.~Lanzoni, {\em et~al.}, ``Batman: a dmd-based multi-object spectrograph on galileo telescope,'' {\em Proc. SPIE} {\bf 9147}, 914713--914713--12  (2014).

\bibitem{Robberto2016}
M.~{Robberto}, M.~{Donahue}, Z.~{Ninkov}, {\em et~al.}, ``{SAMOS: a versatile multi-object-spectrograph for the GLAO system SAM at SOAR},'' in {\em Ground based and Airborne Instrumentation for Astronomy VI},  {\em Proc. SPIE} {\bf 9908}, 99088V  (2016).

\bibitem{Vorobiev2016}
D.~{Vorobiev}, A.~{Travinsky}, A.~D. {Raisanen}, {\em et~al.}, ``{Shock and vibration testing of digital micromirror devices (DMDs) for space-based applications},'' in {\em Advances in Optical and Mechanical Technologies for Telescopes and Instrumentation II},  R.~{Navarro} and J.~H. {Burge}, Eds., {\em Society of Photo-Optical Instrumentation Engineers (SPIE) Conference Series} {\bf 9912}, 99125M  (2016).

\bibitem{Travinsky2017}
A.~{Travinsky}, D.~{Vorobiev}, Z.~{Ninkov}, {\em et~al.}, ``{Evaluation of digital micromirror devices for use in space-based multiobject spectrometer application},'' {\em Journal of Astronomical Telescopes, Instruments, and Systems} {\bf 3}, 035003  (2017).

\bibitem{Fourspring2013}
K.~{Fourspring}, Z.~{Ninkov}, B.~C. {Fodness}, {\em et~al.}, ``{Proton radiation testing of digital micromirror devices for space applications},'' {\em Optical Engineering} {\bf 52}, 091807  (2013).

\bibitem{Travinsky2016a}
A.~{Travinsky}, D.~{Vorobiev}, A.~D. {Raisanen}, {\em et~al.}, ``{The effects of heavy ion radiation on digital micromirror device performance},'' in {\em Emerging Digital Micromirror Device Based Systems and Applications VIII},  M.~R. {Douglass}, P.~S. {King}, and B.~L. {Lee}, Eds., {\em Society of Photo-Optical Instrumentation Engineers (SPIE) Conference Series} {\bf 9761}, 976108  (2016).

\bibitem{Travinsky2016b}
A.~{Travinsky}, D.~{Vorobiev}, Z.~{Ninkov}, {\em et~al.}, ``{Effects of heavy ion radiation on digital micromirror device performance},'' {\em Optical Engineering} {\bf 55}, 094107  (2016).

\bibitem{Oram2019}
K.~{Oram}, D.~{Vorobiev}, Z.~{Ninkov}, {\em et~al.}, ``{The effects of gamma radiation on digital micromirror devices},'' in {\em Emerging Digital Micromirror Device Based Systems and Applications XI},  M.~R. {Douglass}, J.~{Ehmke}, and B.~L. {Lee}, Eds., {\em Society of Photo-Optical Instrumentation Engineers (SPIE) Conference Series} {\bf 10932}, 109320V  (2019).

\bibitem{Oram2020}
K.~{Oram} and Z.~{Ninkov}, ``{Digital micromirror device enabled integral field spectroscopy with the Hadamard transform technique},'' in {\em Optical and Infrared Interferometry and Imaging VII},  P.~G. {Tuthill}, A.~{M{\'e}rand}, and S.~{Sallum}, Eds., {\em Society of Photo-Optical Instrumentation Engineers (SPIE) Conference Series} {\bf 11446}, 114462Y  (2020).

\bibitem{Wells2015}
M.~{Wells}, J.~W. {Pel}, A.~{Glasse}, {\em et~al.}, ``{The Mid-Infrared Instrument for the James Webb Space Telescope, VI: The Medium Resolution Spectrometer},'' {\em PASP} {\bf 127}, 646  (2015).

\bibitem{Sukegawa2023}
T.~{Sukegawa}, H.~{Lin}, and M.~{Bonnet}, ``{Ultra-compact machined slicer IFU},'' in {\em Society of Photo-Optical Instrumentation Engineers (SPIE) Conference Series},  K.~{Minoglou}, N.~{Karafolas}, and B.~{Cugny}, Eds., {\em Society of Photo-Optical Instrumentation Engineers (SPIE) Conference Series} {\bf 12777}, 127773V  (2023).

\bibitem{Danforth2001}
C.~W. {Danforth}, W.~P. {Blair}, and J.~C. {Raymond}, ``{A Detailed Analysis of a Cygnus Loop Shock-Cloud Interaction},'' {\em \aj} {\bf 122}, 938--953  (2001).

\bibitem{DevaudUVContam2018}
G.~Devaud, R.~B. Schweickart, B.~Walther, {\em et~al.}, ``{Europa Ultraviolet Spectrograph: contamination transport modeling to predict mission performance},'' in {\em Systems Contamination: Prediction, Control, and Performance 2018},  C.~E. Soares, E.~M. Wooldridge, and B.~A. Matheson, Eds.,  {\bf 10748}, 107480B, International Society for Optics and Photonics, SPIE  (2018).

\bibitem{Tuttle2014}
S.~{Tuttle}, H.~{Lee}, C.~{Froning}, {\em et~al.}, ``{Builders Instead of Consumers: Training Astronomers in Instrumentation \& Observation},'' {\em arXiv e-prints} , arXiv:1410.2856  (2014).

\bibitem{france2024}
K.~{France}, J.~{Tumlinson}, B.~{Fleming}, {\em et~al.}, ``{The Smallsat Technology Accelerated Maturation Platform-1 (STAMP-1): A Proposal to Advance Ultraviolet Science, Workforce, and Technology for the Habitable Worlds Observatory},'' {\em Journal of Astronomical Telescopes, Instruments, and Systems}   (2024).

\bibitem{pfisterer}
R.~{Pfisterer}, J.~{Harvey}, and J.~{Breckinridge}, ``{The role of narrow-angle forward surface scatter and particulate scatter in exoplanet exploration},'' {\em \prl} {\bf 119}, 131802  (2017).

\end{thebibliography}
\bibliographystyle{spiejour}   


\vspace{1ex}
\noindent Biographies and photographs of the other authors are not available.

\listoffigures
\listoftables

\end{document}